\RequirePackage{fix-cm}

\documentclass[numbook, envcountsect, envcountsame, envcountreset, runningheads, smallextended]{svjour3}       % onecolumn (second format)
\smartqed  % flush right qed marks, e.g. at end of proof
\RequirePackage{bbm,amsmath,amsfonts,amssymb,mathtools}
\RequirePackage[numbers]{natbib}

\usepackage{bbm,eurosym,mathrsfs,enumitem,etoolbox}
\usepackage[english]{babel}
\usepackage[utf8]{inputenc}

\usepackage{graphicx,caption,mathptmx,latexsym,hyperref,xcolor,float}

\usepackage[capitalize]{cleveref}
\usepackage{rccol}

\newtheorem{ass}{Assumption}[section]
\Crefname{ass}{Assumption}{Assumptions}
\Crefname{asslisti}{Assumption}{Assumptions}

\usepackage{booktabs}
\usepackage{siunitx}

\newcommand{\flrn}[1]{\left\lfloor #1 \right\rfloor_n }
\newcommand{\E}{\mathbb E}
\newcommand{\R}{\mathbb R}
\newcommand{\Z}{\mathbb Z}
\newcommand{\N}{\mathbb N}
\newcommand{\Q}{\mathbb Q}
\newcommand{\1}{\mathbbm 1}
\newcommand{\F}{\mathcal F}
\renewcommand{\P}{\mathbb P}
\renewcommand{\d}{\mathrm d}

\usepackage[textsize=footnotesize]{todonotes}

\begin{document}

\title{Second-Order Approximation of Limit Order Books in a Single-Scale Regime
}

\titlerunning{Second-Order Approximation of Limit Order Books}      

\author{Ulrich Horst         \and
        D\"orte Kreher \and Konstantins Starovoitovs %etc.
}

\institute{U.~Horst \at
              Humboldt University Berlin; Department of Mathematics / School of Business and Economics\\
              \email{ulrich.horst@hu-berlin.de}          
           \and
           D.~Kreher \at
              Humboldt University Berlin; Department of Mathematics\\
\email{doerte.kreher@hu-berlin.de}
\and 
K.~Starovoitovs \at
     Humboldt University Berlin; Department of Mathematics\\
\email{starovoitovs@gmail.com}
}

\date{Received: date / Accepted: date}

\maketitle

\begin{abstract}
We establish a first- and second-order approximation for an infinite dimensional limit order book model in a single (``critical'') scaling regime where market and limit orders arrive at a common time scale. With our choice of scaling we obtain non-degenerate first- and second-order approximations for the price and volume dynamics. While the first-order approximation is given by a %standard 
coupled ODE-PDE system, the second-order approximation is %non-standard and 
described in terms of an infinite-dimensional stochastic evolution equation driven by a cylindrical Brownian motion. The driving noise processes exhibit a non-trivial correlation in terms of the model parameters. We prove that the evolution equation has a unique solution and that the sequence of standardized limit order book models converges weakly to the solution of the evolution equation. The proof uses a non-standard martingale problem. We calibrate a linearized model to market data and explain how our model can be used for deriving confidence intervals of portfolio liquidation values.
\end{abstract}

\section{Introduction}

In modern financial markets the overwhelming majority of transactions are settled through electronic limit order books. A limit order book (LOB) is a record, maintained by a specialist or an exchange that contains unexecuted limit orders offering to buy or sell an asset at pre-specified prices that are multiples of a certain tick size.
Incoming buy/sell limit orders can be placed at any price level above/below the best ask/bid price, and incoming market orders are usually matched against standing volume according to a set of precedence rules. In view of its inherent complexity, the detailed working of a LOB is difficult to describe mathematically. 

Scaling limits allow for a tractable description of key macroscopic LOB quantities like prices and aggregate volumes in terms of the underlying (microscopic) dynamics of individual order arrivals and cancellations in a ``high frequency'' regime, when the number of order arrivals and cancellations tends to infinity. Compared to ad hoc PDE or SPDE order book models, scaling-limit approaches single out models with an explicit microscopic foundation. The resulting model coefficients can often be expressed in terms of conditional moments of order arrivals or volume placements that are directly observable from high-frequency market data. This not only improves the interpretability of the models, but also strengthens their empirical relevance and facilitates calibration and validation.

For a Markovian queuing model that describes an order book with finitely many price levels, diffusive high frequency limits for prices and/or volumes have been obtained in\cite{ContPriceDynamicsMarkovian2013, Cont2, HuangErgodicityDiffusivityMarkovian2015a,JedidiStabilityPriceScaling2013} among others. 
A more macroscopic perspective has been adopted in \cite{Keller-ResselStefantypeStochasticMoving2016} and \cite{ZhengZ} where the order book dynamics is described in terms of a coupled system of SPDEs separated by a random interface that can be interpreted as the price (stochastic Stefan problem). An approximation by a reflected SPDE has been obtained in \cite{hambly2020limit}.
To provide microscopic foundations for PDE or SPDE order book models one has to consider order books with a large number of price levels growing to infinity together with a tick size that converges to zero in the high frequency limit. Depending on the scaling assumptions fluid limits \cite{GaoHydrodynamicLimitOrderbook2018, HorstWeakLawLarge2017, HorstScalingLimitLimit2019} and diffusion limits \cite{BayerFunctionalLimitTheorem2017, HorstSecondOrderApproximations2018, HorstDiffusionApproximationLimit2019, HorstScalingLimitLimit2019, KreherJumpDiffusionApproximation2022} have been derived in the literature. 

In this paper, we establish a first- and second-order approximation for LOBs where the second-order approximation can be described in terms of an infinite dimensional stochastic evolution equation driven by correlated noise processes. The correlation operator can be identified by analyzing a novel infinite-dimensional martingale problem and has an explicit representation in terms of the model parameters. 

Following \cite{HorstSecondOrderApproximations2018} we allow the LOB dynamics to depend on volume indicators such as the volume at the top of the book or volume imbalances between the bid and ask side. In that paper, starting from the first-order approximation derived in \cite{HorstWeakLawLarge2017}, the authors derive a second-order approximation for LOBs under the standing assumption that the proportion of market orders and spread placements among all order book events converges to zero in the scaling limit. 
If the first-order approximation of the price process is constant, they were able to rescale by a fast rate corresponding to volume changes, in which case the limiting volume fluctuations could be described by an infinite dimensional SDE. With the first-order approximation of the price being constant, this choice of scaling describes the dynamics when the LOB is observed on a very fast time scale and over very short time periods during which volumes change but prices do not. Allowing for a non-degenerate first-order approximation of the price process required a slower scaling rate. In the slower scaling regime, the second-order approximation for the price and volume density process followed a coupled SPDE system driven by a one-dimensional Brownian motion originating solely from the price fluctuations. 
This choice of scaling describes the LOB dynamics when the book is observed on a much slower timescale and longer time periods, where volume fluctuations average out but price fluctuations persist.  

In contrast to our previous work \cite{HorstWeakLawLarge2017,HorstSecondOrderApproximations2018}, this paper accounts for the empirically well-documented fact that the proportion of market orders among all events does not vanish on intermediate time scales,
and consider a one-sided (for simplicity) LOB model in a single (``critical'') scaling regime in which prices and volumes change on a common time scale; in particular we allow for a fixed ratio of market to limit orders in the scaling limit. Our choice of scaling describes the LOB dynamics on intermediate time scales where neither price nor volume fluctuations average out. 

The first-order approximation of the price-volume process is standard: the process converges to the solution of a coupled system of an ODE that describes the limiting price process, and a first-order PDE that describes the limiting dynamics of the standing volume. Our convergence concept is convergence in probability in a space of distributions, and is hence weaker than the one in \cite{HorstWeakLawLarge2017,HorstLawLargeNumbers2017}, which accounts for the change in the scaling assumptions. The second-order approximation is non-standard. In our setting the rescaled price-volume process converges weakly to the unique solution of an infinite-dimensional linear stochastic evolution equation driven by additive noise with a time-dependent covariance structure. We identify the drift and the covariance operator that both depend on the first-order approximation by analyzing the limit of a sequence of martingale problems associated with the approximating LOB models, and prove that the limiting equation has a unique mild solution induced by a time-dependent evolution family.   
  
To arrive at the second-order approximation, we encounter several mathematical challenges originating from the fact that volume functions shift in space when prices change. This prevents an immediate invocation of standard Grönwall-type arguments to establish moment estimates and hence tightness for our price-volume dynamics. To bypass this problem, by analogy to the linear transport equation, we rewrite the dynamics of the volume function along the stochastic characteristic corresponding to the dynamics of the best bid price process. We first establish pathwise growth estimates for the volume fluctuations starting from time zero by linearizing the dynamics of the residual terms and only then invoke Grönwall arguments. Subsequently, we establish suitable moment estimates that allow us to deduce tightness bounds. Finally, an application of Mitoma's theorem guarantees the weak convergence of the second-order approximation to a distribution-valued process in the Skorokhod space. 

% Structure
The remainder of this paper is structured as follows: in Subsections \ref{subsec: model} and \ref{subsec: results} we introduce the modelling framework and establish the first-order approximation. In Subsection \ref{subsec: fluctuations and main result}, we introduce the stochastic evolution equation corresponding to the second-order approximation, prove its well-posedness and state our main convergence theorem. In Section \ref{subsec:empirics} , we calibrate a linear model for the drift and diffusion parameters to market data and illustrate how our model can be applied to portfolio liquidation problems with endogenous order book shape functions. In Section \ref{sec: preliminaries} we derive preliminary estimates for the discrete dynamics, which are used in Section \ref{sec: moment estimates} to deduce first pathwise growth estimates and then the aforementioned moment estimates. In Section \ref{sec: martingale problem} we prove tightness of the sequence of discrete LOB models and establish martingale relations under the limiting measure, which establishes convergence to the unique mild solution of the system describing the second-order approximation.\\

{\sl Notation.} For real numbers $a$ and $b$ we write $a\lesssim b$ whenever there exists a positive constant $C<\infty$ such that $a \leq Cb$. If the constant depends on some other parameter $n$, then we write $a\lesssim_n b$. For a Banach space $E$ we denote by $C([0,T], E)$ and $D([0,T], E)$ the space of continuous and càdlàg $E$-valued processes, respectively. For $s\in\N$ and $U:=[-L,L]\subset\R$ we denote by $H^s(U)$ the classical Sobolev space $W^{s, 2}(U)$. For $s\in\N$ the Sobolev spaces of negative order $H^{-s}(U)$ are dual spaces of $H^s(U)$ endowed with the norm
\[
\|f\|^2_{-s} = \sum_{k=1}^\infty f(e_k)^2,
\]
where $\{e_k:k\in\N\}$ is a complete orthonormal system of $H^s(U)$. We also consider Sobolev spaces of fractional order $s>0$ defined on the torus $\R / 2\pi \Z$ with the norm
\[
\|f\|^2_s = \sum_{k\in\Z} \left(1 + k^2\right)^s \langle f, e^{i k x} \rangle^2.
\]
Also note that
\[
C^\infty \subset \mathcal E := \bigcap_{n\in\Z} H^n \subset \dots  H^1 \subset L^2 \subset H^{-1} \subset \dots \subset \bigcup_{n\in\Z} H^n =: \mathcal E'.
\]
For a scaling parameter $\Delta^{(n)}$ and a given time interval $[0,T]$ we put $\delta_{n,k} := k\Delta^{(n)}$, $T_{n}:=\lfloor T / \Delta^{(n)}\rfloor$ and consider the price grid $x^{(n)}_k:=k\Delta^{(n)}$ for $k\in\Z$. For $s \in [0,T]$ we write $\flrn{s} := \Delta^{(n)} \lfloor s / \Delta^{(n)}\rfloor$. Moreover, for all $n\in\N$ and $x\in\R$ we define an interval 
\[
    I^{(n)}(x):=\left(x_j^{(n)}, x_{j+1}^{(n)}\right] \quad\text{for}\quad x_j^{(n)}<x \leq x_{j+1}^{(n)}
\]
Furthermore, we introduce the translation operators $T^{(n)}_\pm$ which act on a function $ f:U\rightarrow U$ according to
\begin{equation}
\label{eq: translation operators}
T_{-}^{(n)}(f)(\cdot):=f\left(\cdot-\Delta ^{(n)}\right)\1_{[-L+\Delta^{(n)},L]}(\cdot), \quad T_{+}^{(n)}(f)(\cdot):=f\left(\cdot+\Delta ^{(n)}\right)\1_{[-L,L-\Delta^{(n)}]}(\cdot).
\end{equation}and introduce the left/right finite differences $\nabla^{(n)}_\pm := \pm \frac{\left( T^{(n)}_{\pm} - 1 \right)}{\Delta^{(n)}}$. For any $\varphi\in H^2(U)$,
\begin{equation}
\label{eq: varphi discrete derivative}
\left\|\nabla^{(n)}_\pm \varphi - \varphi'\right\|_{L^2} \lesssim \Delta^{(n)} \|\varphi\|_{H^2}.
\end{equation}

Throughout, we will assume that all test functions in $H^s(U),\ s\in\N,$ satisfy zero boundary conditions. We denote by $C^2_b(\R^k,\R)$ the space of twice continuously differentiable functions $f:\R^k\to\R$ with uniformly bounded derivatives up to order $2$. 

For a stochastic process in discrete time $(X^{(n)}_k)_{k=0,\dots,T_n}$ we will denote its increments by $\delta X^{(n)}_k := X^{(n)}_k-X^{(n)}_{k-1},\ k=1,\dots,T_n$. Finally, we will write $C^{(n)}$ and $K^{(n)}$ for any generic random variables such that $C^{(n)} \rightarrow 0$ in probability and $K^{(n)}\rightarrow K$ in probability for some $0 < K < \infty$, respectively, and $C$ for any generic constant. Note that $C^{(n)},\ K^{(n)},$ and $C$ may vary from line to line.

%%%%%%%%%%%%%%%%%%%%%%%%%%%%%%%%%%%%%%%%%%%%%%%%%%%%%%%%%%%%%%
%%%%%%%%%%%%%%%%%%%%%%%%%%%%%%%%%%%%%%%%%%%%%%%%%%%%%%%%%%%%%%
%%%%%%%%%%%%%%%%%%%%%%%%%%%%%%%%%%%%%%%%%%%%%%%%%%%%%%%%%%%%%%

\section{Model and main results}

Our goal is to derive a second-order approximation for the sequence of LOB models introduced in \cite{HorstWeakLawLarge2017} in a single-scale regime where the order inter-arrival times, the tick size, and the average size of a limit order placement/cancellation are all scaled by the \textit{same} scaling parameter $\Delta^{(n)}$. We prove that the second-order approximation can be described by a stochastic evolution equation with linear drift and time-dependent diffusion operator. 

%%%%%%%%%%%%%%%%%%%%%%%%%%%%%%%%%%%%%%%%%%%%%%%%%%%%%%%%%%%%%%
%%%%%%%%%%%%%%%%%%%%%%%%%%%%%%%%%%%%%%%%%%%%%%%%%%%%%%%%%%%%%%
%%%%%%%%%%%%%%%%%%%%%%%%%%%%%%%%%%%%%%%%%%%%%%%%%%%%%%%%%%%%%%

\subsection{The microstructure model}\label{subsec: model}

Following \cite{HorstWeakLawLarge2017,HorstSecondOrderApproximations2018}, we assume that the dynamics of the buy side of the LOB in the $n$-th model is described by a c\`adl\`ag, $\R\times L^2(U)$-valued stochastic process 
\[
    S^{(n)}=\left(S^{(n)}(t)\right)_{0\leq t\leq T} 
\]    
supported on a probability space $(\Omega^{(n)}, \mathcal F^{(n)}, \P^{(n)})$, where $U:=[-L,L]$ for $L:=M+T$ and some $M>0$. The $\R$-valued component of the process corresponds to the best bid price, while the $L^2(U)$-valued component corresponds to the relative buy side volume density function which indicates the liquidity available at different price levels, relative to the best bid price. 

At $t=0$ the state of the LOB is deterministic and supported on $[-M,M]$, i.e.
\[s_0^{(n)}=\left(B_0^{(n)},u^{(n)}_{0}\right) \in \mathbb{R} \times L^2(U).\]
The state of the book changes due to incoming market and limit orders and cancellations. The model evolves in event time, i.e.~in the $n$-th model there are $T_n=\lfloor T/\Delta^{(n)}\rfloor$ order book events taking place at times
\[
    t_k^{(n)}:=k\Delta^{(n)},\quad k=1,\dots,T_n,
\]
where we set $t_0^{(n)}=0$. The state of the book after $k$ events is denoted by $S^{(n)}_k:=(B_k^{(n)},u_k^{(n)});$ the piecewise constant, continuous-time interpolation is given by
\[
    S^{(n)}(t):=\left(B^{(n)}(t),u^{(n)}(t)\right):=\left(B_k^{(n)},u_k^{(n)}\right)\quad\text{for} \quad t\in\left[t_k^{(n)},t_{k+1}^{(n)}\right)\cap[0,T].
\]

\begin{remark}
The assumption that the model evolves in event time is made for convenience only. Indeed, as in \cite{HorstLawLargeNumbers2017} and \cite{KreherJumpDiffusionApproximation2022}, one could instead work in calendar time and define the event times through an additional stochastic process $\tau_k^{(n)}=\tau_{k-1}^{(n)}+\Delta^{(n)} \varphi_k^{(n)}$. Under suitable assumptions on the first two conditional moments of $\varphi_k^{(n)}$, one could then prove a first and second order approximation for the enlarged state process $\bar S^{(n)}_k:=(\tau_k^{(n)},B_k^{(n)},u_k^{(n)})$ by first showing convergence of the time-changed process evolving in event time (as done in this paper) and afterwards applying the time-change theorem as in \cite{KreherJumpDiffusionApproximation2022}. This would require the joint convergence of the inverse time-change and the time-changed order book process, but not their independence. 
\end{remark}

Orders can be submitted at the {\sl relative} price levels $x^{(n)}_j=j\Delta^{(n)}\in[-M,M]$ for suitable $j\in\Z$. The $L^2(U)$-valued volume density function $u_k^{(n)}$ is a c\`agl\`ad step function on the price grid $\{x_j^{(n)},\ j\in\Z\}\cap U$. The standing volume available at time $t_k^{(n)}$ at the relative price level $x_j^{(n)}$, i.e.~at the price level $j$ ticks below the best bid is given by 
\begin{equation*}
\int_{x_{j-1}^{(n)}}^{x_j^{(n)}}u_k^{(n)}(x)\d x=\Delta^{(n)}u^{(n)}\left(x_j^{(n)}\right).
\end{equation*}

\begin{remark}
We emphasize that the volume density functions are defined on the whole interval $[-L,L]$ to model the arrival of spread placements. 
The restriction of the function $u^{(n)}(t,\cdot)$ to the interval $[-L, 0]$ corresponds to the actual buy side of the order book at time $t$; the restriction to $(0,L]$ corresponds to the \emph{shadow book}, which specifies volumes placed into the spread should such events occur next; we refer to \cite{HorstWeakLawLarge2017,HorstLawLargeNumbers2017} for further details on the working of the shadow book. 
\end{remark}

At each time $t^{(n)}_k$, one of the following three events occurs, changing the buy side of the LOB:
\begin{enumerate}[label=(\Alph*)]
\item arrival of a market sell order of size $\Delta^{(n)}u_{k-1}^{(n)}(0)$, which removes the entire volume at the top of the book. In this case the best bid price decreases by one tick and the relative volume density function shifts one tick to the right.
\item placement of a buy limit order of the size $\Delta^{(n)}u_{k-1}^{(n)}(\Delta ^{(n)})$ into the spread one tick above the best bid price. In this case the best bid price increases by one tick, and the relative volume density function shifts one tick to the left.
\item placement/cancellation of a buy limit order of size $\Delta^{(n)}\omega^{(n)}_k$ at the relative price level $\eta^{(n)}_k$ (cancellation corresponds to $\omega^{(n)}_k < 0$).
\end{enumerate}

\begin{remark}
  For simplicity we only consider orders of size $O(\Delta^{(n)})$. As there are $T_n:=\lfloor T/\Delta^{(n)}\rfloor$ order book events, this will guarantee that a law of large number type results can be derived as first order approximation. One could also allow for limit orders of larger size $O(\Delta^{(n)})^\alpha$ with $0<\alpha<1$, if one assumes that they only arrive with probability $O(\Delta^{(n)})^{1-\alpha}$ at each time $t_k^{(n)}$, i.e.~if their proportion among all order book events vanishes in the high-frequency limit.   
\end{remark}

While the assumption that incoming market orders match precisely against the liquidity at the top of the book has also been made in \cite{BayerFunctionalLimitTheorem2017, HorstSecondOrderApproximations2018, HorstDiffusionApproximationLimit2019, HorstWeakLawLarge2017, HorstScalingLimitLimit2019, KreherJumpDiffusionApproximation2022}, it is made primarily for mathematical convenience. There is some empirical evidence, though, that this assumption is not too restrictive. In an empirical study the authors of \cite{Farmer} found that in their data sample around 85\% of the sell market orders that lead to price changes match exactly the size of the volume standing at the best bid price.\\

The randomness in the $n$-th model is coming from a field of random variables 
\[
    \left(\phi^{(n)}_k, \eta^{(n)}_k, \omega^{(n)}_k \right)_{k,n\in\N} \in \{A,B,C\}\times\left\{x^{(n)}_j:j\in\Z\right\}\cap [-M,M]\times \R.
\]
The random variable $\phi^{(n)}_k$ corresponds to the type of the event taking place at time $t^{(n)}_k$. 
The random variable $\eta^{(n)}_k$ determines the relative price level of the placement respectively cancellation of a limit order of size $\Delta^{(n)} \omega^{(n)}_k$ if $\phi^{(n)}_k = C$. It is more convenient to work with a real-valued random variable $\pi^{(n)}_k$ s.t.~ $\eta_{k}^{(n)}:=\Delta x^{(n)}\lceil \pi_{k}^{(n)} / \Delta x^{(n)}\rceil\wedge M$, so that the discrete dynamics can be equivalently described in terms of
\[
    \left(\phi^{(n)}_k, \pi^{(n)}_k, \omega^{(n)}_k \right)_{k,n\in\N} \in \{A,B,C\} \times [-M,M] \times \R.
\]
For $I\in\{A,B,C\}$ we introduce the shorthand notation $\1^{(n),I}_k := \1_I\left(\phi^{(n)}_k\right)$. The volume density increments can thus be described in terms of the placement operator
\[
    J_{k}^{(n),C}(\cdot):=\1^{(n),C}_k \frac{\omega_{k}^{(n)}}{\Delta ^{(n)}} \1_{I^{(n)}\left(\pi_{k}^{(n)}\right)}(\cdot).
\]
In terms of the translation operators defined in \eqref{eq: translation operators} the discrete dynamics can thus be summarized by the following stochastic difference equations:
\begin{equation}
\label{eq: discrete dynamics}
\begin{aligned}
B_{k}^{(n)} &=B_{k-1}^{(n)}+\Delta^{(n)} \left(\1^{(n),B}_k-\1^{(n),A}_k\right) \\
u_{k}^{(n)} &=u_{k-1}^{(n)}+\left(T_{-}^{(n)}-I\right)\left(u_{k-1}^{(n)}\right) \1^{(n),A}_k+\left(T_{+}^{(n)}-I\right)\left(u_{k-1}^{(n)}\right) \1^{(n),B}_k+\Delta^{(n)} J_{k}^{(n), C}
\end{aligned}
\end{equation}

We allow the order flow dynamics to depend on a \emph{volume indicator} of the form
\[
    Y^{(n)}_k := \left\langle h, u^{(n)}_k\right\rangle,
\]
where $h \in H^4(U)$ is supported on $[-L,0]$. For example, the volume indicator can extract the volume standing at the top of the book. In this way one can account for the empirically well-documented fact that volumes at the top of the book (and volume imbalances in a two-sided model) are important determinants of price movements. For simplicity, we only consider one volume indicator in the following, but the analysis can easily be extended to multiple indicators. 

\begin{remark}
In our model, volume indicators are key determinants of order book dynamics. There is substantial empirical evidence supporting this approach. Biais et al. \cite{Biais1995}, Cont at al.~\cite{ContPriceImpactOrder2014}, Griﬃths et al. \cite{Griffiths2000}, Hautsch and Huang \cite{HautschHuang2012} among many others show that order display aﬀects trading dynamics as market participants observe changes in the order book and adapt their trading strategies accordingly. Cont et al. \cite{Cont2} consider a queueing-theoretic order book model within which they estimate the probability of various events, conditional on the state of the order book, such as mid-price changes and top-of-the book order execution before the next price movement. Ranaldo \cite{Ranaldo2004}, Cao et al. \cite{CaoHanschWang2009}, and Esser and M\"onch \cite{EsserMoench2007}, among others, provide empirical evidence that changes in order-imbalances at the top of the book are a particularly important driver of market dynamics, especially of incoming order flows and price movements. Following up on these insights, Cebiroglu and Horst \cite{CH}, Chen et al. \cite{CHEN201889}, 
and Hollifield et al. \cite{HollifieldMillerSandasSlive2006} consider models of optimal order display where traders face a tradeoff between displaying orders (incurring adverse market impact such as spread placements) and hiding parts of their orders at the expense of losing time priority.    
\end{remark}

%%%%%%%%%%%%%%%%%%%%%%%%%%%%%%%%%%%%%%%%%%%%%%%%%%%%%%%%%%%%%%
%%%%%%%%%%%%%%%%%%%%%%%%%%%%%%%%%%%%%%%%%%%%%%%%%%%%%%%%%%%%%%
%%%%%%%%%%%%%%%%%%%%%%%%%%%%%%%%%%%%%%%%%%%%%%%%%%%%%%%%%%%%%%

\subsection{Assumptions and first-order approximation}\label{subsec: results}

Our key assumption is that the main model parameters -- the tick size, the time between order arrivals, and the average limit order size -- scale by the same parameter $\Delta^{(n)}$. In particular, we do \textit{not} require a vanishing proportion of A,B-events in the scaling limit as in \cite{HorstLawLargeNumbers2017,HorstWeakLawLarge2017,HorstSecondOrderApproximations2018}. 

\begin{remark}
The assumption of a vanishing proportion of A,B-events is very convenient from a mathematical perspective (cf.~\cite{HorstLawLargeNumbers2017,HorstWeakLawLarge2017,HorstSecondOrderApproximations2018}, but is difficult to justify empirically as shown by Figure \ref{spread}. This figure displays the intraday evolution of the proportion of spread placements among all orders for all NASDAQ traded stocks for the month of March 2016. We see that the proportion of spread placements is quite high when markets open and then varies around 5\% for the rest of the day. 
\begin{figure}[h]
\centering
\includegraphics[width=7cm]{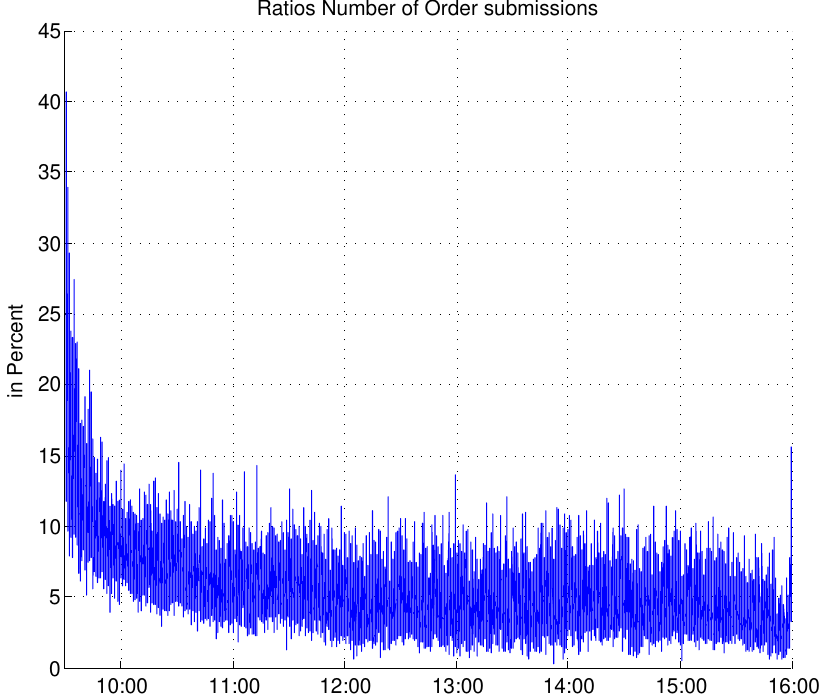}
\caption{Average percentage of spread placement per second (source: \cite{BayerFunctionalLimitTheorem2017}).}
\label{spread}
\end{figure}
\end{remark}
We also require a series of assumptions on the conditional order arrival dynamics. To state our standing assumptions we define, for every $n\in \N$ and $k = 0, 1, \dots, T_n$ the $\sigma$-fields $\F ^{(n)}_k = \sigma ( S^{(n)}_j: j\leq k )$ and denote the corresponding filtration $\mathbb F^{(n)} := (\F^{(n)}_k)_{k\in\N}$.\footnote{We note that $\mathbb F^{(n)}$ is not the market filtration as it also contains information about the shadow book and thus about future spread placements. However, Assumptions \ref{assum: p}, \ref{assum: f}, and \ref{assum: g} below guarantee that this extra information does not enter the LOB dynamics.} 
We start with the following standard assumption on the initial state.

\begin{ass}\label{assum: initial values}
For all $n\in\N$ the initial volume density functions $u^{(n)}_0:U\to\R_+$ are step functions with step size $\Delta^{(n)}$, which satisfy the following conditions:
\begin{enumerate}[label=(\roman*)]
\item\label[ass]{assum: initial values: compact support} $u_0^{(n)}$ are supported on $[-M,M]$ for all $n\in\N$; 
\item $u^{(n)}_0$ are uniformly bounded in the uniform norm, i.e. $\sup_{n\in\N} \left\|u^{(n)}_0\right\| _{\infty}<\infty$;
\item\label{assum: initial values: u^n_0 - u_0 and B^n_0 - B_0} there exist $B_0 \in \R$ and a function $u_0\in L^2(U)\cap C^1(U)$ such that
\[
\begin{aligned}
\left\| u^{(n)}_0 - u_0 \right\|_{L^2} = O\left(\Delta^{(n)}\right)\quad\text{and}\quad|B^{(n)}_0-B_0|&=o(\Delta^{(n)})^{1/2}.
\end{aligned}
\]
\end{enumerate}
\end{ass}

We assume that placements and cancellations take place at a finite distance from the best bid price and that placed volumes are almost surely bounded.

\begin{ass}\label{assum: bound on pi and omega}
For all $n \in \N$ and $k\leq T_n$, 
\begin{enumerate}[label=(\roman*)]
\item \label{assum: bound on pi and omega: pi} the order placement price is bounded a.s., i.e.~$\P\left(\left|\pi^{(n)}_k\right|>M\right)= 0$;
\item \label{assum: bound on pi and omega: omega} 
the order volume is bounded a.s., i.e.~there is $K>0$ such that 
$\P\left(\left|\omega^{(n)}_k\right|>K\right)= 0.$
\end{enumerate}
\end{ass}

Note that the discrete volume functions $u^{(n)}$ defined in \eqref{eq: discrete dynamics} is supported on 
\[
\mathcal T = \{(t,x)\in \R^2:t\in[0,T], x\in[-M-t,M+t]\}\subset [0,T]\times U,
\]
since the initial volume density function $u^{(n)}_0$ is supported on $[-M,M]$ due to Assumption \cref{assum: initial values}~\ref{assum: initial values: compact support}, the order placement price $\pi^{(n)}_k$ is also supported on $[-M, M]$ due to \cref{assum: bound on pi and omega}~\ref{assum: bound on pi and omega: pi} and the volume density function $u^{(n)}$ shifts at most by one tick  $\Delta^{(n)}$ over the time increment $\Delta^{(n)}$. 

The next assumption states that conditional distribution of the next event type depends on the history of the LOB only through the current best bid price and the volume indicator.
\begin{ass}
\label{assum: p}
For $I\in\{A,B\}$ there exist functions $p^{(n),I},p^{I}:\R\times\R\rightarrow[0,1]$ s.t.
\begin{enumerate}[label=(\roman*)]
\item \label{assum: p: definition of p} $\P\left(\left.\phi^{(n)}_k=I\right|\F^{(n)}_{k-1}\right)=p^{(n),I} \left(B^{(n)}_{k-1}, Y^{(n)}_{k-1}\right)$;
\item \label{assum: p: sup} the function $p^I$ is twice continuously differentiable with uniformly bounded derivatives, i.e.~$p^I\in C^2_b(\R\times\R,\R)$; 
\item \label{assum: p: p^n-p} for any $n\in\N$ it holds that 
\[
    \sup_{(b,y)\in\R^2}\left|p^{(n),I}(b,y) - p^{I}(b,y)\right| = O\left(\Delta^{(n)}\right)^{1/2}.
\]

\end{enumerate}
We introduce the notation 
\[
    p^{B\pm A}:=p^{B} \pm p^{A} \quad \text{\emph{and}} \quad p^{(n), B\pm A}:=p^{(n),B} \pm p^{(n),A}. 
\]
If there is no confusion, we drop the superscript $B-A$ and denote the conditional mean price imbalances by 
\begin{equation}
\label{def: p}
    p:=p^{B-A} \quad \mbox{\emph{and}} \quad p^{(n)}:=p^{(n),B-A}.
\end{equation}
\end{ass}

We emphasize that \cref{assum: p} allows for a non-vanishing proportion of A,B-events, i.e.~price changes and volume changes evolve on the same time scale. This is different from \cite{HorstWeakLawLarge2017, HorstSecondOrderApproximations2018} and will allow us to derive a non-degenerate first- and second-order approximation for prices and volumes, simultaneously. 

We need an analogous assumption for the first conditional moment of the order flow.

\begin{ass}\label{assum: f}
There exist functions $f^{(n)},f:\R\times\R\rightarrow L^2(U)$ such that
\begin{enumerate}[label=(\roman*)]
\item \label{assum: f: definition of f} $\E \left[\left.\1_C\left(\phi^{(n)}_k\right)\omega^{(n)}_k\1_{I^{(n)}\left(\pi^{(n)}_k\right)}(\cdot)  \right|\F^{(n)}_{k-1}\right] =\Delta^{(n)} f^{(n)}\left(B^{(n)}_{k-1}, Y^{(n)}_{k-1}\right)$;
\item \label{assum: f: sup of L^infty norm of f^n} the functions $f^{(n)}$ are uniformly bounded in the uniform norm, i.e. \[ \sup_{n\in\N} \sup _{(b, y) \in \R^2 }\left\|f^{(n)}(b, y)\right\| _{\infty} < \infty;\]
\item \label{assum: f: sup of L^2 norms of f} the function $f$ is twice continuously differentiable and 
\begin{align*}
&\sup _{b, y}\left\{\|f(b, y)\|_{L^{2}}+\left\|f_{b}(b, y)\right\|_{L^{2}}+\left\|f_{y}(b, y)\right\|_{L^{2}}\right.\\
&\qquad\qquad\qquad\qquad\left.+\left\|f_{b b}(b, y)\right\|_{L^{2}}+\left\|f_{b y}(b, y)\right\|_{L^{2}}+\left\|f_{y y}(b, y)\right\|_{L^{2}}\right\}<\infty;
\end{align*}
\item \label{assum: f: convergence of f^n to f} for any $n \in \N$ it holds that \[\sup _{(b, y) \in \mathbb{R}^{2}}\left\|f^{(n)}(b, y)-f(b, y)\right\|_{L^{2}}=O\left(\Delta ^{(n)}\right).\]
\end{enumerate}
\end{ass}

%%%%%%%%%%%%%%%%%%%%%%%%%%%%%%%%%%%%%%%%%%%%%%%%%%%%%%%%%%%%%%
%%%%%%%%%%%%%%%%%%%%%%%%%%%%%%%%%%%%%%%%%%%%%%%%%%%%%%%%%%%%%%
%%%%%%%%%%%%%%%%%%%%%%%%%%%%%%%%%%%%%%%%%%%%%%%%%%%%%%%%%%%%%%

The above assumptions readily allow us to establish the following first-order approximation of our LOB model.

\label{subsec: first-order approximation}

\begin{proposition}
\label{prop: law of large numbers}
For any $\varepsilon > 0$ and any test function $\zeta \in \R \times H^{1}(U)$, it holds that
\begin{equation*}
\lim _{n \rightarrow \infty} \mathbb{P}\left(\sup _{0 \leq t \leq T} \left\langle S^{(n)}(t)-S(t), \zeta \right\rangle > \varepsilon\right)=0,
\end{equation*}
where the deterministic process $S:[0,T]\rightarrow \R\times L^2(U)$ satisfying
\begin{equation*}
\sup_{t\in[0,T]}\|S(t)\|^2_{\R\times L^2} < \infty
\end{equation*}
is the unique classical solution to the coupled ODE-PDE system
\begin{equation}
\label{eq: first-order approximation}
\left\{\begin{array}{l}
\dot B(t) =p\big(B(t),Y(t)\big) \\
u_{t}(t, x) =p\big(B(t),Y(t)\big) u_{x}(t, x)+f\big(B(t),Y(t)\big)(x)
\end{array}\right.
\end{equation}
with initial condition $S(0)=s_0$ and $Y(t) = \langle u(t,\cdot), h \rangle$ is the volume indicator function. 
\end{proposition}

\begin{proof}

Our choice of scaling corresponds to the case $\alpha=1$ in \cite{HorstWeakLawLarge2017}. The proof is identical to the proof therein, except for the calculation at the bottom of p.~14, which needs to be adapted: for any function $\varphi \in H^1(U)$ by Kondrachov's embedding theorem it holds that
\begin{equation*}
\|\varphi\|_{L^\infty} \lesssim \|\varphi\|_{H^1}.
\end{equation*}
Thus, we obtain
\begin{equation*}
\begin{aligned}
\left(\frac{1}{\Delta^{(n)}} \int _{I^{(n)}\left(\pi^{(n)}_k\right) } \varphi(x) \d x \right)^2 \leq \left(\frac{1}{\Delta^{(n)}} \int \|\varphi\|_{L^\infty} \1_{I^{(n)}\left(\pi^{(n)}_k\right) }(x) \d x\right)^2 \leq  \| \varphi \|^2_{L^\infty} \lesssim \|\varphi\|^2_{H^1}.
\end{aligned}
\end{equation*}
Using that $\left|\omega^{(n)}_k\right|\leq K$ a.s., we conclude that
\[
\begin{aligned}
&\E\left[\left(\Delta^{(n)} \left\langle \left(T_{-}^{(n)}\right)^{\sum_{i=0}^{k-1} \1_{i}^{(n),B-A}} \left(J^{(n),C}_k - f^{(n)}\right) \left(S^{(n)}_{k-1}\right), \varphi \right\rangle \right)^2 \right] \\
&\quad\leq \left(\Delta^{(n)}\right)^2 \E \left[\1^{(n),C}_k \left(\omega_k^{(n)}\right)^2 \left(\frac{1}{\Delta^{(n)}} \int _{I^{(n)}\left(\pi^{(n)}_k\right) } \varphi(x) \d x\right)^2\right] \leq \left(\Delta^{(n)}\right)^2 K^2 \|\varphi\|_{H_1}^2.
\end{aligned}
\]
Hence, the volume fluctuations vanish as $\Delta^{(n)} \to 0$ and so one can proceed further as in \cite{HorstWeakLawLarge2017} by applying a weak law of large numbers for triangular martingale difference arrays.\qed
\end{proof}

Finally, we specify an assumption for the second conditional moment of the order flow, which will be needed to derive the second order approximation. 

\begin{ass}
\label{assum: g}
There exist functions $g^{(n)}, g: \R\times\R \rightarrow L^1(U)$ such that
\begin{enumerate}[label=(\roman*)]
\item \label{assum: g: definition of g} $\E \left[\left.\1_C\left(\phi^{(n)}_k\right)\left(\omega^{(n)}_k\right)^2\1_{I^{(n)}\left(\pi^{(n)}_k\right)}(\cdot) \right|\F^{(n)}_{k-1}\right] = \Delta^{(n)} g^{(n)}\left(B^{(n)}_{k-1},Y^{(n)}_{k-1}\right)$;
\item \label{assum: g: sup} the functions $g^{(n)}$ are uniformly bounded in the uniform norm \[\sup_{n\in\N}\,\sup_{(b, y)\in\R^2} \left\|g^{(n)}(b, y)\right\|_{\infty} < \infty;\]
\item \label{assum: g: convergence of g^n to g} it holds that 
\[
    \sup _{(b, y)\in\R^2} \int_{\mathbb{R}}\left|g^{(n)}(b, y ; x)-g(b, y ; x)\right| \d x \rightarrow 0;
\]
\item \label{assum: g: fractional sobolev norm} the function $g$ belongs to $H^s(U)$ for some $s>\frac12$.
\end{enumerate}
\end{ass}

%%%%%%%%%%%%%%%%%%%%%%%%%%%%%%%%%%%%%%%%%%%%%%%%%%%%%%%%%%%%%%
%%%%%%%%%%%%%%%%%%%%%%%%%%%%%%%%%%%%%%%%%%%%%%%%%%%%%%%%%%%%%%
%%%%%%%%%%%%%%%%%%%%%%%%%%%%%%%%%%%%%%%%%%%%%%%%%%%%%%%%%%%%%%

\subsection{Second-order approximation}
\label{subsec: fluctuations and main result}

In this section, we describe the second-order approximation for our LOB model. To this end, we introduce the rescaled discrete fluctuation processes 

\begin{equation}
\label{eq: definition of Z^B and Z^u}
Z_{k}^{(n), B}:=\frac{B_{k}^{(n)}-B\left(t_{k}^{(n)}\right)}{\left(\Delta^{(n)}\right)^{1 / 2}}, \quad Z_{k}^{(n), u}(\cdot):=\frac{u_{k}^{(n)}(\cdot)-u\left(t_{k}^{(n)}, \cdot\right)}{\left(\Delta^{(n)}\right)^{1 / 2}}, \quad k=0, 1, \dots, T_{n},
\end{equation}
along with the fluctuations of the volume indicator
\begin{equation}
\label{eq: definition of Z^Y}    
Z_{k}^{(n), Y}:=\frac{Y_{k}^{(n)}-Y\left(t_{k}^{(n)}\right)}{\left(\Delta^{(n)}\right)^{1 / 2}}=\frac{\left\langle h, u_{k}^{(n)}-u\left(t_{k}^{(n)}\right)\right\rangle}{\left(\Delta^{(n)}\right)^{1 / 2}}, \quad k=0, 1, \dots, T_{n}.
\end{equation}
The corresponding continuous time interpolations on $[0,T]$ are denoted by
\[
Z^{(n), B}(t):=\frac{B^{(n)}(t)-B(t)}{\left(\Delta^{(n)}\right)^{1 / 2}}, \quad Z^{(n), u}(t, \cdot):=\frac{u^{(n)}(t, \cdot)-u(t, \cdot)}{\left(\Delta^{(n)}\right)^{1 / 2}}.
\]
Moreover, we set 
$$
Z^{(n), Y}(t):=\left\langle h, Z^{(n), u}(t)\right\rangle\quad\text{and}\quad Z^{(n)}(t):=\left(Z^{(n),B}(t), Z^{(n),u}(t)\right).
$$

%%%%%%%%%%%%%%%%%%%%%%%%%%%%%%%%%%%%%%%%%%%%%%%%%%%%%%%%%%%%%%
%%%%%%%%%%%%%%%%%%%%%%%%%%%%%%%%%%%%%%%%%%%%%%%%%%%%%%%%%%%%%%
%%%%%%%%%%%%%%%%%%%%%%%%%%%%%%%%%%%%%%%%%%%%%%%%%%%%%%%%%%%%%%

\subsubsection{Limiting dynamics}\label{sec:limit}

In what follows we set $\mathcal H := \R \times L^2(U)$. Our goal is to establish the weak convergence of the sequence $\{Z^{(n)}: n \in \mathbb N\}$ in the Skorokhod space $D([0,T],\R\times H^{-4}(U))$ to the unique weak solution of the infinite-dimensional linear stochastic differential equation of the form
\begin{equation}
\label{eq: abstract spde}
\begin{split}
\d Z_t &= A(t) Z_t  \d t + F(t, Z_t)  \d t + \sigma(t)  \d W_t, \\
Z_0 &= z_0,
\end{split}
\end{equation}
where $z_0 = 0$ and $\{W_t: t\in[0,T]\}$ is an $\mathcal H$-valued cylindrical Brownian motion. 

Since there are no canonical candidates for the drift and volatility operators, identifying the respective operators is challenging and will be achieved by an approximation argument in later sections. To specify the operators we use the following shorthand notation: given the first-order approximation $(B,Y)$ of the price and volume indicator process process, we put 
\[
\begin{aligned}
    p &= p(t) := p\big(B(t),Y(t)\big),& \quad f &= f(t) := f\big(B(t),Y(t)\big), \\
    p_b &= p_b(t) := \frac{\partial}{\partial b}p\big(B(t),Y(t)\big),& \quad
    p_y &= p_y(t) := \frac{\partial}{\partial y}p\big(B(t),Y(t)\big), \\   
    f_b& = f_b(t) := \frac{\partial}{\partial b}f\big(B(t),Y(t)\big), &\quad
    f_y &= f_y(t) := \frac{\partial}{\partial y}f\big(B(t),Y(t)\big),
\end{aligned}
\]
where $p$ was defined in \eqref{def: p} and $f$ was introduced in \cref{assum: f}. It turns out that the drift operators in \eqref{eq: abstract spde} are defined as follows: for all $t\in[0,T]$,
$$
A(t): D(A) \rightarrow \mathcal H, \qquad Z \mapsto A(t)Z := \left( \begin{matrix} 0\\ p \partial_x Z^u \end{matrix}\right),
$$ 
is an unbounded operator with time-invariant domain $D(A) := \R\times H^1(U) \subset \mathcal H$, and
$$
    F(t): \mathcal H \rightarrow \mathcal H, \qquad Z \mapsto F(t,Z) := \left(\begin{matrix}U(t,Z) \\ V(t,Z) \end{matrix}\right),
$$ 
is a linear operator given by
\[
\left(\begin{matrix}U(t,Z) \\ V(t,Z) \end{matrix}\right) := \left(\begin{matrix} p_b Z^B + p_y \langle Z^u, h\rangle \\
p_b Z^B \partial_x u + p_y \langle Z^u, h \rangle \partial_x u + Z^B f_b + \langle Z^u, h \rangle f_y \end{matrix}\right).
\]

We prove in Proposition \ref{prop: properties of the variance and volatility operators} that the volatility operator $\sigma$ belongs to the space $L^2([0,T], \mathcal L_2(\mathcal H))$. To explicitly define the operator we introduce the function
\[
    \sigma_0^2 := \sigma_0^2(t) := p^{B+A}(B(t),Y(t)) - p(B(t),Y(t))^2.
\]
It turns out that the diffusion operator $\sigma$ is the square-root of the non-negative operator 
\[
  \Sigma(t): \mathcal H \to \mathcal H; \qquad \left(\begin{matrix}a \\ \varphi \end{matrix} \right) \mapsto \left(\begin{matrix} P(t)&Q(t) \\ Q(t)&R(t) \end{matrix}\right) \left(\begin{matrix}a \\ \varphi \end{matrix} \right),
\]
whose components are defined in terms of the functions
\[
\begin{aligned}
    & P: [0,T] \to \mathbb R; \qquad \qquad t \mapsto \sigma_0^2(t) \\
    & Q: [0,T] \to L^2(U); \qquad t \mapsto \sigma_0^2(t) \partial_x u(t;\cdot) - p(t)f(t)
\end{aligned}
\]
and the family of linear operators for $t \in [0,T]$,
\[
\begin{aligned}
R(t): L^2(U) &\to L^2(U); \\ \varphi &\mapsto \sigma_0^2\left\langle\partial_x u, \varphi\right\rangle \partial_x u - p\langle f, \varphi\rangle \partial_x u - pf \left\langle\partial_x u, \varphi\right\rangle + g \varphi - \langle f, \varphi\rangle f.
\end{aligned}
\]
For later reference, we also define for $t\in[0,T]$ the family of linear operators
\[
\begin{aligned}
r(t): L^2(U) &\to L^2(U); \\ \varphi &\mapsto g \varphi - \langle f, \varphi\rangle f.
\end{aligned}
\]

\begin{remark}
    We note that the drift operators $A$ and $F$ are the same as in \cite[Theorem 6.9]{HorstSecondOrderApproximations2018}, i.e.~when rescaling their model with the slow time scale of price changes, while the covariance operator $\Sigma$ is the same as in \cite[Theorem 5.10]{HorstSecondOrderApproximations2018}, i.e.~when rescaling their model with the fast time scale of volume changes. Unifying both time scales makes both operators appear in the limiting dynamics, giving rise to the full dynamics, where price and volume fluctuations survive in the limit and neither the first order nor the second order approximation degenerates as in \cite{HorstSecondOrderApproximations2018}.
\end{remark}

%%%%%%%%%%%%%%%%%%%%%%%%%%%%%%%%%%%%%%%%%%%%%%%%%%%%%%%%%%%%
%%%%%%%%%%%%%%%%%%%%%%%%%%%%%%%%%%%%%%%%%%%%%%%%%%%%%%%%%%%%
%%%%%%%%%%%%%%%%%%%%%%%%%%%%%%%%%%%%%%%%%%%%%%%%%%%%%%%%%%%%

\subsubsection{Well-posedness of the stochastic evoluation equation and main result}\label{sec: well-posedness}

In what follows, we show that the limiting equation \eqref{eq: abstract spde} is well-posed and has a unique weak solution. To this end, we first formalize the notion of a solution to \eqref{eq: abstract spde}.

\begin{definition}
\label{def: probabilistically weak solution}
An $\mathcal H$-valued continuous adapted process $Z$ supported on the filtered probability space $(C([0,T], \mathcal H), \mathcal F, \mathbb F, \Q)$ is called a \emph{weak solution} to \eqref{eq: abstract spde} if its trajectories are $\Q$-a.s.~Bochner integrable and for all $\ell \in D(A^*)$ and $t\in[0,T]$ we have $\Q$-a.s.~that
\begin{equation}
\label{eq: equation for the weak solution}
\langle \ell, Z_t \rangle = \langle \ell, z_0 \rangle + \int _0^t \langle A(s)^* \ell, Z_s \rangle \d s + \int_0^t \langle \ell, F(s) Z_s \rangle \d s + \int_0^t \langle \ell, \sigma(s) \d W_s \rangle.
\end{equation}
\end{definition}

Note that if we had $p \equiv \operatorname{const}$, the operator $p \partial_x$ would generate a $C_0$-semigroup, which would amount to a translation at constant speed $p$. The equation \eqref{eq: equation for the weak solution} would then be solvable, for example, in the classical framework of \cite[Section 7.1]{DaPratoStochasticEquationsInfinite2014}. However, the differential operator $p(t)\partial_x$ in our equation varies over time, leading to a non-autonomous stochastic Cauchy problem. 

Instead, our operator $A(t)Z = (0, p(t)\partial_x Z^u)'$ with the domain $D(A) = \R\times H^1(U)$ generates a unique strongly continuous evolution family in the sense of \cite{VeraarNONAUTONOMOUSSTOCHASTICCAUCHY}. 
We are hence required to work in the framework of \emph{evolution families} instead of semigroups. In the absence of the drift $F$, the existence, uniqueness and continuity of the unique mild solution as well as the equivalence between weak and mild solutions was established in \cite[Section 3]{VeraarNONAUTONOMOUSSTOCHASTICCAUCHY}. Making use of those results, we can accommodate the drift $F$ by a fixed-point argument.

We introduce the translation evolution family on $\mathcal H$ given for $(a,\varphi)\in\mathcal H$ by
\[
S(t,s)(a,\varphi)' = \left(a,\varphi \left(\cdot+\int_s^t p(u) \d u\right)\right),\quad 0\leq s\leq t\leq T.
\]
This is a strongly continuous evolution family generated by $A(t)$ in the sense of \cite[Section 2.1]{VeraarNONAUTONOMOUSSTOCHASTICCAUCHY}. Moreover, it holds $\|S(t,s)\| = 1$. In the following proof we make use of Proposition \ref{prop: properties of the variance and volatility operators} below, which establishes properties of the volatility operator, resulting from the martingale relations proved later.

\begin{proposition}
\label{prop: well-posedness}
There exists a unique weak solution to equation \eqref{eq: equation for the weak solution}.
\end{proposition}

\begin{proof}
It is well known that the bounded operator from $H$ into a Hilbert space $E$ is $\gamma$-Radonifying if and only if it is Hilbert–Schmidt. Since the covariance operator $\sigma$ is Hilbert–Schmidt by virtue of Proposition \ref{prop: properties of the variance and volatility operators}, it satisfies the hypothesis of \cite[Theorem 3.3]{VeraarNONAUTONOMOUSSTOCHASTICCAUCHY}. Therefore, it follows that the stochastic integral
\[
S(t,0) z_0 + \int_0^t S(t, s) \sigma(s) \d W_s
\]
is integrable and has a modification with continuous paths in $\mathcal H$. We now fix $g\in C([0,T], \mathcal H)$ and define a map $K_{g,T}:C([0,T], \mathcal H) \rightarrow C([0,T], \mathcal H)$ by
\[
K_{g,T}Z(t) := \int_0^t S(t,s)F(s,Z_s) \d s + g(t),\quad t\in[0,T].
\]
Note that since $\|S(t,s)\| = 1$ and $F$ is uniformly Lipschitz in $t\in[0,T]$ with some constant $L$, we have
\[
\sup_{t\in[0,T]}\left\|K_{g,T}u(t) - K_{g,T}v(t) \right\| \leq LT \sup_{t\in[0,T]}\|u(t)-v(t)\|,
\]
so for $T$ small enough the map $K_{g,T}$ is a contraction on the space $C([0,T], \mathcal H)$.  In this case, taking
\[
g(t):=S(t, 0) z_0+\int_0^t S(t, s) \sigma(s) \d W_s \in C([0,T],\mathcal H),
\]
the unique fixed point yields the global mild solution with continuous paths:
\[
Z_t = S(t,0)z_0 + \int_0^t S(t,s) F(s,Z_s) \d s + \int_0^t S(t,s)\sigma(s) \d W_s,\qquad t\in[0,T].
\]
Since the Lipschitz constant does not depend on the initial condition, we can iterate to obtain a mild solution on $[0,T]$ for arbitrary $T$.

Hence, it suffices to prove that every weak solution is a mild solution. We proceed along the lines of \cite[Theorem 5.4]{DaPratoStochasticEquationsInfinite2014}. Note that we have $A^*(t)Z = (0, -p(t) \partial_x Z^u)'$, i.e.~for all $t\in[0,T]$ the domain of the adjoint is given by $D(A^*) = \R\times H^1(U)$, and the adjoint semigroup is given by $S^*(t,s)(a, \varphi)' = \left(a, \varphi\left(x-\int_s^t p(u) \d u\right)\right)'$. Consider first $\psi(s) = \varphi(s) \ell_0$ for some $\varphi \in C^1([0,T], \R)$ and $\ell_0\in D(A^*)$. We get by It\^o's formula
\[
\begin{aligned}
\langle \psi(t), Z_t \rangle &= \langle \ell_0 , S(t, 0)z_0 \rangle + \int_0^t \langle \psi'(s), Z_s \rangle \d s + \int _0^t \langle A^*(s) \psi(s), Z_s \rangle \d s \\
&\qquad\qquad\qquad\qquad+ \int_0^t \langle \psi(s), F(s, Z_s) \rangle \d s + \int_0^t \langle \psi(s), \sigma(s)  \d W_s \rangle \\
&= \langle \ell_0 , S(t, 0)z_0 \rangle + \int_0^t \langle S^*(t,s)\ell_0, F(s, Z_s) \rangle \d s + \int _0^t \langle S^*(t,s)\ell_0, \sigma(s)  \d W_s \rangle. \\
\end{aligned}
\]
Since the functions $\psi(s) = \varphi(s) \ell_0$ are dense in $C^1\left([0, T], D\left(A^*\right)\right)$, the equality above holds for any $\psi\in C^1\left([0, T], D\left(A^*\right)\right)$. We fix $\ell \in D(A^*) = H^1(U)$ and consider $\psi(s) = S^*(t,s) \ell$, so that $\psi'(s) + A^*(s) \psi(s) = 0$. We get by stochastic Fubini theorem 
\[
\begin{aligned}
\langle \ell, Z_t \rangle = \langle \psi(t), Z_t \rangle &= \langle \ell , S(t, 0)z_0 \rangle + \int_0^t \langle S^*(t,s)\ell, F(s, Z_s) \rangle \d s + \int _0^t \langle S^*(t,s)\ell, \sigma(s)  \d W_s \rangle \\
&= \left\langle \ell, S(t, 0)z_0 + \int_0^t S(t,s) F(s, Z_s) \d s + \int _0^t S(t,s) \sigma(s)  \d W_s \right\rangle.
\end{aligned}
\]
Since $H^1(U)$ is dense in $L^2(U)$, it follows that $Z$ is also mild solution; hence each weak solution $Z$ is unique.\qed
\end{proof}

Armed with the preceding existence and uniqueness result, we are now ready to state the following second-order approximation that is the main result of this paper. It states that the fluctuations of the price-volume process converge in law to the weak solution of the said stochastic evolution equation. Note that convergence takes place in a negative Sobolev norm due to our tightness estimates.

\begin{theorem}
\label{thm: main theorem}
The sequence $\{Z^{(n)}: n \in \mathbb N\}$ converges weakly in $D([0,T], \R\times H^{-4}(U))$ to the unique solution of equation \eqref{eq: abstract spde} in the sense of Definition \ref{def: probabilistically weak solution}.
\end{theorem}

The proof of the above theorem is split into several steps and is carried out in Sections 3–5. Before turning to the proof, we provide empirical evidence supporting our scaling assumptions. 

%%%%%%%%%%%%%%%%%%%%%%%%%%%%%%%%%%%%%%%%%%%%%%%%%%%%%%%%%%%%
%%%%%%%%%%%%%%%%%%%%%%%%%%%%%%%%%%%%%%%%%%%%%%%%%%%%%%%%%%%%
%%%%%%%%%%%%%%%%%%%%%%%%%%%%%%%%%%%%%%%%%%%%%%%%%%%%%%%%%%%%

\subsection{Empirical implementation and application to portfolio liquidation}\label{subsec:empirics}

As pointed out at the end of subsection \ref{sec:limit}, the second order approximation derived in this paper under a single-scale regime can be understood as the full dynamics one obtains when both, price and volume fluctuations survive in the limit. The two dynamics obtained in \cite{HorstSecondOrderApproximations2018} under different scaling assumptions correspond to the cases where either $p=0$, cf.~\cite[Theorem 5.10]{HorstSecondOrderApproximations2018}, or $r=0$, cf.~\cite[Theorem 6.9]{HorstSecondOrderApproximations2018}. 

While it is notoriously difficult to empirically verify scaling assumptions, one can estimate $p^A,p^B,f,g$ from real data and test whether $p=0$ or $r=0$. The outcome of that test provides empirical justification for choosing either the full dynamics or a restricted version as an approximate model for the price-volume fluctuations. 

In what follows we calibrate our model to market data for Microsoft (MSFT) and Verizon (VZ) using a simple linear model for the coefficient functions. Our results suggest that for MSFT the full SPDE dynamics obtained in Theorem \ref{thm: main theorem} is empirically better justified than the restricted one derived in \cite{HorstSecondOrderApproximations2018}, while for VZ the restricted dynamics of \cite[Theorem 5.10]{HorstSecondOrderApproximations2018} may already yield a reasonable approximation. 
In what follows we calibrate our model to market data for Microsoft (MSFT) and Verizon (VZ) using a simple linear model for the coefficient functions. Our results suggest that for MSFT the full SPDE dynamics obtained in Theorem \ref{thm: main theorem} is empirically better justified than the restricted one derived in \cite{HorstSecondOrderApproximations2018}, while for VZ the restricted dynamics of \cite[Theorem 5.10]{HorstSecondOrderApproximations2018} may already yield a reasonable approximation. 

Subsequently, we briefly discuss how our model can be applied to compute confidence intervals for the portfolio value of buy-and-hold strategies.

%%%%%%%%%%%%%%%%%%%%%%%%%%%%%%%%%%%%%%%%%%%%%%%%%%%%%%%%%%%%
%%%%%%%%%%%%%%%%%%%%%%%%%%%%%%%%%%%%%%%%%%%%%%%%%%%%%%%%%%%%
%%%%%%%%%%%%%%%%%%%%%%%%%%%%%%%%%%%%%%%%%%%%%%%%%%%%%%%%%%%%

\subsubsection{Empirical implementation}

An advantage of our model over ad-hoc SPDE models for order books is that our model lends itself to easier empirical calibrations as it provides a-priori information on the structural form of the coefficients $A,F,\sigma$ in \eqref{eq: equation for the weak solution}. Indeed, the drift and diffusion operators of the first and second order approximation are all defined in terms of the first two conditional moments of the price respectively volume process. Hence, one only has to estimate the functions $p^A,p^B,f,$ and $g$ defined in Assumptions \ref{assum: p}-\ref{assum: g}. 

Since the empirical volume function is piecewise constant, we may choose as test functions $\varphi_k(x):=\1_{(-(k+1)\Delta^{(n)},-k\Delta^{(n)}]}(x)$, even though they are not in $H^4$. Indeed, integrating the empirical volume density against $\varphi_k$ gives the same value as integrating it against any non-negative $\hat\varphi_k\in H^4$ with compact support in $(-k\Delta^{(n)},-(k-1)\Delta^{(n)}]$ and $L^1$-norm $\|\hat\varphi_k\|_{L^1}=\Delta^{(n)}$. Therefore, it suffices to consider as test functions $\varphi_k,\ k=0,\dots,K_L,$ with $(K_L+1)\Delta^{(n)}=L$ to estimate $f$ and $g$ on $[-L,0]$. Moreover, we choose as volume indicator the top of the book, i.e.~$h=\varphi_0$ and $Y_t=\langle u(t),\varphi_0\rangle$.

We use NASDAQ ITCH MBP-10 data for MSFT and VZ on 03/02/2025, which provides the full limit order book up to $K_L=10$ price levels, reconstructed from the raw message feed. Each record reflects the state of the book after processing all messages up to that point in time. We restrict to a single hour of the trading day (2-3 pm) to limit the effect of intraday non-stationarity and resample to 1-second intervals by taking the last snapshot in each bin and forward-filling gaps. 
All features and targets are clipped to $\pm 3$ standard deviations to reduce outlier influence. The tick size is $\Delta^{(n)}=0.01$ USD for both stocks. 

First, we fit the full (visible) dynamics \ref{eq: abstract spde} for MSFT and VZ, i.e.~we estimate the functions $p^A,p^B,f^k:=\langle f,\varphi_k\rangle,g^k:=\langle g,\varphi_k\rangle,\ k=0,\dots,K_L,$ using the following regression model: 
\begin{equation}\label{eq: regression model}
\begin{aligned}
p^{A}(B,Y) &= \hat p^A_c + \hat p^A_b B + \hat p^A_y Y + \varepsilon^{A},\\
p^{B}(B,Y) &= \hat p^B_c + \hat p^B_b B + \hat p^B_y Y + \varepsilon^{B},\\
f^k(B,Y) &= \hat f^k_{c} + \hat f^k_{b}B + \hat f^k_{y}Y + \varepsilon^{f,k},\quad k=0,\dots,K_L\\
g^k(B,Y) &= \hat g^k_{c} + \hat g^k_{b}B + \hat g^k_{y}Y^2 + \varepsilon^{g,k},\quad k=0,\dots,K_L.
\end{aligned}
\end{equation}
To estimate the linear model \eqref{eq: regression model} we use ridge regression with a common penalty $\lambda=0.1$ applied to all parameters. We also compute nonparametric bootstrap confidence intervals using $N=100$ resamples, each consisting of 3600 data points drawn with replacement. The results are reported in Appendix \ref{app:empirics}.

Second, to test the applicability of our model, we also directly estimate $p:=p^B-p^A$ and $r^k:=\langle r,\varphi_k\rangle$, $k=0,\dots,K_L$, using the linear model

\begin{equation}\label{eq: regression model 2}
\begin{aligned}
p(B,Y) &= \hat p_c + \hat p_b B + \hat p_y Y + \varepsilon^{p},\\
r^k(B,Y) &= \hat r^k_{c} + \hat r^k_{b}B + \hat r^k_{y}Y^2 + \varepsilon^{r,k},\quad k=0,\dots,K_L.
\end{aligned}
\end{equation}

%%%%%%%%%%%%%%%%%%%%%%%%%%%%%%%%%%%%%%%%%%%%%%%%%%%%%%%
%%%%%%%%%%%%%%%%%%%%%%%%%%%%%%%%%%%%%%%%%%%%%%%%%%%%%%%

\begin{figure}[h]
    \centering
    \includegraphics[width=0.447\textwidth]{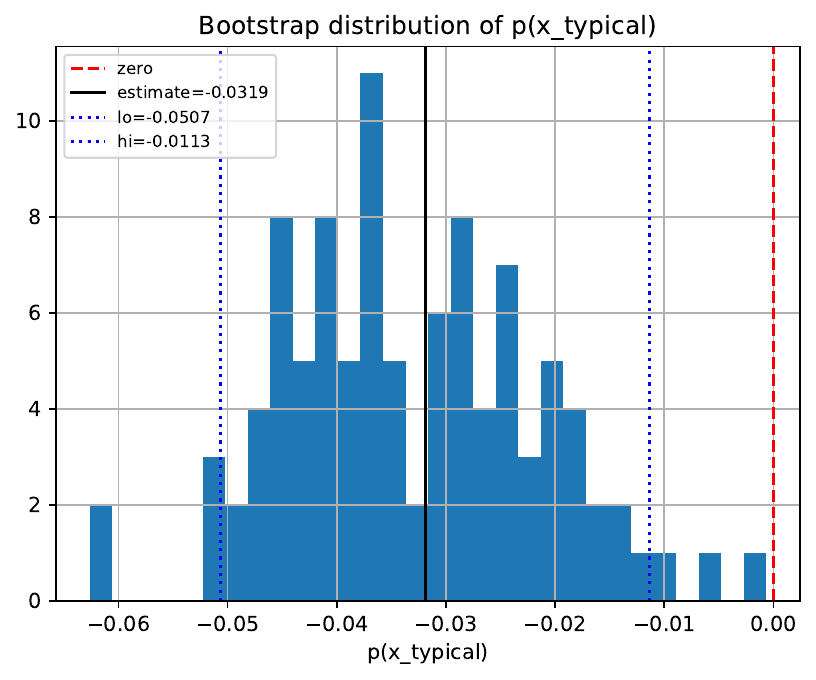}
    \includegraphics[width=0.44\textwidth]{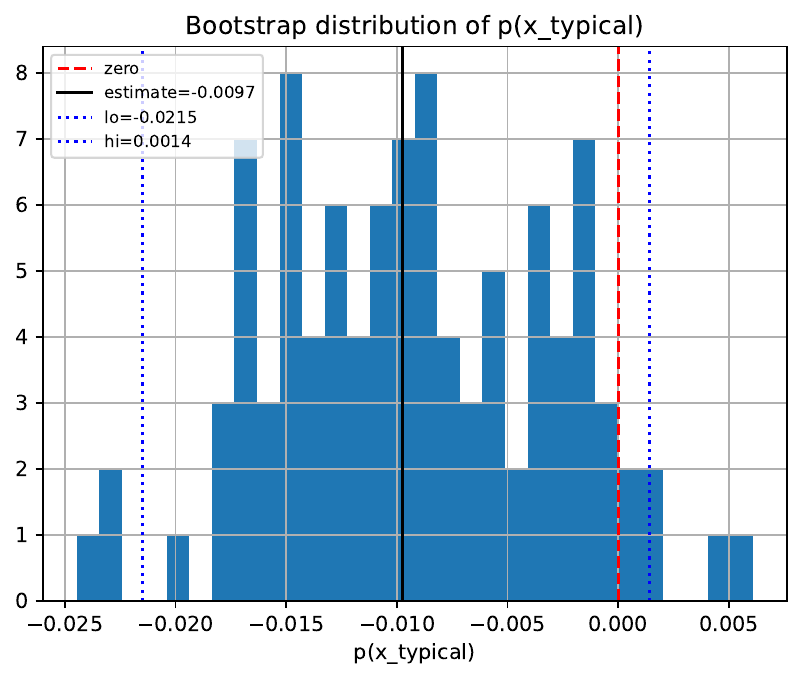}
    \caption{Bootstrap distributions of $\hat p(B^*,Y^*)$ for MSFT (left) and VZ (right) (black line: point estimate using the original data set; blue dotted lines: 95\% bootstrap CI)}
    \label{fig:MSFT/VZ}
\end{figure}

\begin{figure}[h]
    \centering
    \includegraphics[width=0.95\textwidth]{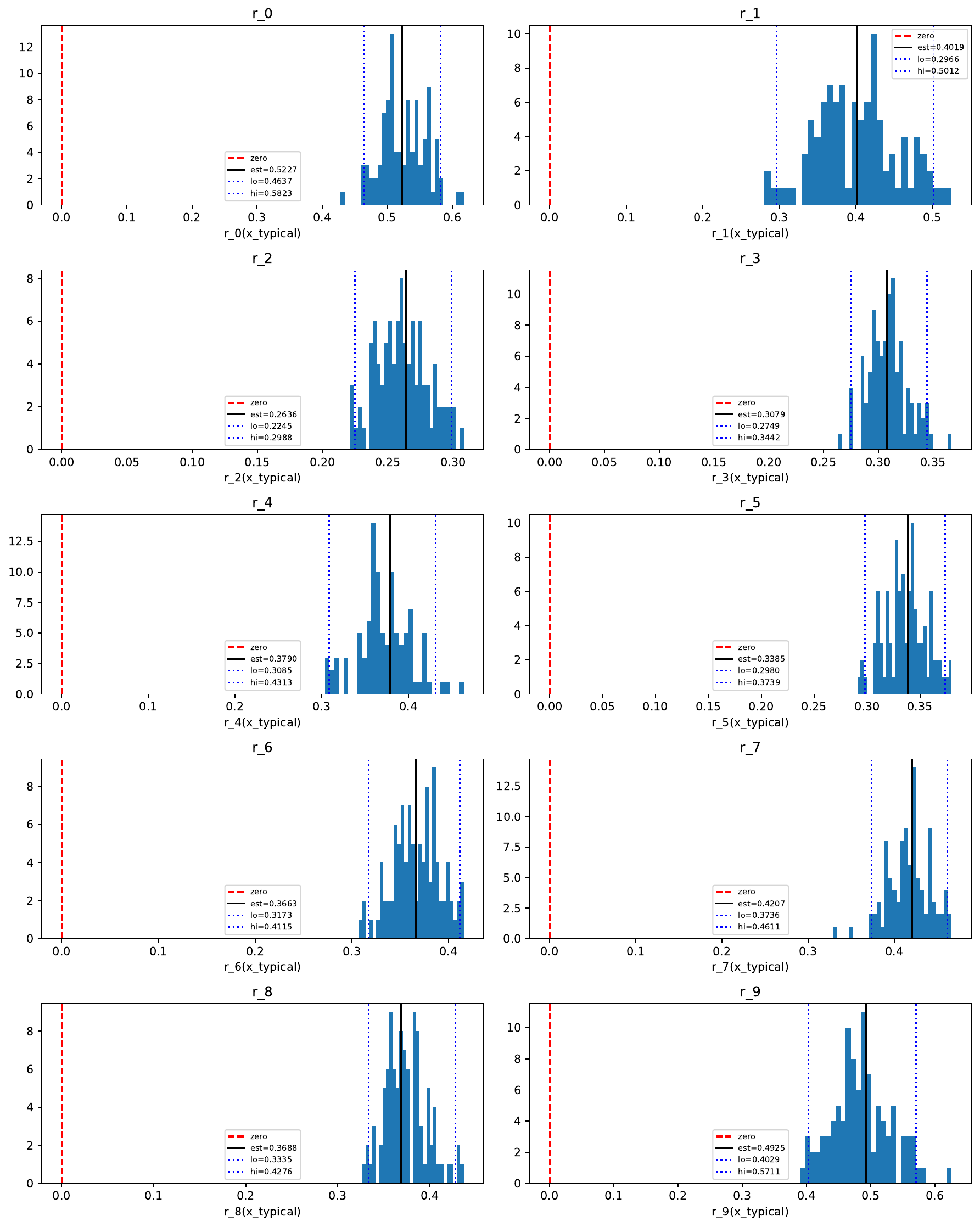}
    \caption{Bootstrap distributions of $\hat r^k(B^*,Y^*),\ k=0,\dots,9,$ for MSFT (black line: point estimate using the original data set; blue dotted lines: 95\% bootstrap CI)}
    \label{fig:MSFT2}
\end{figure}

\begin{figure}[h]
    \centering
    \includegraphics[width=0.95\textwidth]{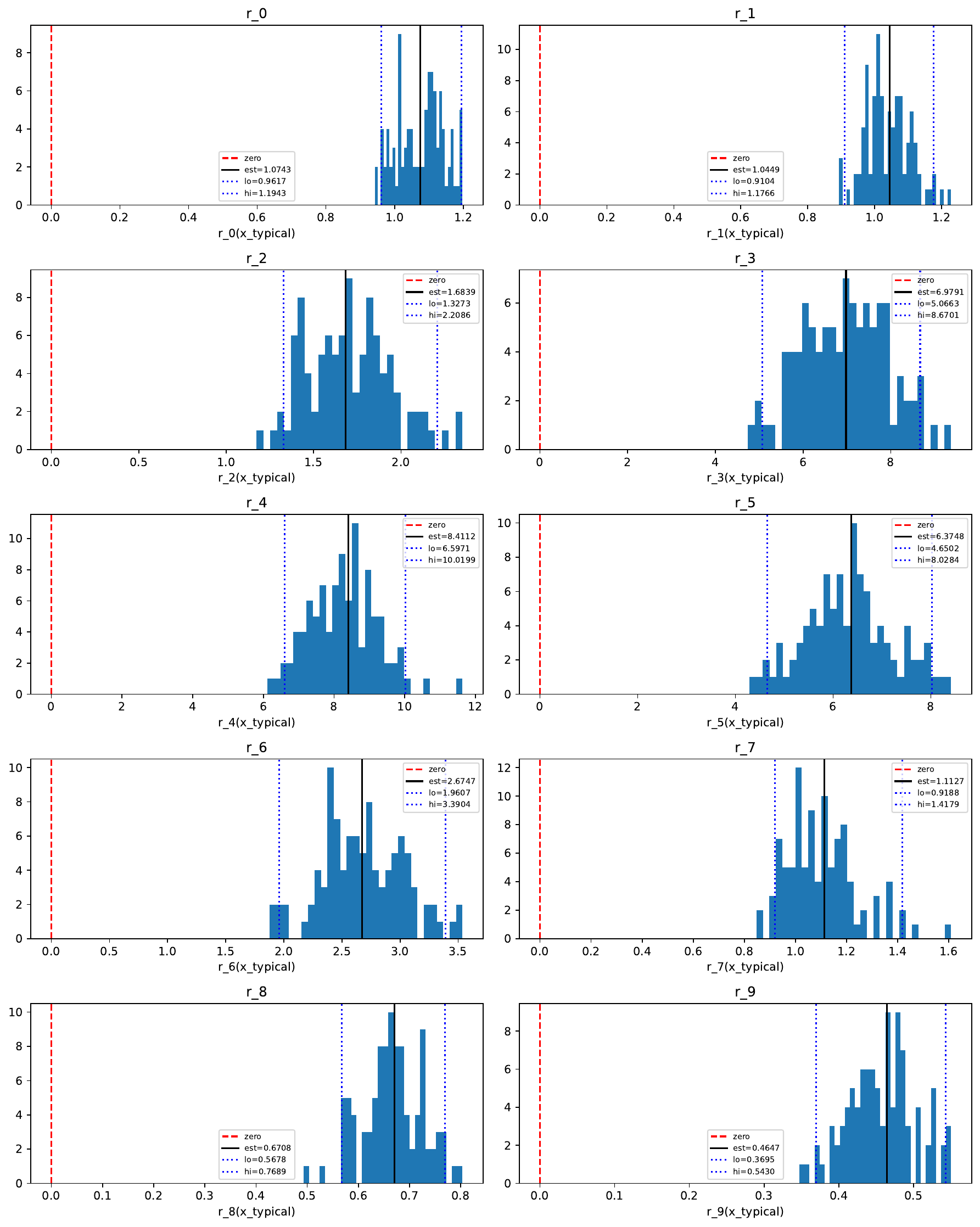}
    \caption{Bootstrap distributions of $\hat r^k(B^*,Y^*),\ k=0,\dots,9,$ for VZ (black line: point estimate using the original data set; blue dotted lines: 95\% bootstrap CI)}
    \label{fig:VZ2}
\end{figure}

As $r^k$ is not directly observable, we calculate it from $f^k$ and $g^k$ estimated in \eqref{eq: regression model}. 
To see whether $p$ and/or $r$ are distinguishable from zero at a typical market state, we report in Figures \ref{fig:MSFT/VZ} -\ref{fig:VZ2} the bootstrap distribution of 
\begin{align*}
\hat p(B^*,Y^*)&:=\hat p_c + \hat p_b B^* + \hat p_y Y^*,  \\ 
\hat r^k(B^*,Y^*) : &= \hat r^k_{c} + \hat r^k_{b}B^* + \hat r^k_{y}(Y^*)^2,\quad k=0,\dots,K_L.
\end{align*}
where $B^*$ resp.~$Y^*$ are the empirical mean of $B$ resp.~$Y$ of the clipped data sets. 

Figures \ref{fig:MSFT2} and \ref{fig:VZ2} suggest that $r$ is significantly different from zero for both MSFT and VZ. 
On the other hand, Figure \ref{fig:MSFT/VZ} suggests that on a significance level of $\alpha=5\%$ the parameter $p$ is different from zero for MSFT, while this is not the case for VZ. This suggests that for MSFT it is important to consider the full SPDE  model \eqref{eq: abstract spde} analyzed in this paper. On the other hand, for VZ one does not necessarily have to consider the full model, but can work with the reduced model studied in \cite[Theorem 5.10]{HorstSecondOrderApproximations2018}, assuming that changes in prices occur much less frequently than changes in volume.

\subsubsection{Application: confidence intervals for buy-and-hold portfolio values}

The results of this paper may also be used to compute confidence intervals for the revenues of a simple buy-and-hold strategy - or more generally, for the liquidation value of a large stock position using endogenous order book shape functions. 

Let us consider a simple buy-and-hold strategy where a trader buys at time $t=0$ a number $X$ of shares and holds them until some later time $T$. The trader's profit equals
\begin{align*}
    V_T^{(n)} & =\int_{-D^{(n)}}^0(B_T^{(n)}+x)u^{(n)}(T,x)dx-B_0^{(n)}X \\
    & =(B_T^{(n)}-B_0^{(n)})X+\int_{-D^{(n)}}^0xu^{(n)}(T,x)dx
\end{align*}
with the random variable $D^{(n)}>0$ defined via 
\[X=\int_{-D^{(n)}}^0u^{(n)}(T,x)dx.\]
As $n$ is fixed, there is for any $d>0$ a function $\varphi_d\in H^4$ such that for all $k\in\Z$,
\[\int_{-k\Delta^{(n)}}^{-(k-1)\Delta^{(n)}}\varphi_d(x)dx=\int_{-k\Delta^{(n)}}^{-(k-1)\Delta^{(n)}}x\1_{[-d,0]}(x)dx. \]
Since $u^{(n)}$ is piecewise constant, we have
\[V^{(n)}_T=\left(B_T^{(n)}-B_0^{(n)}\right)X+\left\langle u^{(n)}(T),\varphi_{D^{(n)}}\right\rangle.\]
According to our second order approximation, we have for any $t\in[0,T]$ and $d>0$,
\begin{align*}
    B_t^{(n)}&\approx B_t+\sqrt{\Delta^{(n)}}Z_t^B,\\
    \langle u^{(n)}(t),\varphi_d\rangle &\approx \langle u(t),\varphi_d\rangle+\sqrt{\Delta^{(n)}}\langle Z^u_t,\varphi_d\rangle,
\end{align*}
where $S=(B,u)$ given by \eqref{eq: first-order approximation} is deterministic and $Z=(Z^B,Z^u)$ is the unique solution of \eqref{eq: abstract spde}. The profit of the buy-and-hold strategy can thus be approximated by
\begin{align*}
 V^{(n)}_T&\approx\left(B_T-B_0+\sqrt{\Delta^{(n)}}Z^B_T\right)X+\left\langle u(T)+\sqrt{\Delta^{(n)}}Z^u_T,\varphi_{D}\right\rangle\\
 &\approx   \left(B_T-B_0+\sqrt{\Delta^{(n)}}Z^B_T\right)X+\int_{-D}^0xu(T,x)dx+ \sqrt{\Delta^{(n)}}\left\langle Z^u_T,\varphi_D\right\rangle,
\end{align*}
where $D>0$ is determined by 
\[X=\int_{-D}^0u(T,x)dx.\]
Knowing the dynamics of $Z$, the above approximation can be used to construct confidence intervals for $V_T^{(n)}$. Technically, the approximation depends on the choice of $\varphi_d$. In practice, however, this should not be a problem as long as one chooses any reasonable, smooth approximation of $\1_{[-d,0]}$. Using similar ideas, one may also compute confidence intervals for portfolio liquidation problems as done in \cite[Section 4]{HorstSecondOrderApproximations2018}.

%%%%%%%%%%%%%%%%%%%%%%%%%%%%%%%%%%%%%%%%%%%%%%%%%%%%%%%%%%%%%%
%%%%%%%%%%%%%%%%%%%%%%%%%%%%%%%%%%%%%%%%%%%%%%%%%%%%%%%%%%%%%%
%%%%%%%%%%%%%%%%%%%%%%%%%%%%%%%%%%%%%%%%%%%%%%%%%%%%%%%%%%%%%%

\section{Properties of the discrete dynamics}
\label{sec: preliminaries}

In this section, we establish some auxiliary results for the discrete models that will be used repeatedly in our subsequent analysis. 
Since the discrete fluctuations of the volume function defined in \eqref{eq: definition of Z^B and Z^u} satisfy
\begin{equation*}
\sqrt{\Delta^{(n)}} \, \nabla^{(n)}_+ Z^{(n),u}_k = \nabla^{(n)}_+ u^{(n)}_k - \nabla^{(n)}_+ u\left(t_k^{(n)}\right),
\end{equation*}
the increments of the volume fluctuation process can be rewritten as
\begin{equation}
\label{eq: discrete increment of volume}
\begin{aligned}
&\delta Z^{(n),u}_k = \left\{ \1^{(n),B}_k \left(T^{(n)}_+ - 1\right) + \1^{(n),A}_k \left(T^{(n)}_- - 1\right) \right\} Z^{(n),u}_{k-1} \\
&\quad+ \sqrt{\Delta^{(n)}} \left\{ \1^{(n),B}_k \nabla ^{(n)}_+ - \1^{(n),A}_k \nabla ^{(n)}_- \right\} u\left(t_{k-1}^{(n)} \right)- \left(\Delta^{(n)}\right)^{-1/2} \int_{t_{k-1}^{(n)}} ^{t_{k}^{(n)}} p\left(S(r)\right) u_x(r) \d r\\
&\quad + \left(\Delta^{(n)}\right)^{-1/2} \left( \Delta^{(n)} J_k^{(n),C} -\int_{t_{k-1}^{(n)}} ^{t_{k}^{(n)}} f\left(S(r)\right) \d r \right).
\end{aligned}
\end{equation}

The above representation can be used to rewrite the dynamics of the volume fluctuations along the stochastic characteristic given by the best bid price process. The proof of the following result follows easily by induction on $\ell>k$ and is omitted. 

\begin{proposition}
\label{prop: decomposition of discrete dynamics of volume fluctuations}
For any $k,\ell \in \N$ with $k<\ell$, the discrete scheme for the volume fluctuations can be rewritten as
\begin{equation*}
\left\langle Z^{(n),u}_\ell - Z^{(n),u}_k, \varphi \right\rangle = A_1^{(n), k, \ell} + A_2^{(n), k, \ell} + A_3^{(n), k, \ell},
\end{equation*}
where
\begin{equation*}
\begin{aligned}
A_1^{(n), k, \ell}&=\left\langle Z^{(n),u}_k, \varphi\left(\cdot - \left(B^{(n)}_\ell - B^{(n)}_k\right) \right) \right\rangle - \left\langle Z^{(n),u}_k, \varphi \right\rangle, \\
A_2^{(n), k, \ell}&= \sum_{i=k+1}^{\ell} \sqrt{\Delta^{(n)}}\left\langle \left\{ \1^{(n),B}_i\nabla^{(n)}_+ - \1^{(n),A}_i \nabla^{(n)}_- \right\} u\left(t_{i-1}^{(n)}\right) \right. \\
& \quad \left. - \frac{1}{\Delta^{(n)}} \int_{t_{i-1}^{(n)}}^{t_{i}^{(n)}} p\left(S(r)\right) \partial_x u(r) \d r , \varphi\left(\cdot-\left(B^{(n)}_\ell - B^{(n)}_i\right) \right) \right\rangle, \\
A_3^{(n), k, \ell}&= \sum_{i=k+1}^{\ell} \sqrt{\Delta^{(n)}} \left\langle J_{i}^{(n),C} - 
\frac{1}{\Delta^{(n)}} \int_{t_{i-1}^{(n)}}^{t_{i}^{(n)}} f\left(S(r)\right) \d r , \varphi \left(\cdot-\left(B^{(n)}_\ell - B^{(n)}_i\right) \right) \right\rangle.
\end{aligned}
\end{equation*}
\end{proposition}

The term $A_1^{(n), k, \ell}$ corresponds to the difference of the volume fluctuation profile $Z_k^{(n), u}$ and its shift along the stochastic characteristic $B_{\ell}^{(n)}-B_k^{(n)}$. The term $A_2^{(n), k, \ell}$ corresponds to the difference between the transport term of the microscopic dynamics and the one of the deterministic first-order approximation. The term $A_3^{(n), k, \ell}$ corresponds to the difference in the source contributions, i.e.~the limit order placements, of the microscopic dynamics and those of the deterministic first-order approximation. To apply standard tightness criteria, we will have to derive moment estimates for all three terms. While this is unproblematic for $A_2^{(n), k, \ell}$ and $A_3^{(n), k, \ell}$, the challenge is to control the term $A_1^{(n), k, \ell}$ as it apparently requires a priori bounds on $Z^{(n),u}_k$ itself. We will deal with this issue in Section \ref{sec: moment estimates}.

The following lemma will be used repeatedly throughout the paper. It allows to pass from discrete coefficient functions to their continuous counterparts.

\begin{lemma}
\label{lemma: taylor expansion of p, f and g}
There exist real-valued random variables $C^{(n),1}_i$ and $C^{(n),2}_i$ $(i \in \N)$, which converge to zero in probability uniformly in $i\in\N$, s.t.~for $I\in\{A,B\}$ it holds that
\[
\begin{aligned}
&\frac{1}{\sqrt{\Delta^{(n)}}}\left(\Delta^{(n)} p^{(n),I}\left(S^{(n)}_{i}\right) - \int_{t_{i}^{(n)}}^{t_{i+1}^{(n)}} p^{I}\left(S(r)\right) \d r\right)\\
&\qquad-\frac{1}{\sqrt{\Delta^{(n)}}}\left(\Delta^{(n)} p_b^{I}\left(S\left(t_i^{(n)}\right)\right) \sqrt{\Delta^{(n)}} Z^{(n),B}_i + \Delta^{(n)} p^{I}_y \left(S\left(t_i^{(n)}\right)\right) \sqrt{\Delta^{(n)}} Z^{(n),Y}_i \right)\\
&\qquad= \Delta^{(n)} C^{(n),1}_i Z^{(n),B}_i + \Delta^{(n)} C^{(n),2}_i Z^{(n),Y}_i + C^{(n)}.
\end{aligned}
\]
Likewise, there exist $L^2(\R)$-valued random variables $\widetilde C^{(n),1}_i$ and $\widetilde C^{(n),2}_i,\ i\in\N,$ converging to zero in probability uniformly in $i\in\N$, s.t.~for $(h, h^{(n)}) \in \{(f, f^{(n)}, (g, g^{(n)})\}$ it holds that
\[
\begin{aligned}
&\frac{1}{\sqrt{\Delta^{(n)}}}\left(\Delta^{(n)} h^{(n)}\left(S^{(n)}_{i}\right) - \int_{t_{i}^{(n)}}^{t_{i+1}^{(n)}} h\left(S(r)\right) \d r\right)\\
&\qquad-\frac{1}{\sqrt{\Delta^{(n)}}}\left(\Delta^{(n)} h_b\left(S\left(t_i^{(n)}\right)\right) \sqrt{\Delta^{(n)}} Z^{(n),B}_i + \Delta^{(n)} h_y\left(S\left(t_i^{(n)}\right)\right) \sqrt{\Delta^{(n)}} Z^{(n),Y}_i \right)\\
&\qquad= \Delta^{(n)} \widetilde C^{(n),1}_i Z^{(n),B}_i + \Delta^{(n)} \widetilde C^{(n),2}_i Z^{(n),Y}_i + C^{(n)}.
\end{aligned}
\]
\end{lemma}
\begin{proof}
We show the proof for the price drift $p$ only, the volume drift $f$ is handled analogously. Let
\[
F(s) := p\left(B(t_i) + s(B^{(n)}_i - B(t_i)), Y(t_i) + s(Y^{(n)}_i - Y(t_i))\right).
\]
Applying a Taylor expansion yields that
\[
\begin{aligned}
&\frac{1}{\sqrt{\Delta^{(n)}}}\left(\Delta^{(n)} p^{(n)}\left(S^{(n)}_{i}\right) - \int_{t_{i}^{(n)}}^{t_{i+1}^{(n)}} p\left(S(r)\right) \d r\right) = R^{(n),1}_i+R^{(n),2}_i+R^{(n),3}_i+R^{(n),4}_i,
\end{aligned}
\]
where
\[
\begin{aligned}
R^{(n),1}_i &= \sqrt{\Delta^{(n)}} \left(p^{(n)}\left(S^{(n)}_{i}\right) - p\left(S^{(n)}_{i}\right)\right), & R^{(n),3}_i &= \sqrt{\Delta^{(n)}} F'(0), \\
R^{(n),2}_i &= \frac{1}{\sqrt{\Delta^{(n)}}} \int_{t_i^{(n)}}^{t_{i+1}^{(n)}} \int _{t_i^{(n)}}^s \frac{\partial}{\partial r} p\left(S(r)\right) \d r \d s, &
R^{(n),4}_i &= \sqrt{\Delta^{(n)}} \int_0^1(1-s) F''(s) \d s. \\
\end{aligned}
\]
It follows from \cref{assum: p}\ref{assum: p: p^n-p} that $R^{(n),1}_i = o(1)$. Moreover, due to \cref{assum: p}~\ref{assum: p: sup} and the regularity of $B$ and $Y$, it follows that
\begin{equation}
\label{eq: remainder of second-order in the taylor expansion}
R^{(n),2}_i=O\left(\Delta^{(n)}\right)^{3/2}.
\end{equation}
Using the definition of $F$ we get that
\begin{equation*}
%\label{eq: taylor expansion of p}
R^{(n),3}_i = \Delta^{(n)} p_b\left(S\left(t_i^{(n)}\right)\right) Z^{(n),B}_i + \Delta^{(n)} p_y\left(S\left(t_i^{(n)}\right)\right) Z^{(n),Y}_i.
\end{equation*}
Lastly, similarly to the proof of \cite[Lemma  5.9]{HorstSecondOrderApproximations2018}, there exist random variables $C^{(n),1}_i$ and $C^{(n),2}_i$ converging to zero in probability uniformly in $i\in\N$ such that
\[
R^{(n),4}_i = \Delta^{(n)} \left(C^{(n),1}_i Z^{(n),B}_i + C^{(n),2}_i Z^{(n),Y}_i\right).
\]\qed
\end{proof}

%%%%%%%%%%%%%%%%%%%%%%%%%%%%%%%%%%%%%%%%%%%%%%%%%%%%%%%%%%%%%%
%%%%%%%%%%%%%%%%%%%%%%%%%%%%%%%%%%%%%%%%%%%%%%%%%%%%%%%%%%%%%%
%%%%%%%%%%%%%%%%%%%%%%%%%%%%%%%%%%%%%%%%%%%%%%%%%%%%%%%%%%%%%%

\section{Growth estimates}
\label{sec: moment estimates}

In this section we derive moment estimates for the fluctuation processes. More precisely, our goal is to show that for any $0 \leq k<\ell\leq T_n$,
\begin{equation}
\label{eq: moment estimate formal intro}
\E \left[\left| Z^{(n),B}_\ell - Z^{(n),B}_k \right| ^m\right] + \E\left[\left\langle Z^{(n),u}_\ell - Z^{(n),u}_k, \varphi \right\rangle ^m\right] \lesssim_{\|\varphi\|} \delta^{m/2}_{n,\ell-k} + \delta^{m}_{n,\ell-k}.
\end{equation}
This estimate for arbitrary $k<\ell$ will be used in the tightness proof in Section \ref{subsec: tightness}. Moreover, for $k=0$, the estimate yields a growth estimate for the moments of the fluctuations, which will be used to establish martingale relations in Section \ref{subsec: martingale relations}.

To derive the estimate \eqref{eq: moment estimate formal intro} for any $0 \leq k<\ell\leq T_n$, one would like to derive estimates on the increments of $Z^{(n)}$, which allow for an application of Gr\"onwall inequality. However, the presence of $A_1^{(n),k,\ell}$ in the decomposition of the volume dynamics (cf.~Proposition \ref{prop: decomposition of discrete dynamics of volume fluctuations}), which arises from the transport dynamics of the system, precludes us from directly applying Gr\"onwall inequality to derive the above moment estimate for $k>0$, since the growth of the volume fluctuation profile $Z^{(n),u}_k$ for $k>0$ cannot be controlled a priori. 

To address this problem, we first provide pathwise estimates of $A_2^{(n),k,\ell}$ and $A_3^{(n),k,\ell}$ (cf.~Lemmas \ref{lemma: pathwise estimate for transport contribution} and \ref{lemma: pathwise estimate for source contribution}). Subsequently we derive a pathwise growth estimate for $k=0$ in Proposition \ref{prop: pathwise estimate for Zs}, relying on the assumption that the initial volume fluctuation profile $Z^{(n),u}_0$ is vanishing. This will allow us to control the initial value contribution $A_1^{(n),k,\ell}$ and, ultimately, to obtain the moment estimate \eqref{eq: moment estimate formal intro} in Proposition \ref{prop: moment estimate for Z}.

We first introduce shorthand notations for some processes that will arise later in the decomposition of the fluctuation dynamics. For $ 0 \leq k \leq T_n$, we define 
\begin{equation*}
\begin{aligned}
\delta N^{(n),1}_i := \delta Z^{(n),B}_i- \E \left[\left.\delta Z^{(n),B}_i\right|\F^{(n)}_{i-1}\right] , \qquad N^{(n),1}_k := \sum_{i=1} ^k \delta N^{(n),1}_i .
\end{aligned}
\end{equation*}
Furthermore, for a fixed function $\varphi\in L^2(U)$ and $\ell\leq T_n$, we define for $ 0 \leq k \leq \ell$,
\begin{equation}
\label{eq: definition of N_2 martingales}
\begin{aligned}
\delta N^{(n),\ell,2}_i(\varphi) &:= \sqrt{\Delta^{(n)}} \left(\1^{(n),B-A}_i - p^{(n)}\left(S^{(n)}_{i-1}\right)\right) \left\langle \nabla^{(n)}_+ u\left(t_{i-1}^{(n)}\right) , \varphi\left(\cdot-B^{(n)}_\ell + B^{(n)}_{i-1} \right) \right\rangle , \\
N^{(n),\ell,2}_k (\varphi) &:= \sum_{i=1} ^k \delta N^{(n),\ell,2}_i(\varphi) ,
\end{aligned}
\end{equation}
and
\begin{equation}
\label{eq: definition of N_3 martingales}
\begin{aligned}
\delta N^{(n),\ell,3}_i(\varphi) &:= \sqrt{\Delta^{(n)}} \left\langle J^{(n),C}_i - f^{(n)}\left(S^{(n)}_{i-1}\right), \varphi\left(\cdot-\left(B^{(n)}_\ell - B^{(n)}_{i-1}\right)\right)\right\rangle,\\
N^{(n),\ell,3}_k(\varphi) &:= \sum_{i=1} ^k \delta N^{(n),\ell,3}_i(\varphi).
\end{aligned}
\end{equation}
While $N^{(n),1}$ is a martingale, the processes $N^{(n),\ell,2}$ and $N^{(n),\ell,3}$ are {\it not} martingales due to the presence of the spatial shift by $B_\ell^{(n)}$ in their definition. For the sum of the $m$-th powers of the increments of these processes we write
\[\widetilde{\mathcal N}^{(n),\ell,m}_k(\varphi)  := \left|N^{(n),1}_\ell-N^{(n),1}_k\right|^m + \sum_{r=2,3}\left|N^{(n),\ell,r}_\ell(\varphi)-N^{(n),\ell,r}_k(\varphi)\right|^m .\]

%%%%%%%%%%%%%%%%%%%%%%%%%%%%%%%%%%%%%%%%%%%%%%%%%%%%%%%%%%%%
%%%%%%%%%%%%%%%%%%%%%%%%%%%%%%%%%%%%%%%%%%%%%%%%%%%%%%%%%%%%
%%%%%%%%%%%%%%%%%%%%%%%%%%%%%%%%%%%%%%%%%%%%%%%%%%%%%%%%%%%%

\subsection{Pathwise estimates}

We first establish a pathwise estimate of the price fluctuations in terms of a discrete integral, a submartingale, and a residual term with the desired scaling.

\begin{lemma}
\label{lemma: pathwise estimate for price fluctuations}
Let $m\in 2\N\cup\{1\}$ and $0 \leq k < \ell \leq T_n$. Then almost surely
\[
\begin{aligned}
&\left|Z^{(n),B}_\ell - Z^{(n),B}_k\right|^m \\
&\quad\lesssim  K^{(n)} \delta_{n,\ell-k}^{m-1} \sum_{i=k}^{\ell-1} \Delta^{(n)} \left( \left|Z_{i}^{(n),B}\right|^{m} +\left|Z_{i}^{(n),Y}\right|^{m} \right) + \left|N^{(n),1}_\ell - N^{(n),1}_k\right|^m + C^{(n)}\delta^m_{n,\ell-k} .
\end{aligned}
\]
\end{lemma}

\begin{proof}

Using Lemma \ref{lemma: taylor expansion of p, f and g} and Assumption \ref{assum: p: sup}, we obtain 
\[
\begin{aligned}
&\sum_{i=k+1}^\ell \E \left[\left.\delta Z^{(n),B}_i \right|\F^{(n)}_{i-1}\right] \\
& = \sum_{i=k}^{\ell-1} \left\{ \Delta^{(n)} \left(p_b\left(S\left(t_{i}^{(n)}\right)\right) + C^{(n)}_i\right) Z^{(n),B}_{i} + \Delta^{(n)} \left(p_y\left(S\left(t_{i}^{(n)}\right)\right) + C^{(n)}_i\right) Z^{(n),Y}_{i} + C^{(n)}_i \right\}  \\
&\leq 
K^{(n)}\sum_{i=k}^{\ell-1} \left(\Delta^{(n)} Z^{(n),B}_i + \Delta^{(n)} Z^{(n),Y}_i + C^{(n)}_i \right).
\end{aligned}
\]
Therefore, by H\"older's inequality we have
\[
\begin{aligned}
&\left|Z^{(n),B}_\ell - Z^{(n),B}_k\right|^m
 \lesssim
\left|N^{(n),1}_\ell - N^{(n),1}_k\right|^m + \left(\sum_{i=k+1}^\ell \E \left[\left.\delta Z^{(n),B}_i \right|\F^{(n)}_{i-1}\right]\right)^m \\
&\quad\leq \left|N^{(n),1}_\ell - N^{(n),1}_k\right|^m + K^{(n)} \delta_{n,\ell-k}^{m-1} \sum_{i=k}^{\ell-1} \Delta^{(n)} \left( \left|Z_{i}^{(n),B}\right|^{m} +\left|Z_{i}^{(n),Y}\right|^{m}\right) + C^{(n)}\delta^m_{n,\ell-k}.
\end{aligned}
\]
\qed
\end{proof}

Next, we establish a pathwise estimate for the transport contribution $A_2^{(n),k,\ell}$ in the decomposition of the volume fluctuations in terms of a discrete integral, the process $N^{(n),\ell,2}$ defined in \ref{eq: definition of N_2 martingales}, and a residual term with the desired scaling.

\begin{lemma}
\label{lemma: pathwise estimate for transport contribution}
Let $m\in 2\N\cup\{1\}$, $\varphi\in H^3$, and $0 \leq k < \ell \leq T_n$. Then almost surely
\[
\begin{aligned}
\left|A_2^{(n), k, \ell}\right|^m &\lesssim  \delta_{n,\ell-k}^{m-1}  \|\varphi\|^m_{H^2} \sum_{i=k}^{\ell-1} \Delta^{(n)} \left( \left|Z^{(n),B}_i\right|^m + \left| Z^{(n),Y}_i \right|^m \right )\\
&\qquad\qquad + \left|N^{(n),\ell,2}_\ell(\varphi) - N^{(n),\ell,2}_k(\varphi)\right|^m  + \delta_{n,\ell-k}^m \left(\|\varphi\|^m_{H^3} + C^{(n)}\right).
\end{aligned}
\]
\end{lemma}

\begin{proof}
We consider the decomposition
\[
\begin{aligned}
&A_2^{(n), k, \ell}=\sqrt{\Delta^{(n)}}(K_1+K_2+K_3+K_4+K_5):=\\
&\sqrt{\Delta^{(n)}}\left[\sum_{i=k+1}^{\ell} \left\langle \sqrt{\Delta^{(n)}} \left( \1^{(n),B-A}_i - p^{(n)}\left(S^{(n)}_{i-1}\right) \right) \nabla^{(n)}_+ u\left(t_{i-1}^{(n)}\right) , \varphi\left(\cdot-B^{(n)}_\ell + B^{(n)}_{i-1} \right) \right\rangle\right. \\
&+ \sum_{i=k+1}^{\ell} \left\langle \left( \1^{(n),B-A}_i - p^{(n)}\left(S^{(n)}_{i-1}\right) \right) \nabla^{(n)}_+ u\left(t_{i-1}^{(n)}\right) , \varphi\left(\cdot-B^{(n)}_\ell +B^{(n)}_i \right) - \varphi\left(\cdot-B^{(n)}_\ell +B^{(n)}_{i-1} \right) \right\rangle \\
&+ \sum_{i=k+1}^{\ell} \left\langle  \1^{(n),A}_{i}\left(\nabla_{+}^{(n)}-\nabla_{-}^{(n)}\right) u\left(t_{i-1}^{(n)}\right) , \varphi\left(\cdot-B^{(n)}_\ell +B^{(n)}_i \right) \right\rangle \\
&+ \sum_{i=k+1}^{\ell} \left\langle \left(p^{(n)}\left(S_{i-1}^{(n)}\right)-\frac{1}{\Delta^{(n)}} \int_{t_{i-1}^{(n)}} ^{t_{i}^{(n)}} p\left(S(r)\right) \d r \right)\nabla^{(n)}_+u\left(t^{(n)}_{i-1}\right) , \varphi\left(\cdot-B^{(n)}_\ell + B^{(n)}_i \right) \right\rangle \\
&+ \left.\sum_{i=k+1}^{\ell} \left\langle \frac{1}{\Delta^{(n)}} \int_{t_{i-1}^{(n)}} ^{t_{i}^{(n)}} p\left(S(r)\right) \left(\nabla^{(n)}_+u\left(t^{(n)}_{i-1}\right)-\partial_x u(r)\right) \d r  , \varphi\left(\cdot-B^{(n)}_\ell + B^{(n)}_i \right) \right\rangle\right].
\end{aligned}
\]
Note that by definition of $N^{(n),\ell,2}(\varphi)$, 
\begin{equation*}
\begin{aligned}
\left|\sqrt{\Delta^{(n)}}K_1\right|^m = \left|N^{(n),\ell,2}_\ell(\varphi) - N^{(n),\ell,2}_k(\varphi)\right|^m. \\
\end{aligned}
\end{equation*}
Using Hölder inequality and \eqref{eq: varphi discrete derivative}, we can write
\[
\begin{aligned}
\left|K_2\right|^m &\leq  (\ell-k)^{m-1}\sum_{i=k+1}^\ell \left( \1^{(n),B-A}_i - p^{(n)}\left(S^{(n)}_{i-1}\right) \right)^m \left\|u\left(t_{i-1}^{(n)}\right)\right\|^m_{L^2}\left\|\varphi\right\|^m_{H^3}\left(B^{(n)}_i-B^{(n)}_{i-1}\right)^{m} \\
& \lesssim (\ell-k)^{m-1} \left(\Delta^{(n)}\right)^{m} \sum_{i=k+1}^\ell \|\varphi\|^m_{H^3} \leq \delta_{n,\ell-k}^m \|\varphi\|^m_{H^3}.
\end{aligned}
\]
Using Hölder inequality and \eqref{eq: varphi discrete derivative}, we obtain
\[
\begin{aligned}
\left|K_3\right|^m &\leq (\ell-k)^{m-1} \sum_{i=k+1}^{\ell} \left(\1_i^{(n),A}\right)^m \left\langle u\left(t^{(n)}_{i-1}\right), \left(\nabla^{(n)}_+ - \nabla^{(n)}_-\right) \varphi \left(\cdot-B^{(n)}_\ell + B^{(n)}_i\right)\right\rangle^m \\
&\leq (\ell-k)^{m-1} \sup_{t\in [0,T]} \| u(t) \|^m_{L^2} \sum_{i=k+1}^{\ell} \left\|\left(\nabla^{(n)}_+ - \nabla^{(n)}_-\right) \varphi \left(\cdot-B^{(n)}_\ell + B^{(n)}_i\right) \right\|^m_{L^2} \\
&\lesssim (\ell-k)^{m-1} \sum_{i=k+1}^\ell \|\varphi\|^m_{H^2} \left(\Delta^{(n)}\right)^{m} = \delta_{n,\ell-k}^m \|\varphi\|^m_{H^2}. 
\end{aligned}
\]
Using Hölder and Cauchy-Schwarz inequalities, Lemma \ref{lemma: taylor expansion of p, f and g}, \eqref{eq: varphi discrete derivative}, and \cref{assum: p}~\ref{assum: p: sup}, we have
\[
\begin{aligned}
&\left|\sqrt{\Delta^{(n)}}K_4\right|^m \leq (\ell-k)^{m-1} \left(\Delta^{(n)}\right)^{m/2} \cdot\\
&\quad\sum_{i=k}^{\ell-1} \left(p^{(n)}\left(S_{i}^{(n)}\right)-\frac{1}{\Delta^{(n)}}\int_{t_{i-1}^{(n)}}^{t_i^{(n)}}p\left(S(r)\right)\d r\right)\left\langle  \nabla_{+}^{(n)} u\left(t_{i-1}^{(n)}\right) , \varphi\left(\cdot-B^{(n)}_\ell +B^{(n)}_i \right) \right\rangle ^m \\
&\lesssim \delta_{n,\ell-k}^{m-1} \sup_{t\in [0,T]}\|u(t)\|^m_{L^2} \|\varphi\|^m_{H^2} \sum_{i=k}^{\ell-1} \Delta^{(n)} \left(\left(p_b\left(S\left(t_i^{(n)}\right)\right) + C^{(n)}_i\right)Z^{(n),B}_i \right)^m \\
&\quad+ \delta_{n,\ell-k}^{m-1} \sup_{t\in [0,T]}\|u(t)\|^m_{L^2} \|\varphi\|^m_{H^2} \sum_{i=k}^{\ell-1} \Delta^{(n)} \left( \left(p_y \left(S\left(t_i^{(n)}\right)\right) + C^{(n)}_i\right) Z^{(n),Y}_i + C^{(n)}\right)^m \\
&= K^{(n)} \delta_{n,\ell-k}^{m-1} \|\varphi\|^m_{H^2} \sum_{i=k}^{\ell-1} \Delta^{(n)} \left(\left|Z^{(n),B}_i\right|^m + \left| Z^{(n),Y}_i \right|^m \right )+ C^{(n)}\delta^m_{n,\ell-k}\|\varphi\|^m_{H^2} .
\end{aligned}
\]
Finally, using H\"older inequality, \eqref{eq: varphi discrete derivative}, and Cauchy-Schwarz inequality, we estimate
\begin{align*}
&\left|\sqrt{\Delta^{(n)}}K_5\right|^m\leq (\ell-k)^{m-1}\left(\Delta^{(n)}\right)^{m/2} \cdot\\
&\,\quad\sum_{i=k}^{\ell-1}\left\langle  \frac{1}{\Delta^{(n)}}\int_{t_{i-1}^{(n)}}^{t_i^{(n)}}p\left(S(r)\right)\left(\nabla_{+}^{(n)} u\left(t_{i-1}^{(n)}\right)-\partial_xu(r)\right)\d r , \varphi\left(\cdot-B^{(n)}_\ell + B^{(n)}_i \right) \right\rangle ^m \\
&\lesssim \delta^m_{n,\ell-k} \|\varphi\|_{H^2}^m\left(\Delta^{(n)}\right)^{m/2}+(\ell-k)^{m-1}\left(\Delta^{(n)}\right)^{m/2} \cdot\\
&\qquad\sum_{i=k}^{\ell-1}\left(\frac{1}{\Delta^{(n)}}\int_{t_{i-1}^{(n)}}^{t_i^{(n)}}p(S(r))\left\langle u(r)-u\left(t^{(n)}_{i-1}\right),\varphi'\left(\cdot-B_\ell^{(n)}+B_i^{(n)}\right)\right\rangle \d r\right)^m\\
&\lesssim \left(\Delta^{(n)}\right)^{m/2}\left[\delta^m_{n,\ell-k} \|\varphi\|_{H^2}^m+\right.\\
&\quad\left.(\ell-k)^{m-1}\sum_{i=k}^{\ell -1}\left(\frac{1}{\Delta^{(n)}}\int_{t_{i-1}^{(n)}}^{t_i^{(n)}}p(S(r))\int_{t_{i-1}^{(n)}}^r\left\langle u(q),\varphi''\left(\cdot-B_\ell^{(n)}+B_i^{(n)}\right)\right\rangle\d q \d r\right)^m\right]\\
&\lesssim \delta^m_{n,\ell-k} \|\varphi\|_{H^2}^m\left(\Delta^{(n)}\right)^{m/2}+\delta_{n,\ell-k}^{m}\left(\Delta^{(n)}\right)^{m/2}\|\varphi\|_{H^3}^m\sup_{t\in[0,T]}\|u(t)\|_{L^2}^m.
\end{align*}

Putting all estimates together yields the statement.\qed
\end{proof}

It remains to establish a pathwise estimate for the quantity $A_3^{(n),k,\ell}$ in the decomposition of the volume fluctuations in terms of a discrete integral, the process $N^{(n),\ell,3}$ defined in \eqref{eq: definition of N_3 martingales}, and a residual term with the desired scaling. 

\begin{lemma}
\label{lemma: pathwise estimate for source contribution}
Let $m\in 2\N\cup\{1\}$. $\varphi\in L^2$, and $0 \leq k < \ell \leq T_n$. Then almost surely
\[
\begin{aligned}
\left|A_3^{(n), k, \ell}\right|^m &\lesssim K^{(n)}\delta_{n,\ell-k}^{m-1} \|\varphi\|^m_{L^2} \sum_{i=k}^{\ell-1} \Delta^{(n)} \left(\left|Z^{(n),B}_i\right|^m + \left| Z^{(n),Y}_i \right|^m \right ) \\
&\qquad\qquad+ \left| N^{(n),\ell,3}_\ell(\varphi) - N^{(n),\ell,3}_k(\varphi) \right|^m+ K^{(n)} \delta_{n,\ell-k}^m \|\varphi\|^m_{L^2}.
\end{aligned}
\]
\end{lemma}

\begin{proof}
We consider the decomposition
\[
\begin{aligned}
&A_3^{(n), k, \ell}=\sum_{i=k+1}^{\ell} \left\langle \sqrt{\Delta^{(n)}} \left( J_i^{(n),C} - f^{(n)}\left(S^{(n)}_{i-1}\right) \right) , \varphi\left(\cdot-B^{(n)}_\ell + B^{(n)}_{i-1} \right) \right\rangle \\
&\quad+ \sum_{i=k+1}^{\ell} \left\langle \sqrt{\Delta^{(n)}} \left( J_i^{(n),C} - f^{(n)}\left(S^{(n)}_{i-1}\right) \right), \varphi \left(\cdot- B^{(n)}_\ell+B^{(n)}_{i-1}\right) - \varphi \left(\cdot- B^{(n)}_\ell+B^{(n)}_{i}\right) \right\rangle \\
&\quad+ \sum_{i=k+1}^{\ell} \left\langle \sqrt{\Delta^{(n)}} \left( f^{(n)}\left(S^{(n)}_{i-1}\right) - \frac{1}{\Delta^{(n)}} \int _{t^{(n)}_{i-1}} ^{t^{(n)}_{i}} f\left(S(r)\right) \d r \right) , \varphi\left(\cdot-B^{(n)}_\ell + B^{(n)}_i \right) \right\rangle \\
&=: L_1 + L_2 + L_3.
\end{aligned}
\]
By the definition of $N^{(n),\ell,3}(\varphi)$, we have    
\[
\begin{aligned}
|L_1|^m = \left| N^{(n),\ell,3}_\ell(\varphi) - N^{(n),\ell,3}_k(\varphi) \right|^m.
\end{aligned}
\]
Using Cauchy-Schwarz and Hölder inequalities, the fact that $J^{(n)}_i=0$ if $\1^{(n),B-A}_i \neq 0$, and \cref{assum: bound on pi and omega}\ref{assum: bound on pi and omega: omega}, we observe that
\[
\begin{aligned}
|L_2|^m &\leq (\ell-k)^{m-1} \left(\Delta^{(n)}\right)^{m/2} \sum_{i=k+1}^\ell \left\|J^{(n),C}_i - f^{(n)}\left(S^{(n)}_{i-1}\right) \right\|^m_{L^2} \|\varphi\|^m_{H^1} \left(B^{(n)}_i - B^{(n)}_{i-1}\right)^{m}\\
&= (\ell-k)^{m-1} \left(\Delta^{(n)}\right)^{m/2} \sum_{i=k+1}^\ell \left\|f^{(n)}\left(S^{(n)}_{i-1}\right) \right\|^m_{L^2} \|\varphi\|^m_{H^1} \left(B^{(n)}_i - B^{(n)}_{i-1}\right)^{m}\\
& \lesssim (\ell-k)^{m-1} \left(\Delta^{(n)}\right)^{m/2} \sum_{i=k+1}^\ell \|\varphi\|^m_{H^1} \left(\Delta^{(n)}\right)^{m} =
 \left(\Delta^{(n)}\right)^{m/2}\delta_{n,\ell-k}^m \|\varphi\|^m_{H^1}.
\end{aligned}
\]
Using Lemma \ref{lemma: taylor expansion of p, f and g}, Hölder inequality, and \cref{assum: f}~\ref{assum: f: sup of L^2 norms of f}, we obtain
\[
\begin{aligned}
&|L_3|^m \lesssim \delta_{n,\ell-k}^{m-1} \left[\sum_{i=k}^{\ell-1} \Delta^{(n)} \left\langle f_b\left(S\left(t_i^{(n)}\right)\right) + C^{(n)}_i, \varphi\left(\cdot-B^{(n)}_\ell + B^{(n)}_i\right) \right\rangle^m \left|Z^{(n),B}_i\right|^m  \right.\\
&\qquad+ \left. \sum_{i=k}^{\ell-1} \Delta^{(n)}  \left\langle f_y\left(S\left(t_i^{(n)}\right)\right) + C^{(n)}_i, \varphi\left(\cdot-B^{(n)}_\ell + B^{(n)}_i\right)\right\rangle^m \left|Z^{(n),Y}_i\right|^m + C^{(n)}\|\varphi\|_{L^2}^m\right]\\
&\quad\leq K^{(n)}\delta_{n,\ell-k}^{m-1} \|\varphi\|^m_{L^2} \sum_{i=k}^{\ell-1} \Delta^{(n)}  \left( \left|Z^{(n),B}_i\right|^m + \left|Z^{(n),Y}_i\right|^m  \right) + C^{(n)}\delta^m_{n,\ell-k}\|\varphi\|_{L^2}^m.
\end{aligned}
\]
Putting all estimates together yields the statement.\qed
\end{proof}

Combining the three lemmata proven above, we obtain the following pathwise growth estimate for the fluctuations of both, price and volume.

\begin{proposition}
\label{prop: pathwise estimate for l-k}
Let $m\in 2\N\cup\{1\}$, $\varphi\in H^3$, and $0 \leq k < \ell \leq T_n$. Then almost surely
\begin{equation*}
\begin{aligned}
&\left| Z^{(n),B}_\ell - Z^{(n),B}_k \right| ^m + \left|\left\langle Z^{(n),u}_\ell - Z^{(n),u}_k, \varphi \right\rangle \right|^m \lesssim 
 \widetilde{\mathcal N}^{(n),\ell,m}_k(\varphi) + \left|A_1^{(n), k, \ell}\right|^m
\\
&\qquad\qquad+ K^{(n)} \delta_{n,\ell-k}^{m-1} \|\varphi\|^m_{H^3} \sum_{i=k}^{\ell-1} \Delta^{(n)} \left(\left|Z^{(n),B}_i\right|^m+ \left| Z^{(n),Y}_i\right|^m \right) + K^{(n)}\delta_{n,\ell-k}^m \|\varphi\|^m_{H^3}.
\end{aligned}
\end{equation*}
\end{proposition}

\begin{proof}
Follows from Proposition \ref{prop: decomposition of discrete dynamics of volume fluctuations} and Lemmas \ref{lemma: pathwise estimate for price fluctuations}, \ref{lemma: pathwise estimate for transport contribution}, \ref{lemma: pathwise estimate for source contribution}.\qed
\end{proof}

We emphasize that the estimate in Proposition \ref{prop: pathwise estimate for l-k} contains the term $|A_1^{(n), k, \ell}|^m$ on the RHS, which depends on the volume fluctuation $Z^{(n),u}_k$. This precludes us from applying Grönwall inequality directly. To bypass this problem, we first consider the pathwise estimate for $\varphi = h$ and $k = 0$. Then, we can apply Grönwall inequality, since we know that the initial volume fluctuations $Z^{(n),u}_0$ are vanishing in the limit due to \cref{assum: initial values}\ref{assum: initial values: u^n_0 - u_0 and B^n_0 - B_0}. 

\begin{proposition}
\label{prop: pathwise estimate for Zs}

Let $m \in 2\N$, $\varphi\in H^3$, and $\ell \leq T_n$. Then almost surely
\begin{align*}
\left| Z^{(n),B}_\ell\right| ^m + \left| Z^{(n),Y}_\ell \right| ^m &\lesssim K^{(n)} + \sup_{j\in\{0,\dots \ell\}} \widetilde{\mathcal N}^{(n),j,m}_0 (\varphi), \\
\left|\left\langle Z_\ell^{(n),u}, \varphi\right\rangle\right|^m &\lesssim \widetilde{\mathcal N}^{(n),\ell,m}_0(\varphi) + K^{(n)} \|\varphi\|^m_{H^3} \left(1+\sup_{j\in\{0,\dots \ell\}} \widetilde{\mathcal N}^{(n),j,m}_0 (h) \right) .
\end{align*}
\end{proposition}

\begin{proof}
Due to \cref{assum: initial values}~\ref{assum: initial values: u^n_0 - u_0 and B^n_0 - B_0} it holds that
\begin{equation*}
\begin{split}
\label{eq: z0 estimate}
\left|A_1^{(n), 0, \ell}\right|^m & = \left|\int_0^{B^{(n)}_\ell-B^{(n)}_0} \left\langle Z^{(n),u}_0, \varphi'(\cdot+r) \right\rangle \d r \right|^m \lesssim \left(\Delta^{(n)}\right)^{m/2} \|\varphi\|_{H^1}^m \left(B^{(n)}_\ell - B^{(n)}_0 \right)^m. \\
\end{split}
\end{equation*}
By Proposition \ref{prop: pathwise estimate for l-k},  the fact that $\left| B^{(n)}_\ell - B^{(n)}_0 \right| \leq T$ a.s.~and $\delta_{n,\ell} = \ell \Delta^{(n)} \leq T$, we obtain
\begin{equation}
\label{eq: gronwall hypothesis for Z^B_k abd Z^Y_k}
\begin{aligned}
&\left| Z^{(n),B}_\ell\right| ^m + \left|\left\langle Z^{(n),u}_\ell, \varphi \right\rangle\right| ^m\\
&\quad \lesssim C^{(n)}+ K^{(n)}\|\varphi\|^m_{H^3} + \widetilde{\mathcal N}^{(n),\ell,m}_0(\varphi) + \|\varphi\|^m_{H^3} \sum_{i=0}^{\ell-1} \Delta^{(n)} \left( \left| Z^{(n),B}_i \right|^m + \left| Z^{(n),Y}_i \right|^m \right ).
\end{aligned}
\end{equation}
Plugging $\varphi = h$ into \eqref{eq: gronwall hypothesis for Z^B_k abd Z^Y_k}, we have 
\[
\begin{aligned}
&\left| Z^{(n),B}_\ell \right| ^m + \left| Z^{(n),Y}_\ell \right| ^m \lesssim K^{(n)}+ \widetilde{\mathcal N}^{(n),\ell,m}_0(h) + \sum_{i=0}^{\ell-1} \Delta^{(n)} \left(\left|Z^{(n),B}_i\right|^m + \left| Z^{(n),Y}_i\right|^m \right).
\end{aligned}
\]
Using Grönwall inequality, we obtain
\begin{equation}
\label{eq: gronwall estimates from Z^B_k Z^Y_k}    
\begin{aligned}
\left| Z^{(n),B}_\ell\right| ^m + \left| Z^{(n),Y}_\ell \right| ^m &\lesssim K^{(n)} + \widetilde{\mathcal N}^{(n),\ell,m}_0(h) + \sum_{j=0}^{\ell-1} \Delta^{(n)} \widetilde{\mathcal N}^{(n),j,m}_0(h)\\
& \lesssim K^{(n)} + \sup_{j\in\{0,\dots, \ell\}} \widetilde{\mathcal N}^{(n),j,m}_0 (h) .
\end{aligned}
\end{equation}
Using Proposition \ref{prop: pathwise estimate for l-k}, \eqref{eq: z0 estimate}, \eqref{eq: gronwall estimates from Z^B_k Z^Y_k}, and the fact that $\delta_{n,\ell} = \ell\Delta^{(n)} \leq T$, we observe that
\begin{equation*}
\begin{aligned}
&\left|\left\langle Z_{\ell}^{(n),u}, \varphi\right\rangle\right|^m \lesssim \widetilde{\mathcal N}^{(n),\ell,m}_0(\varphi) + K^{(n)}\|\varphi\|^m_{H^3}\left(1+ \sum_{i=0}^{\ell-1} \Delta^{(n)} \left(\left|Z^{(n),B}_i\right|^m + \left| Z^{(n),Y}_i\right|^m \right ) \right)\\
&\quad\lesssim \widetilde{\mathcal N}^{(n),\ell,m}_0(\varphi) + K^{(n)}\|\varphi\|^m_{H^3}\left(1+ \sum_{i=0}^{\ell-1} \Delta^{(n)} \left(K^{(n)} + \sup_{j\in\{0,\dots, i\}} \widetilde{\mathcal N}^{(n),j,m}_0 (h)\right) \right)\\
&\quad\lesssim \widetilde{\mathcal N}^{(n),\ell,m}_0(\varphi) + K^{(n)} \|\varphi\|^m_{H^3} \left(1+\sup_{j\in\{0,\dots \ell\}} \widetilde{\mathcal N}^{(n),j,m}_0 (h) \right).
\end{aligned}
\end{equation*}\qed
\end{proof}

%%%%%%%%%%%%%%%%%%%%%%%%%%%%%%%%%%%%%%%%%%%%%%%%%%%%%%%%%%%%
%%%%%%%%%%%%%%%%%%%%%%%%%%%%%%%%%%%%%%%%%%%%%%%%%%%%%%%%%%%%
%%%%%%%%%%%%%%%%%%%%%%%%%%%%%%%%%%%%%%%%%%%%%%%%%%%%%%%%%%%%

\subsection{Moment estimates}

We first give a moment estimate for the term $\widetilde{\mathcal{N}}_0^{(n),k,m}$  appearing in the pathwise estimates above.

\begin{lemma}
\label{lemma: bdg estimate of the martingale parts}

For $m\in 2\N\cup\{1\}$, $\varphi\in H^2$, and $0\leq k<\ell \leq T_n$, it holds that
\begin{equation*}
\E\left[\sup_{j\in\{k,\dots,\ell\}}\widetilde{\mathcal N}^{(n),j,m}_k(\varphi) \right] \lesssim \delta^{m/2}_{n,\ell-k} \left(1+\|\varphi\|^m_{H^2}\right).
\end{equation*}
\end{lemma}

\begin{proof}
As $N^{(n),1}$ is a martingale, we may apply Burkholder-Davis-Gundy inequality and afterwards H\"older to obtain
\begin{align*}
\E\left[\sup_{j\in\{k,\dots,\ell\}}\left|N^{(n),1}_j-N^{(n),1}_k\right|^m\right]&\lesssim \E\left[\left| \sum_{i=k+1}^\ell \left(\delta N^{(n),1}_i\right)^2 \right|^{m/2}\right]\\
&\leq (\ell-k)^{m/2-1} \sum_{i=k+1}^\ell\E\left[\left(\delta N^{(n),1}_i\right)^m \right]
\leq \delta^{m/2}_{n,\ell-k} .\end{align*}
Next consider the $L^2$-valued martingale 
\[N^{(n),2}_k:=\sum_{i=1}^k\sqrt{\Delta^{(n)}}\left(\1_i^{(n),B-A} - p^{(n)} \left(S^{(n)}_{i-1}\right)\right)u\left(t^{(n)}_{i-1}\right),\quad k\in\N.\] 
 Applying first the Cauchy-Schwarz inequality to the process $N_j^{(n),k,2}(\varphi),\ j\leq k,$ and afterwards the Burkholder-Davis-Gundy inequality for Hilbert-space valued martingales (cf.~\cite[Theorem 1.1]{marinelli2016}) to $N^{(n),2}$ we obtain
\begin{align*}
&\E\left[\sup_{j\in\{k,\dots,\ell\}}\left|N^{(n),j,2}_j(\varphi)-N^{(n),j,2}_k(\varphi)\right|^m\right]\lesssim
\E\left[\sup_{j\in\{k,\dots,\ell\}}\left\|N_j^{(n),2}-N_k^{(n),2}\right\|_{L^2}^m\right]%(1+\Delta^{(n)})
\|\varphi\|_{H^2}^m \\
&\lesssim
\E\left[\left|\sum_{i=k+1}^\ell\Delta^{(n)}\left(\1_i^{(n),B-A} - p^{(n)} \left(S^{(n)}_{i-1}\right)\right)^2\left\|u\left(t^{(n)}_{i-1}\right)\right\|^2_{L^2}\right|^{m/2}\right]\|\varphi\|_{H^2}^m \\
&\lesssim\left(\Delta^{(n)}\right)^{m/2}(\ell-k)^{m/2-1}  \|\varphi\|_{H^2}^m\sum_{i=k+1}^\ell \E \left[\left(\1_i^{(n),B-A} - p^{(n)} \left(S^{(n)}_{i-1}\right)\right)^m\right] \leq \delta_{n,\ell-k}^{m/2} \|\varphi\|_{H^2}^m.
\end{align*}
Next, we define the martingale
\[N^{(n),3}_k({\varphi}):=\sum_{i=1}^k\sqrt{\Delta^{(n)}}\left\langle J_i^{(n),C}-f^{(n)}\left(S^{(n)}_{i-1}\right),{\varphi}\left(\cdot+B^{(n)}_{i-1}\right)\right\rangle,\quad k\in\N.\]
Using Assumption \ref{assum: f}, we obtain 
\[
\begin{aligned}
&\E\left[\left(\delta N^{(n),3}_i(\varphi)\right)^m \right] = \left(\Delta^{(n)}\right)^{m/2} \E \left[\left\langle J_{i}^{(n),C}-f^{(n)}\left(S_{i-1}^{(n)}\right), \varphi\left(\cdot+B_{i-1}^{(n)}\right)\right\rangle ^m\right] \\
&\quad\lesssim \left(\Delta^{(n)}\right)^{m/2} \E\left[\left(\frac{M}{\Delta^{(n)}} \int_{I^{(n)}\left(\pi^{(n)}_{i}\right)} \varphi\left(x+B_{i-1}^{(n)}\right) d x \right)^m +\left\|f^{(n)}\left(S^{(n)}_{i-1}\right)\right\|_{L^2}^m\left\|\varphi\right\|_{L^2}^m\right]\\
&\quad\lesssim \left(\Delta^{(n)}\right)^{m/2} \left(\|\varphi\|^m_{L^{\infty}}+\|\varphi\|^m_{L^2}\right)\lesssim \left(\Delta^{(n)}\right)^{m/2} \|\varphi\|^m_{H^1}.
\end{aligned}
\]
Then applying Burkholder-Davis-Gundy and H\"older inequalities, we have
\begin{equation}\label{N3}
\begin{split}
&\E\left[\sup_{j\in\{k,\dots,\ell\}}\left|N_j^{(n),3}(\varphi)-N_k^{(n),3}(\varphi)\right|^m\right]\lesssim\E\left[\left|\sum_{i=k+1}^\ell\left(\delta N_i^{(n),3}(\varphi)\right)^2\right|^{m/2}\right]\\
&\qquad\qquad\qquad\qquad\leq (\ell-k)^{m/2-1}\sum_{i=k+1}^\ell\E\left[\left(\delta N_i^{(n),3}(\varphi)\right)^m\right]\leq \delta^{m/2}_{n,\ell-k}\|\varphi\|^m_{H^1}.
\end{split}
\end{equation}
As
\begin{align*}
&\E\left[\sup_{j\in\{k,\dots,\ell\}}\left|N_k^{(n),j,3}(\varphi)\right|^m\right]\lesssim
\E\left[\sup_{j\in\{k,\dots,\ell\}}\left|N_j^{(n),3}-N_k^{(n),3}(\varphi)\right|^m\right]
+\\
&\E\left[\sup_{j\in\{k,\dots,\ell\}}\left|\sum_{i=k+1}^j\sqrt{\Delta^{(n)}}\left\langle J_i^{(n)}-f^{(n)}\left(S^{(n)}_{i-1}\right),\varphi\left(\cdot-B_j^{(n)}+B^{(n)}_{i-1}\right)-{\varphi}\left(\cdot+B^{(n)}_{i-1}\right)\right\rangle\right|^m\right],
\end{align*}
it remains to estimate the term in the line above. To this end, note that $B_k^{(n)}\in[-L,L]$ a.s.~and thus
\begin{align*}
&\E\left[\sup_{j\in\{k,\dots,\ell\}}\left|\sum_{i=k+1}^j\sqrt{\Delta^{(n)}}\left\langle J_i^{(n),C}-f^{(n)}\left(S^{(n)}_{i-1}\right),\varphi\left(\cdot-B_j^{(n)}+B^{(n)}_{i-1}\right)-\varphi\left(\cdot+B^{(n)}_{i-1}\right)\right\rangle\right|^m\right]\\
&=\E\left[\sup_{j\in\{k,\dots,\ell\}}\left|\sum_{i=k+1}^j\sqrt{\Delta^{(n)}}\left\langle J_i^{(n),C}-f^{(n)}\left(S^{(n)}_{i-1}\right),\int_0^{-B_j^{(n)}}\varphi'\left(\cdot+r+B_{i-1}^{(n)}\right)dr\right\rangle\right|^m\right]\\
&\lesssim\int_{-L}^L\E\left[\sup_{j\in\{k,\dots,\ell\}}\left|\sum_{i=k+1}^j\sqrt{\Delta^{(n)}}\left\langle J_i^{(n),C}-f^{(n)}\left(S^{(n)}_{i-1}\right),\varphi'\left(\cdot+r+B_{i-1}^{(n)}\right)\right\rangle \right|^m\right]dr\\
&\lesssim \delta^{m/2}_{n,\ell-k}\|\varphi\|^m_{H^2},
\end{align*}
where the last estimate follows analogously to \eqref{N3}. Putting all the estimates together, the claimed result follows.
\qed
\end{proof}

Finally, combining the previous lemma with Proposition \ref{prop: pathwise estimate for Zs} allows us to give an estimate for $A_1^{(n),k,\ell}$, which still appears on the RHS in Proposition \ref{prop: pathwise estimate for l-k}.

\begin{lemma}
\label{lemma: moments initial value}
It holds for all $m\in 2\N\cup\{1\}$ and $\varphi\in H^4$,
\begin{equation*}
\E\left[\left|A_1^{(n), k, \ell}\right|^m\right] \lesssim K^{(n)} \delta^m_{n,\ell-k} \left(1+\|\varphi\|^m_{H^4} (1+\|h\|^m_{H^2})\right).
\end{equation*}
\end{lemma}

\begin{proof}
Note that since $B^{(n)}_\ell-B^{(n)}_k \leq \delta_{n,\ell-k}$, it holds
\[
A_1^{(n), k, \ell} = \int_0^{B^{(n)}_\ell-B^{(n)}_k} \left\langle Z^{(n),u}_k, \varphi'(\cdot+r) \right\rangle \d r \leq \int_{-\delta_{n,\ell-k}}^{\delta_{n,\ell-k}} \left|\langle Z^{(n),u}_k,  \varphi'(\cdot+r)\rangle \right| \d r.
\]
By invoking Proposition \ref{prop: pathwise estimate for Zs} with $m=1$, we obtain
\begin{equation*}
\begin{split}
&\int_{-\delta_{n,\ell-k}}^{\delta_{n,\ell-k}} \left|\langle Z^{(n),u}_k, \varphi'(\cdot+r)\rangle \right| \d r \\
&\qquad\lesssim \int_{-\delta_{n,\ell-k}}^{\delta_{n,\ell-k}} \widetilde{\mathcal{N}}_0^{(n),\ell,1}(\varphi'(\cdot+r))\d r+K^{(n)} \delta_{n,\ell-k} \|\varphi\|_{H^4} \left(1+\sup_{j\in\{0,\dots \ell\}} \widetilde{\mathcal N}^{(n),j,1}_0 (h) \right).
\end{split}
\end{equation*}
Applying H\"older and Lemma \ref{lemma: bdg estimate of the martingale parts}, we thus get
\begin{align*}
\E\left[\left|A_1^{(n), k, \ell}\right|^m\right] &\lesssim \delta^{m-1}_{n,\ell-k} \int_{-\delta_{n,\ell-k}}^{\delta_{n,\ell-k}}\E\left[\widetilde{\mathcal{N}}_0^{(n),\ell,m}(\varphi'(\cdot+r))\right]dr\\
&\qquad+K^{(n)}\delta^m_{n,\ell-k}\|\varphi\|_{H^4}^m\left(1+\E\left[\sup_{j\in\{0,\dots,\ell\}}\widetilde{\mathcal{N}}_0^{(n),j,m}(h)\right]\right)\\
&\lesssim \delta_{n,\ell-k}^m\delta_{n,\ell}^{m/2}(1+\|\varphi\|^m_{H^3})+K^{(n)}\delta^m_{n,\ell-k}\|\varphi\|^m_{H^4}\left(1+\delta^m_{n,\ell}(1+\|h\|_{H^2}^m)\right)
\end{align*}
As $\delta_{n,\ell}^{m/2} \leq T^{m/2}<\infty$, we obtain
\[
\begin{aligned}
\E\left[\left|A_1^{(n), k, \ell}\right|^m\right] &\lesssim K^{(n)} \delta^m_{n,\ell-k} \left(1+\|\varphi\|^m_{H^4} (1+\|h\|^m_{H^2})\right).
\end{aligned}
\]\qed
\end{proof}

The final moment estimates of the increments of $Z^{(n)}$ now follow from a combination of the previous results.

\begin{proposition}
\label{prop: moment estimate for Z}

For $m\in 2\N\cup\{1\}$, $\varphi\in H^4$, and $0\leq s\leq t \leq T$, it holds that
\begin{align*}
\E\left[ \left| Z^{(n),B}(t) - Z^{(n),B} (s) \right| ^m \right] + \E\left[ \left| Z^{(n),Y}(t) - Z^{(n),Y}(s) \right| ^m \right] &\leq K^{(n)} \left((t-s)^{m/2} + (t-s)^m \right),
\end{align*}
and
\begin{align*}
\E \left[\left|\left\langle Z^{(n),u}(t) - Z^{(n),u}(s), \varphi\right\rangle\right|^m\right] &\leq K^{(n)} \left((t-s)^{m/2} + (t-s)^m \right) \left(\|\varphi\|^m_{H^4} + \|\varphi\|^{2m}_{H^4}\right).
\end{align*}
\end{proposition}

\begin{proof}

We first consider $0\leq k\leq \ell \leq T_n$. Using Proposition \ref{prop: pathwise estimate for Zs} and Lemma \ref{lemma: bdg estimate of the martingale parts}, 
\begin{equation}
\begin{split}
\label{eq: moment estimates for Z^B_k and Z^Y_k}
\E\left[ \left|Z^{(n),B}_k\right|^m \right] + \E\left[ \left| Z^{(n),Y}_k\right|^m \right] & \lesssim K^{(n)} + \E\left[\sup_{j\in\{0,\dots k\}} \widetilde{\mathcal N}^{(n),j,m}_0(\varphi) \right] \\
& \lesssim K^{(n)} + \delta^{m/2}_{n,k} (1+\|\varphi\|^m_{H^2}) \lesssim K^{(n)} + \|\varphi\|^m_{H^2}.
\end{split}
\end{equation}
Plugging $\varphi = h \in H^4(U)$ into Proposition \ref{prop: pathwise estimate for l-k} and using Lemma \ref{lemma: bdg estimate of the martingale parts}, Lemma \ref{lemma: moments initial value}, and \eqref{eq: moment estimates for Z^B_k and Z^Y_k}, we obtain
\[
\begin{aligned}
&\E\left[ \left| Z^{(n),B}_\ell - Z^{(n),B}_k \right| ^m \right] + \mathbb E\left[ \left| Z^{(n),Y}_\ell - Z^{(n),Y}_k \right| ^m \right] \\
&\qquad\leq   K^{(n)} \delta_{n,\ell-k}^{m-1} \sum_{i=k}^{\ell-1} \Delta^{(n)} \E\left[ \left|Z^{(n),B}_i\right|^m+ \left| Z^{(n),Y}_i\right|^m \right]  +\delta^{m/2}_{n,\ell-k} +K^{(n)} \delta_{n,\ell-k}^m 
\\ & \qquad \lesssim K^{(n)} \left(\delta^{m/2}_{n,\ell-k} + \delta_{n,\ell-k}^m \right).
\end{aligned}
\]
Analogously, we obtain
\begin{equation*}
\begin{aligned}
\E \left[\left|\left\langle Z_{\ell}^{(n),u} - Z_{k}^{(n),u}, \varphi\right\rangle\right|^m\right] &\lesssim  K^{(n)} \delta_{n,\ell-k}^{m-1} \|\varphi\|^m_{H^3} \sum_{i=k}^{\ell-1} \Delta^{(n)} \E\left[ \left|Z^{(n),B}_i \right|^m+\left| Z^{(n),Y}_i \right|^m \right] \\
&\qquad+\delta^{m/2}_{n,\ell-k} \left(1+\|\varphi\|^m_{H^2}\right) + K^{(n)} \delta^m_{n,\ell-k} \left(1+\|\varphi\|^m_{H^4} \right) \\
&\lesssim K^{(n)} \left(\delta^{m/2}_{n,\ell-k} + \delta^m_{n,\ell-k}\right) \left(1+\|\varphi\|^m_{H^4} \right).
\end{aligned}
\end{equation*}
If $t^{(n)}_i \leq s \leq t < t^{(n)}_{i+1}$ for some $i\in\{0, \dots, T_n\}$, it follows from Proposition \ref{prop: law of large numbers} that
\[
    \begin{aligned}
        &\E\left[ \left| Z^{(n),B}(t) - Z^{(n),B} (s) \right| ^m \right] = \E\left[\left( \frac{\left|\int_s^t p(S(r)) \d r\right|}{\left(\Delta^{(n)}\right)^{1/2}} \right)^m \right] \lesssim (t-s)^{m/2}, \\
        &\E\left[ \left| \left\langle Z^{(n),u}(t) - Z^{(n),u} (s), \varphi \right\rangle \right| ^m \right] \\
        &\quad= \E\left[\left( \frac{\left|\int_s^t p(S(r)) \left\langle u(S(r)), - \partial_x \varphi \right\rangle + \left\langle f(S(r)), \varphi \right\rangle \d r\right|}{\left(\Delta^{(n)}\right)^{1/2}} \right)^m \right] \lesssim (t-s)^{m/2} \|\varphi\|_{H^1}^m \\
    \end{aligned}
\]
and with the choice $h = \varphi$ also
\[
    \E\left[ \left| Z^{(n),Y}(t) - Z^{(n),Y} (s) \right| ^m \right] \lesssim (t-s)^{m/2} \|h\|_{H^1}^m.
\]
Now, for arbitrary $0\leq s\leq t\leq T$, there is $i \leq  j$ such that $t^{(n)}_i = \Delta^{(n)} \lfloor \frac{s}{ \Delta^{(n)}} \rfloor$ and $t^{(n)}_j = \Delta^{(n)} \lfloor \frac{t}{\Delta^{(n)}} \rfloor$. Therefore,
\[
    \begin{aligned}
        &\E\left[ \left| Z^{(n),B}(t) - Z^{(n),B} (s) \right| ^m \right] \lesssim \E\left[ \left| Z^{(n),B}(t) - Z^{(n),B} \left(t^{(n)}_j \right) \right| ^m \right] \\
        &\qquad\qquad+ \E\left[ \left| Z^{(n),B} \left(t^{(n)}_j\right) - Z^{(n),B} \left(t^{(n)}_i \right) \right| ^m \right] + \E\left[ \left| Z^{(n),B} \left(t^{(n)}_i\right) - Z^{(n),B} (s) \right| ^m \right] \\
        &\qquad\lesssim K^{(n)} \left( (t-s)^{m/2} + (t-s)^{m} \right).
    \end{aligned}
\]
The bounds for $Z^{(n),Y}(t) - Z^{(n),Y}(s)$ and $\left\langle Z^{(n),u}(t) - Z^{(n),u}(s), \varphi \right\rangle$ follow analogously.

\qed
\end{proof}

%%%%%%%%%%%%%%%%%%%%%%%%%%%%%%%%%%%%%%%%%%%%%%%%%%%%%%%%%%%%
%%%%%%%%%%%%%%%%%%%%%%%%%%%%%%%%%%%%%%%%%%%%%%%%%%%%%%%%%%%%
%%%%%%%%%%%%%%%%%%%%%%%%%%%%%%%%%%%%%%%%%%%%%%%%%%%%%%%%%%%%

\section{Martingale problem and weak solution}
\label{sec: martingale problem}

Using the moment estimates established in the previous section, we establish in this section the tightness of the price volume fluctuations and prove that any weak limit is continuous. Subsequently, we prove a series of martingale relations, both in the pre-limit and the limit. This allows us to show that the limiting dynamics can be obtained as a weak solution of our evolution equation by a martingale representation argument.  

%%%%%%%%%%%%%%%%%%%%%%%%%%%%%%%%%%%%%%%%%%%%%%%%%%%%%%%%%%%%
%%%%%%%%%%%%%%%%%%%%%%%%%%%%%%%%%%%%%%%%%%%%%%%%%%%%%%%%%%%%
%%%%%%%%%%%%%%%%%%%%%%%%%%%%%%%%%%%%%%%%%%%%%%%%%%%%%%%%%%%%

\subsection{Tightness}
\label{subsec: tightness}

We start by deriving from Proposition \ref{prop: moment estimate for Z} an estimate for the moments of the dual Sobolev norm, which we will use to infer continuity of the limiting process.

\begin{lemma}
\label{lemma: H^-5 estimate}

Let $0\leq s \leq t \leq T$. Then
\begin{equation*}
\E \left[\left\|Z^{(n),u}(t) - Z^{(n),u}(s) \right\|_{H^{-5}}^{4}\right] \lesssim K^{(n)} \left((t-s)^{4} + (t-s)^{2} \right).
\end{equation*}

\end{lemma}

\begin{proof}

Let $\{e_k:k\in\N\}$ be an orthonormal basis of $H^5(U)$. Note that by Maurin's theorem (e.g. \cite{DonoghueTheoremMaurin1964} or \cite[p.~334]{WalshIntroductionStochasticPartial1986}) there exists $L<\infty$ such that
\begin{equation}
\label{eq: maurin theorem}
\sum_{i=1}^\infty \|e_i\|^2_{H^4} \leq L.
\end{equation}
Also note that $\|e_i\|^m_{H^4}\leq \|e_i\|^m_{H^5}=1$ and thus
\begin{equation}
\label{eq: sum of 2nd and 4th power of the ONB}
\|e_i\|^2_{H^4} + \|e_i\|^{4}_{H^4} \leq 2 \|e_i\|^2_{H^4}.
\end{equation}
Using Proposition \ref{prop: moment estimate for Z}, \eqref{eq: maurin theorem}, and \eqref{eq: sum of 2nd and 4th power of the ONB}, we obtain
\[
\begin{aligned}
&\E\left[\left\|Z^{(n),u}(t) - Z^{(n),u}(s)\right\|_{H^{-5}}^{4}\right] = \E\left[\left(\sum_{i=1}^\infty \left\langle Z^{(n),u}(t) - Z^{(n),u}(s), e_i \right\rangle^2 \right)^{2}\right] \\
&\quad\qquad\leq \sum_{i,j=1}^\infty \E\left[\left\langle Z^{(n),u}(t) - Z^{(n),u}(s), e_i \right\rangle^4\right]^{1/2} \E\left[ \left\langle Z^{(n),u}(t) - Z^{(n),u}(s), e_j \right\rangle^4\right]^{1/2} \\
&\quad\qquad\leq K^{(n)} \left((t-s)^4 + (t-s)^2 \right) \sum_{i,j=1}^\infty \|e_i\|^2_{H^4} \|e_j\|^2_{H^4} \\
&\quad\qquad \leq L^2 K^{(n)} \left( (t-s)^4 + (t-s)^2 \right).
\end{aligned}
\]\qed
\end{proof}

We now proceed to prove tightness of the fluctuation dynamics and continuity of the limit using the moment estimates from the previous section. To this end, we consider probability measures $\Q^{(n)},\ n\in\N,$ on the path space $D([0,T], \R\times L^2(\R))$ with canonical process denoted by $Z=(Z^B, Z^u)$, such that for all $n\in\N$,
\[
\mathscr L^{\Q^{(n)}} (Z) = \mathscr L^{\P^{(n)}} \left(Z^{(n)}\right).
\]
\begin{proposition}
\label{prop: tightness}
The sequence $\Q^{(n)}$ is tight in $D([0,T], \R\times H^{-4}(U))$ and each limiting measure $\Q^*$ is supported on $C([0,T], \R\times H^{-5}(U))$.
\end{proposition}

\begin{proof}

Let us put $K_n(t):=\max \left\{k \in \mathbb{N}: k \Delta^{(n)} \leq t \wedge T\right\}$, and consider an arbitrary sequence of bounded $\mathbb F^{(n)}$-stopping times $(\tau_n)$ and a vanishing sequence $(\delta_n)\subset\R_+$. By the Markov inequality and Proposition \ref{prop: moment estimate for Z}, we obtain for $\varphi\in H^4$,
\begin{equation*}
\begin{aligned}
&\Q^{(n)} \left(\left|Z_{K_n\left(\tau_{n}+\delta^{(n)}\right)}^{B}-Z_{K_n\left(\tau_{n}\right)}^{B}\right|+\left|\left\langle Z_{K_n\left(\tau_{n}+\delta_n\right)}^{u}-Z_{K_n\left(\tau_{n}\right)}^{u}, \varphi\right\rangle\right| > \varepsilon\right)\\
&\quad\leq 2\varepsilon^{-2} \left(\E^{\Q^{(n)}} \left[ \left|Z_{K_n\left(\tau_{n}+\delta_n\right)}^{B}-Z_{K_n\left(\tau_{n}\right)}^{B}\right|^2\right] + \E^{\Q^{(n)}} \left[\left|\left\langle Z_{K_n\left(\tau_{n}+\delta_n\right)}^{u}-Z_{K_n\left(\tau_{n}\right)}^{u}, \varphi\right\rangle\right|^2\right] \right) \\
&\quad\leq \varepsilon^{-2} K^{(n)} \left(\|\varphi\|^2_{H^4} + \|\varphi\|^{4}_{H^4}\right) \left(\delta^2_n + \delta_n \right)
\longrightarrow 0.
\end{aligned}
\end{equation*}
Thus Aldous' criterion \cite[Theorem 6.8]{WalshIntroductionStochasticPartial1986} implies tightness  in $D([0,T], \R^2)$ of the laws of $(Z^{(n), B}, \langle Z^{(n),u}, \varphi \rangle)$. Since $\varphi$ is arbitrary, tightness in $D([0,T], \mathbb R \times \mathcal E')$ follows by Mitoma's theorem \cite[Theorem 6.13]{WalshIntroductionStochasticPartial1986}. Moreover, from Proposition \ref{prop: moment estimate for Z}, we obtain
\begin{equation}
\label{eq: estimate for tightness in H^-3}
\begin{aligned}
\E^{\Q^{(n)}}\left[\left| Z^{(n),B}_k \right|\right] + \E^{\Q^{(n)}}\left[\left|\left\langle Z_{k}^{(n),u}, \varphi\right\rangle\right|\right] &\leq K^{(n)} \left(\|\varphi\|_{H^4}+\|\varphi\|_{H^4}^{2}\right).
\end{aligned}
\end{equation}
Thus tightness in $D([0,T], \R\times H^{-4}(U))$ follows from \eqref{eq: estimate for tightness in H^-3} and \cite[Corollary 6.16]{WalshIntroductionStochasticPartial1986}. 

Let us denote by $\Q^*$ a weak limit of $\Q^{(n)}$. By Skorokhod's representation theorem there exists a filtered probability space $$(D([0,T], \R\times H^{-4}(U)), \F, \mathbb F, \Q),$$ which supports random variables $Z^{(n)} = \left(Z^{(n),B}, Z^{(n),u}\right)$ and $Z^* = \left(Z^{*,B}, Z^{*,u}\right)$ s.t.
\[
    \mathscr L ^{\Q} \left(Z^*\right) = \mathscr L ^{\Q^*} (Z)
\]
and
\[
\begin{aligned}
Z^{(n)} &\longrightarrow Z^* \quad \Q \text{-a.s. in }\quad D([0,T],\R\times H^{-4}(U)).
\end{aligned}
\]

Therefore, there exist continuous bijections $\gamma_n$ on $[0,T]$ such that $\gamma_n(t) \rightarrow t$ and $Z^{(n)}_{\gamma_n(\cdot)}$ converges to $Z$ uniformly. Using Lemma \ref{lemma: H^-5 estimate}, we obtain
\[
\begin{aligned}
&\E^{\Q} \left[\|Z_t - Z_s\|_{\R\times H^{-5}}^4\right] \lesssim \lim_{n\rightarrow\infty} \E^{\Q} \left[\left\|Z_t - Z_{\gamma_n(t)}^{(n)}\right\|_{\R\times H^{-5}}^4\right] \\
&\qquad\qquad+ \lim_{n\rightarrow\infty} \E^{\Q} \left[\left\|Z_s - Z_{\gamma_n(s)}^{(n)}\right\|_{\R\times H^{-5}}^4\right] + \lim_{n\rightarrow\infty} \E^{\Q} \left[\left\| Z_{\gamma_n(t)}^{(n)} - Z_{\gamma_n(s)}^{(n)} \right\|_{\R\times H^{-5}}^4\right] \\
&\qquad\qquad\lesssim \lim_{n\rightarrow\infty} \left(|\gamma_n(t)-\gamma_n(s)|^2 + |\gamma_n(t)-\gamma_n(s)|^4\right) = |t-s|^2 + |t-s|^4,
\end{aligned}
\]
from which C-tightness follows by Kolmogorov's continuity criterion. Moreover, since both processes are C-tight, the joint tightness in $D([0,T],\R\times H^{-4}(U))$ follows from Corollary C.4 in \cite{BayerFunctionalLimitTheorem2017}.
\qed
\end{proof}

%%%%%%%%%%%%%%%%%%%%%%%%%%%%%%%%%%%%%%%%%%%%%%%%%%%%%%%%%%%%
%%%%%%%%%%%%%%%%%%%%%%%%%%%%%%%%%%%%%%%%%%%%%%%%%%%%%%%%%%%%
%%%%%%%%%%%%%%%%%%%%%%%%%%%%%%%%%%%%%%%%%%%%%%%%%%%%%%%%%%%%

\subsection{Martingale relations}
\label{subsec: martingale relations}

In this section, we will prove that for all $\varphi \in H^5(U)$ the following processes are continuous square-integrable $\Q^*$-martingales:
\begin{equation}
\label{eq: continuous martingale problem}
\begin{aligned}
M^B_t &= Z^B_t - Z^B_0 - \int_0^t U(r)Z_r \d r, \\
L^B_t &= \left(M^B_t\right)^2 - \int_0^t P(r) \d r, \\
M^u_t(\varphi) &= \langle Z^u_t, \varphi \rangle - \langle Z^u_0, \varphi \rangle + \int_0^t \langle Z_r, \varphi' \rangle \d r - \int_0^t \langle V(r)Z_r, \varphi \rangle \d r,\\
L^u_t(\varphi) &= M^u_t(\varphi)^2 - \int_0^t \langle R(r) \varphi, \varphi \rangle \d r, \\
N_t(\varphi) &= M^B_t M^u_t(\varphi) - \int _0 ^t \langle Q(r), \varphi \rangle \d r.
\end{aligned}
\end{equation}
To this end, we define discrete martingale relations, which are satisfied in the microscopic models, obtain moment estimates for the discrete martingales, and then deduce martingale relations under the limiting measure. 

Let $\Q^{(n)}$ and $\Q^*$ be as in Section \ref{subsec: tightness} and, as before, denote by $Z=(Z^B, Z^u)$ the canonical process on the space $D([0,T],\R\times L^2(\R))$.

%%%%%%%%%%%%%%%%%%%%%%%%%%%%%%%%%%%%%%%%%%%%%%%%%%%%%%%%%%%%
%%%%%%%%%%%%%%%%%%%%%%%%%%%%%%%%%%%%%%%%%%%%%%%%%%%%%%%%%%%%
%%%%%%%%%%%%%%%%%%%%%%%%%%%%%%%%%%%%%%%%%%%%%%%%%%%%%%%%%%%%

\subsubsection{Martingale relations in the pre-limit}

Our next goal is to define discrete drift and covariance operators for the discrete second-order approximation. To this end, we set
\[
[\varphi]_n(x) := \sum_{k \in \Z} \1_{\left(x^{(n)}_k, x^{(n)}_{k+1}\right]}(x) \int_{x^{(n)}_k}^{x^{(n)}_{k+1}} \varphi(y) \d y.
\]
We introduce the discrete version 
\[
\sigma ^{(n)}_0 (b,y)^2 := p^{(n),B+A}(b,y) - p^{(n)}(b,y)^2
\]
of the variance function introduced in Section 2 and proceed by computing the building blocks of the desired drift and covariance operator. Using \cref{assum: g}~\ref{assum: g: definition of g} we obtain
\begin{equation}
\label{eq: calculation discrete martingale J part}
\begin{aligned}
&\E^{\Q^{(n)}} \left[ \left. \left\langle J^{(n),C}_k, \varphi \right\rangle ^2 \right| \F^{(n)}_{k-1} \right] = \int_\R \int_\R \E\left[ \left. J^{(n),C}_k(x) J^{(n),C}_k(y) \right| \F^{(n)}_{k-1} \right] \varphi(x) \varphi(y) \d x \d y \\
&\qquad= \int_\R \int_\R \E\left[ \left. \1_{k}^{(n),C} \1_{I^{(n)}\left(\pi^{(n)}_k\right)}(x) \1_{I^{(n)} \left(\pi^{(n)}_k\right)}(y) \left(\omega^{(n)}_k\right)^2 \right| \F^{(n)}_{k-1} \right] \varphi(x) \varphi(y) \d x \d y \\
&\qquad= \int_\R \E\left[ \left. \1_{k}^{(n),C} \1_{I^{(n)}\left(\pi^{(n)}_k\right)}(x) \left(\omega^{(n)}_k\right)^2 \right| \F^{(n)}_{k-1} \right] \varphi(x) [\varphi]_n(x) \d x\\
&\qquad= \left\langle g^{(n)} \left(S^{(n)}_{k-1}\right), \varphi \cdot [\varphi]_n \right\rangle.
\end{aligned}
\end{equation}
Moreover, using that A-,B-, and C-events are disjoint, we have
\begin{equation}
\label{eq: calculation discrete martingale cross part}
\begin{aligned}
&\E^{\Q^{(n)}} \left[\left.\left(\1^{(n),B-A}_k - p^{(n)} \left(S^{(n)}_{k-1}\right) \right) \left\langle J^{(n),C}_k - f^{(n)}\left(S^{(n)}_{k-1}\right), \varphi \right\rangle \right| \F^{(n)}_{k-1} \right] \\
&\qquad= \E \left[\left.\1^{(n),B-A}_k \left\langle J^{(n),C}_k, \varphi \right\rangle \right| \F^{(n)}_{k-1} \right] - p^{(n)} \left(S^{(n)}_{k-1}\right) \left\langle f^{(n)}\left(S^{(n)}_{k-1}\right), \varphi \right\rangle \\
& \qquad = - p^{(n)} \left(S^{(n)}_{k-1}\right) \left\langle f^{(n)}\left(S^{(n)}_{k-1}\right), \varphi \right\rangle.
\end{aligned}
\end{equation}
Also note that
\begin{equation}
\label{eq: calculation discrete martingale price part}
\begin{aligned}
\E^{\Q^{(n)}} \left[\left.\left(\1^{(n),B-A}_k - p^{(n)} \left(S^{(n)}_{k-1}\right) \right)^2 \right| \F^{(n)}_{k-1}\right] = \sigma^{(n)}_0 \left(S^{(n)}_{k-1}\right)^2.
\end{aligned}
\end{equation}

Recalling that by definition, $S^{(n)}(t) = S(t) + \sqrt{\Delta^{(n)}} Z(t)$ $\Q^{(n)}$-almost surely, we define the discrete drift operators
\[
\begin{aligned}
U^{(n)}(t, Z) &:= \left(\Delta^{(n)}\right)^{-1/2} \left(p^{(n)}\left(S^{(n)}(t)\right) - p\left(S(t)\right)\right), \\
V^{(n)}(t, Z)(\varphi) &:= \left(\Delta^{(n)}\right)^{-1/2} p^{(n),A} \left(S^{(n)}(t)\right) \left\langle \left(\nabla^{(n)}_+ - \nabla^{(n)}_-\right) u^{(n)}(t) , \varphi \right\rangle \\
&+ \left(\Delta^{(n)}\right)^{-1/2}  \left(p^{(n)}\left(S^{(n)}(t)\right) \left\langle\nabla^{(n)}_+ u(\flrn t) , \varphi \right\rangle - p\left(S(t)\right) \left\langle \partial_x u(t), \varphi \right\rangle \right) \\
&+ \left(\Delta^{(n)}\right)^{-1/2}  \left\langle f^{(n)}\left(S^{(n)}(t)\right) - f\left(S(t)\right), \varphi \right\rangle,
\end{aligned}
\]
and discrete covariance operators
\[
\begin{aligned}
P^{(n)}(t, Z) &:= \sigma_0^{(n)} \left(S^{(n)}(t)\right)^2, \\
Q^{(n)}(t, Z)(\varphi) &:=  \sigma_0^{(n)} \left(S^{(n)}(t)\right)^2 \left\langle \nabla^{(n)}_+ u(\flrn t), \varphi \right\rangle - p^{(n)}\left(S^{(n)}(t)\right) \left\langle f^{(n)}\left(S^{(n)}(t)\right), \varphi \right\rangle, \\
R^{(n)}(t, Z)(\varphi) &:= \sigma_0^{(n)} \left(S^{(n)}(t)\right)^2 \left\langle \nabla^{(n)}_+ u(\flrn t), \varphi \right\rangle^2 \\
&\qquad\qquad- 2p^{(n)}\left(S^{(n)}(t)\right) \left\langle f^{(n)}\left(S^{(n)}(t)\right), \varphi \right\rangle \left\langle \nabla^{(n)}_+ u(\flrn t), \varphi \right\rangle \\
&\qquad\qquad+  \left\langle g^{(n)}\left(S^{(n)}(t)\right), \varphi \cdot [\varphi]_n \right\rangle - \left\langle f^{(n)}\left(S^{(n)}(t)\right), \varphi \right\rangle^2.
\end{aligned}
\]
Altogether, the definition of the discrete dynamics \eqref{eq: discrete dynamics}, the first-order approximation \eqref{eq: first-order approximation}, the definition of the fluctuations \eqref{eq: definition of Z^B and Z^u} and \eqref{eq: definition of Z^Y}, and equations \eqref{eq: discrete increment of volume}, \eqref{eq: calculation discrete martingale J part}, \eqref{eq: calculation discrete martingale cross part}, and \eqref{eq: calculation discrete martingale price part} imply that the following are $\Q^{(n)}$-martingales:
\begin{equation*}
\begin{aligned}
M^{(n),B}_t &= Z^{B}_t - Z^{B}_0 - \int_0^t U^{(n)}(r, Z_r) \d r, \\
L^{(n),B}_t &= \left(M^{(n),B}_t\right)^2 - \int_0^t P^{(n)}(r, Z_r) \d r, 
\end{aligned}
\end{equation*}                                                                    
\begin{equation*}
    \begin{aligned}
M^{(n),u}_t(\varphi) &= \left\langle Z^{u}_t, \varphi \right\rangle - \left\langle Z^{u}_0, \varphi \right\rangle - \int_0^t p^{(n)}\left(S^{(n)}(t)\right) \left\langle\nabla^{(n)}_+ Z^{u}_r, \varphi \right\rangle \d r - \int_0^t V^{(n)}(r, Z_r)(\varphi) \d r, \\
L^{(n),u}_t(\varphi) &= M^{(n),u}_t(\varphi)^2 - \int_0^t R^{(n)}(r, Z_r)(\varphi) \d r + C^{(n)}_t, \\
N^{(n)}_t (\varphi) &= M^{(n),B}_t M^{(n),u}_t(\varphi) - \int_0^t Q^{(n)}(r, Z_r)(\varphi) \d r +C^{(n)}_t, \\
\end{aligned}
\end{equation*}
where the remainder terms are $C_t^{(n)}=O(\Delta^{(n)})^{1/2}$, uniformly in $t\in[0,T]$.

The next proposition proves the integrability of the martingales defined above and their convergence towards the continuous counterparts defined in \eqref{eq: continuous martingale problem}.

\begin{proposition}
\label{prop: martingale relations convergence and bounds}
Let $\varphi\in H^4(\R)$.
\begin{enumerate}[label=(\roman*)]
\item The following uniform bounds hold:
\begin{align}
\label{eq: uniform bound on fourth moment of M}
&\sup_{n \in \N, t \in [0, T]} \E^{\Q^{(n)}}\left[\left|M^{(n),B}_t\right|^4+\left|M^{B}_t\right|^4+\left|M^{(n),u}_t(\varphi)\right|^4+\left|M^{u}_t(\varphi)\right|^4\right] < \infty, \\
\label{eq: uniform bound on fourth moment of L}
&\sup_{n \in \N, t \in [0, T]} \E^{\Q^{(n)}} \left[\left|L^{(n),B}_t\right|^2+\left|L^{B}_t\right|^2+\left|L^{(n),u}_t(\varphi)\right|^2+\left|L^{u}_t(\varphi)\right|^2\right] <\infty, \\
\label{eq: uniform bound on fourth moment of N}
&\sup_{n \in \N, t \in [0, T]} \E^{\Q^{(n)}} \left[\left|N^{(n)}_t(\varphi)\right|^2+\left|N_t(\varphi)\right|^2\right] <\infty.
\end{align}

\item The following convergence results hold:
\begin{align*}
\begin{split}
&\E^{\Q^{(n)}}\left[\left|M^{(n),B}_t - M^B_t\right|^2+\left|M^{(n),u}_t(\varphi) - M^u_t(\varphi)\right|^2\right] \to 0,\\
& \E^{\Q^{(n)}}\left[\left|L^{(n),B}_t - L^B_t\right|^2+\left|L^{(n),u}_t(\varphi) - L^u_t(\varphi)\right|^2\right] \to 0,\\
& \E^{\Q^{(n)}}\left[\left|N^{(n)}_t(\varphi) - N_t(\varphi)\right|^2\right] \to 0.
\end{split}
\end{align*}

\end{enumerate}

\end{proposition}

\begin{proof}

We analyze each term separately.
\smallskip

\textbf{The $M^B$ term.} We write
\begin{align*}
M^B_t - M^{(n),B}_t = \frac{1}{\sqrt{\Delta^{(n)}}} \int_0^t \left(p^{(n)}\left(S^{(n)}(r)\right) - p\left(S(r)\right) - p_b\left(S(r)\right) Z^B_r - p_y\left(S(r)\right) Z^Y_r \right) \d r.
\end{align*}
Using Jensen inequality and Lemma \ref{lemma: taylor expansion of p, f and g}, we conclude by Proposition \ref{prop: moment estimate for Z} that
\begin{equation}
\label{eq: vanishing 4th moments of M^B difference}
\begin{aligned}
&\E^{\Q^{(n)}} \left[\left|M^B_t - M^{(n),B}_t\right|^4\right] \\
&\qquad\leq \int _0 ^t C^{(n),1}(r) \E^{\Q^{(n)}} \left[ \left| Z^{B}_r \right|^4 \right] + C^{(n),2}(r) \E^{\Q^{(n)}} \left[ \left| Z^{Y}_r \right| ^4 \right] \d r + C^{(n),3} \longrightarrow 0.
\end{aligned}
\end{equation}
Using Jensen inequality, Lemma \ref{lemma: taylor expansion of p, f and g}, Proposition \ref{prop: moment estimate for Z} and Assumptions \ref{assum: p}~\ref{assum: p: p^n-p} and \ref{assum: p}\ref{assum: p: sup}, we have for all $n\in\N$ and $t\in[0,T]$,
\begin{equation}
\label{eq: bound on E(M^n,B)^4}
\begin{split}
\E^{\Q^{(n)}} \left[\left|M^{(n),B}_t\right|^4\right] \lesssim \E^{\Q^{(n)}} \left[\left| Z^{(n),B}_t \right|^4\right] + \E^{\Q^{(n)}} \left[\left| Z^{(n),B}_0 \right|^4\right] + \int_0^t \E^{\Q^{(n)}} \left[U^{(n)}(r, Z_r)^4\right] \d r ,
\end{split}
\end{equation}
which is uniformly bounded from above by some constant. 
This establishes the bound for the first summand in \eqref{eq: uniform bound on fourth moment of M}. The bound for the second summand in \eqref{eq: uniform bound on fourth moment of M} follows from \eqref{eq: vanishing 4th moments of M^B difference} and \eqref{eq: bound on E(M^n,B)^4}.
\smallskip

\textbf{The $L^B$ term.} We write
\[
\begin{aligned}
L^B_t - L^{(n),B}_t &= R^{(n)}_1 + R^{(n)}_2
\end{aligned}
\]
with
\[
\begin{aligned}
R^{(n)}_1 &= \left(M^B_t\right)^2 - \left(M^{(n),B}_t\right)^2, &\quad
R^{(n)}_2 &= \int_0^t \sigma_0^{(n)} \left(S^{(n)}(r)\right)^2 \d r - \int_0^t \sigma_0(S(r))^2 \d r. \\
\end{aligned}
\]
Using Cauchy–Schwarz inequality and noting that the first factor below vanishes due to \eqref{eq: vanishing 4th moments of M^B difference} and the second factor is uniformly bounded due to \eqref{eq: uniform bound on fourth moment of M}, we obtain
\[
\E^{\Q^{(n)}} \left[\left|R^{(n)}_1\right|^2\right]^2 \leq \E^{\Q^{(n)}} \left[\left|M^{(n),B}_t - M^{B}_t\right|^4\right] \E^{\Q^{(n)}} \left[\left|M^{(n),B}_t + M^{B}_t \right|^4\right] \longrightarrow 0.
\]
Using Jensen inequality and Lemma \ref{lemma: taylor expansion of p, f and g}, we conclude by Proposition \ref{prop: moment estimate for Z} that
\[
\E^{\Q^{(n)}} \left[\left|R^{(n)}_2\right|^2\right] \leq \int _0 ^t C^{(n),1}(r) \E^{\Q^{(n)}} \left[ \left| Z^{B}_r \right|^2 \right] + C^{(n),2}(r) \E^{\Q^{(n)}} \left[ \left| Z^{Y}_r \right| ^2 \right] \d r + C^{(n),3} \longrightarrow 0.
\]
Overall, we have
\begin{equation}
\label{eq: vanishing 2nd moments of L^B difference}
\E^{\Q^{(n)}} \left[\left|L^B_t - L^{(n),B}_t\right|^2\right] \longrightarrow 0.
\end{equation}
Using Lemma \ref{lemma: taylor expansion of p, f and g}, Proposition \ref{prop: moment estimate for Z}, Assumptions \ref{assum: p}~\ref{assum: p: p^n-p} and \ref{assum: p}\ref{assum: p: sup} and \eqref{eq: bound on E(M^n,B)^4}, we have for all $n\in\N$, $t\in[0,T]$, and some $K<\infty$,
\begin{equation}
\label{eq: bound on E(L^n,B)^4}
\E^{\Q^{(n)}} \left[\left|L^{(n),B}_t\right|^2\right] \lesssim \E^{\Q^{(n)}} \left[\left| M^{(n),B}_t \right|^4\right]^{1/2} + \int_0^t \E^{\Q^{(n)}} \left[P^{(n)}(r, Z_r)^2\right] \d r < K.
\end{equation}
This establishes the bound for the first summand in \eqref{eq: uniform bound on fourth moment of L}. The bound for the second summand in \eqref{eq: uniform bound on fourth moment of L} follows from \eqref{eq: vanishing 2nd moments of L^B difference} and \eqref{eq: bound on E(L^n,B)^4}.
\smallskip

\textbf{The $M^u$ term.} We write
\[
M^u_t(\varphi) - M^{(n),u}_t(\varphi) = \sum_{i} R^{(n)}_i(\varphi)
\]
with
\[
\begin{aligned}
R^{(n)}_1(\varphi) &= \int _0 ^t \left(p^{(n)}\left(S^{(n)}(r)\right) - p\left(S(r)\right) \right) \left\langle \nabla^{(n)}_+ Z^{u}_r, \varphi \right\rangle \d r, \\
R^{(n)}_2(\varphi) &= \int_{0}^{t} p\left(S(r)\right) \left( \left\langle Z_{r}^{u}, \varphi^{\prime}\right\rangle - \left\langle Z^{u}_r, \nabla^{(n)}_+ \varphi \right\rangle \right) \d r, \\
R^{(n)}_3(\varphi) &= \left(\Delta^{(n)}\right)^{-1/2} \int _0 ^t p^{(n),A} \left(S^{(n)}(r)\right) \left\langle \left(\nabla^{(n)}_+ - \nabla^{(n)}_-\right) u^{(n)}_k , \varphi \right\rangle \d r, \\
R^{(n)}_4(\varphi) &= - \int_0^t \left(p_b\left(S(r)\right) Z^B_r + p_y\left(S(r)\right) Z^Y_r \right) \left\langle \partial _x u(r), \varphi \right\rangle \d r, \\
&\quad+ \left(\Delta^{(n)}\right)^{-1/2} \int _0 ^t \left(p^{(n)}\left(S^{(n)}(r)\right) \left\langle\nabla^{(n)}_+ u(\flrn r) , \varphi \right\rangle - p\left(S(r)\right) \left\langle \partial_x u(r) , \varphi \right\rangle \right) \d r, \\
R^{(n)}_5(\varphi) &= - \int_0^t \left\langle f_b\left(S(r)\right), \varphi \right\rangle Z^B_r \d r - \int_0^t \left\langle f_y\left(S(r)\right), \varphi \right\rangle Z^Y_r \d r , \\
&\quad+ \left(\Delta^{(n)}\right)^{-1/2} \int _0 ^t \left\langle f^{(n)}\left(S^{(n)}(r)\right) - f\left(S(r)\right) , \varphi \right\rangle \d r. \\
\end{aligned}
\]
Using Jensen inequality, Lemma \ref{lemma: taylor expansion of p, f and g}, Young inequality and Proposition \ref{prop: moment estimate for Z}, we obtain
\[  
\begin{aligned}
&\E^{\Q^{(n)}} \left[\left|R^{(n)}_1(\varphi)\right|^4\right] \lesssim \int _0 ^t \E^{\Q^{(n)}} \left[ \left(p^{(n)}\left(S^{(n)}(r)\right) - p\left(S(r)\right) \right)^4 \left\langle \nabla^{(n)}_+ Z^{u}_r, \varphi \right\rangle ^4 \right] \d r  \\
&\lesssim \int _0 ^t \E^{\Q^{(n)}} \left[ \left(\sqrt{\Delta^{(n)}} Z^{B}_r + \sqrt{\Delta^{(n)}} Z^{Y}_r + C^{(n)}(r) \right)^4 \left\langle Z^{u}_r, \nabla^{(n)}_+ \varphi \right\rangle ^4 \right] \d r \\
&\lesssim \left(\Delta^{(n)}\right)^2 \int _0 ^t \E^{\Q^{(n)}}\left[\left|Z^{B}_r\right|^8\right] + \E^{\Q^{(n)}}\left[\left|Z^{Y}_r\right|^8\right] + \E^{\Q^{(n)}} \left[\left\langle Z^{u}_r, \nabla^{(n)}_+ \varphi \right\rangle ^8 \right]+ C^{(n)}(r)  \d r \\
&\leq K^{(n)} \left(\Delta^{(n)}\right)^2 \|\varphi\|^8_{H^3}.
\end{aligned}
\]
By Hölder inequality, \eqref{eq: varphi discrete derivative}, Proposition \ref{prop: uniform bound of discrete dynamics}, and Proposition \ref{prop: law of large numbers}, we have
\[
\begin{aligned}
&\E^{\Q^{(n)}} \left[ \left| R^{(n)}_2(\varphi) \right|^4 \right]\\
 &\quad\leq \left(\Delta^{(n)}\right)^{-2} \sup_{(b,y)\in\R^2} |p(b,y)|^4 \int _0^T \E^{\Q^{(n)}} \left[\left\|u^{(n)}(r) - u(r)\right\|_{L^2}^4\right] \left\| \nabla^{(n)}_+ \varphi - \varphi' \right\|^4_{L^2} \d r \\
&\quad\lesssim \int _0^T \E^{\Q^{(n)}} \left[\left\| u^{(n)}(r) - u(r) \right\|_{L^2}^4\right] \left\| \varphi \right\|^4_{H^2} \left(\Delta^{(n)}\right)^2  \d r \\
&\quad\lesssim \left(\Delta^{(n)}\right)^2 \left(\sup_{r\in[0,T]} \E^{\Q^{(n)}} \left[\left\|u^{(n)}(r)\right\|_{L^2}^4\right] + \sup_{r\in[0,T]} \| u(r)\|_{L^2}^4\right) \|\varphi\|^4_{H^2} \longrightarrow 0.
\end{aligned}
\]
Using Jensen inequality, Hölder inequality, and Proposition \ref{prop: uniform bound of discrete dynamics},
\[
\E^{\Q^{(n)}} \left[\left|R^{(n)}_3(\varphi)\right|^4\right] \leq C \left(\Delta^{(n)}\right)^2 \|\varphi\|^4_{H^2} \sup_{r\in[0,T]} \E \left[ \left\|u^{(n)}(r)\right\|^4_{L^2} \right] \longrightarrow 0.
\]
Note that we can write
\[
\begin{aligned}
R_4^{(n)}(\varphi) &= \left(\Delta^{(n)}\right)^{-1/2} \int _0 ^t p^{(n)}\left(S^{(n)}(r)\right) \left\langle u(\flrn r), \left(\partial_x -\nabla^{(n)}_+ \right) \varphi \right\rangle \d r \\
&\quad+ \left(\Delta^{(n)}\right)^{-1/2} \int _0^t p^{(n)}\left(S^{(n)}(r)\right) \left\langle \partial _x u(\flrn r) - \partial _x u(r) , \varphi \right\rangle \d r \\
&\quad+ \left(\Delta^{(n)}\right)^{-1/2} \int _0 ^t \left(p^{(n)}\left(S^{(n)}(r)\right) \left\langle \partial _x u(r) , \varphi \right\rangle - p\left(S(r)\right) \left\langle \partial_x u(r) , \varphi \right\rangle \right) \d r \\
&\quad- \int_0^t \left(p_b\left(S(r)\right) Z^B_r + p_y\left(S(r)\right) Z^Y_r \right) \left\langle \partial _x u(r), \varphi \right\rangle \d r.
\end{aligned}
\]
The terms in the first two lines can be controlled using the regularity and integrability of the first-order approximation, cf.~Proposition \ref{prop: law of large numbers}. The terms in the last two lines can be controlled with Lemma \ref{lemma: taylor expansion of p, f and g} and \eqref{eq: varphi discrete derivative}. Overall, we get from Proposition \ref{prop: moment estimate for Z} that
\[
\E^{\Q^{(n)}} \left[\left|R_4^{(n)}(\varphi)\right|^4\right] \leq \int _0 ^t C^{(n),1}(r) \E^{\Q^{(n)}} \left[ \left| Z^{B}_r \right|^4 \right] + C^{(n),2}(r) \E^{\Q^{(n)}} \left[ \left| Z^{Y}_r \right| ^4 \right] \d r + C^{(n),3} \longrightarrow 0.
\]
$R^{(n)}_5(\varphi)$ vanishes analogously to $R^{(n)}_4(\varphi)$. Altogether, we obtain
\begin{equation}
\label{eq: vanishing 4th moments of M^u difference}
\E^{\Q^{(n)}} \left[\left|M^u_t(\varphi) - M^{(n),u}_t(\varphi)\right|^4\right] \longrightarrow 0.
\end{equation}
From here, the bound for $M^{(n),u}$ and $M^u$ in \eqref{eq: uniform bound on fourth moment of M} follows analogously as the bound for $M^{(n),B}$ and $M^B$.
\smallskip

\textbf{The $L^u$ term.} We can write
\[
L^u_t(\varphi) - L^{(n),u}_t(\varphi) = \sum_{i} R^{(n)}_i(\varphi)
\]
with
\[
\begin{aligned}
R^{(n)}_1(\varphi) &= M^u_t(\varphi)^2 - M^{(n),u}_t(\varphi)^2, \\
R^{(n)}_2(\varphi) &= \int _0^t \left( \left\langle g^{(n)}\left(S^{(n)}(r)\right), \varphi \cdot [\varphi]_n \right\rangle - \left\langle g^{(n)}\left(S^{(n)}(r)\right), \varphi^2 \right\rangle \right) \d r, \\
R^{(n)}_3(\varphi) &= \int _0^t \left( \left\langle g^{(n)}\left(S^{(n)}(r)\right), \varphi^2 \right\rangle- \left\langle g(S(r)), \varphi^2 \right\rangle \right) \d r, \\
R^{(n)}_4(\varphi) &= \int_0^t \sigma_0^{(n)} \left(S^{(n)}(r)\right)^2 \left\langle u(\flrn r), \nabla^{(n)}_+ \varphi \right\rangle^2 \d r - \int_0^t \sigma_0(S(r))^2 \left\langle u(r), \varphi' \right\rangle ^2 \d r, \\
R^{(n)}_5(\varphi) &= \int _0^t \left(\left\langle f(S(r)), \varphi \right\rangle ^2 - \left\langle f^{(n)}\left(S^{(n)}(r)\right), \varphi \right\rangle ^2 \right) \d r, \\
R^{(n)}_6(\varphi) &= \int_0^t 2p(S(r)) \left\langle f(S(r)), \varphi \right\rangle \left\langle \partial _x u(r), \varphi\right\rangle \d r, \\
&\qquad- \int_0^t 2p^{(n)}\left(S^{(n)}(r)\right) \left\langle f^{(n)}\left(S^{(n)}(r)\right), \varphi \right\rangle \left\langle \nabla^{(n)}_+ u(\flrn r), \varphi \right\rangle \d r.
\end{aligned}
\]
Using Cauchy–Schwarz inequality, noting that the difference term vanishes due to \eqref{eq: vanishing 4th moments of M^u difference} and that the sum term is uniformly bounded due to \eqref{eq: uniform bound on fourth moment of M}, we obtain
\[
\begin{aligned}
&\E^{\Q^{(n)}} \left[\left|R^{(n)}_1(\varphi)\right|^2\right]^2\leq \E^{\Q^{(n)}} \left[\left|M^u_t(\varphi) - M^{(n),u}_t(\varphi)\right|^4\right] \E^{\Q^{(n)}}\left[\left|M^u_t(\varphi) + M^{(n),u}_t(\varphi)\right|^4\right],
\end{aligned}
\]
which converges to zero as $n\rightarrow\infty$. Moreover, it holds by Assumption \ref{assum: g} \ref{assum: g: sup} that
\[
    \E\left[\left|R_2^{(n)}(\varphi)\right|^2\right] \lesssim \left\| \varphi \cdot [\varphi]_n - \varphi^2 \right\|_{L^1}^2 \longrightarrow 0,
\]
Other residuals vanish via calculations similar to the proof of the $M^u$ part. Overall, we obtain
\[
\E^{\Q^{(n)}} \left[\left|L^u_t(\varphi) - L^{(n),u}_t(\varphi) \right| ^2\right]\longrightarrow 0.
\]
The bound for the third and fourth summand in \eqref{eq: uniform bound on fourth moment of L} follows by analogy with the first and second summand in \eqref{eq: uniform bound on fourth moment of L}.
\smallskip

\textbf{The $N$ term.} We can write
\[
N_{t}(\varphi)-N_{t}^{(n)}(\varphi)=\sum_{i} R_{i}^{(n)}(\varphi)
\]
with
\[
\begin{aligned}
R^{(n)}_1(\varphi) &= M^{B}_t M^{u}_t(\varphi) - M^{(n),B}_t M^{(n),u}_t(\varphi), \\
R^{(n)}_2(\varphi) &= \int_0^t \sigma_0^{(n)} \left(S^{(n)}(r)\right)^2 \left\langle \nabla^{(n)}_+ u(\flrn r), \varphi \right\rangle \d r - \int_{0}^{t}\left\langle\partial_{x} u(r), \varphi\right\rangle \sigma_{0}(S(r))^{2} \d r, \\
R^{(n)}_3(\varphi) &= \int_{0}^{t} p(S(r)) \langle f(S(r)), \varphi \rangle \d r - \int_0^t p^{(n)} \left(S^{(n)}(r)\right) \left\langle f^{(n)}\left(S^{(n)}(r)\right), \varphi \right\rangle \d r.
\end{aligned}
\]
Using Cauchy–Schwarz inequality and $ab-cd = \frac 12(a-c)(b+d) + \frac 12(a+c)(b-d)$, noting that the difference term vanishes due to \eqref{eq: vanishing 4th moments of M^B difference} and \eqref{eq: vanishing 4th moments of M^u difference} and that the sum term is uniformly bounded due to \eqref{eq: uniform bound on fourth moment of M}, we obtain
\[
\begin{aligned}
&\E^{\Q^{(n)}} \left[\left|R^{(n)}_1(\varphi)\right|^2\right] \lesssim \E^{\Q^{(n)}} \left[\left|M^u_t(\varphi) - M^{(n),u}_t(\varphi)\right|^4\right]^{1/2} \E^{\Q^{(n)}}\left[\left|M^B_t + M^{(n),B}_t\right|^4\right]^{1/2} \\
&\qquad+ \E^{\Q^{(n)}}\left[\left|M^B_t - M^{(n),B}_t\right|^4\right]^{1/2} \E^{\Q^{(n)}}\left[\left|M^u_t(\varphi) + M^{(n),u}_t(\varphi)\right|^4\right]^{1/2} \longrightarrow 0.
\end{aligned}
\]
Again, other residuals vanish with calculations similar to the proof of the $M^u$ part. Overall, we obtain
\[
\E^{\Q^{(n)}} \left[\left|N_t(\varphi) - N^{(n) }_t(\varphi) \right| ^2\right]\longrightarrow 0.
\]
The bound \eqref{eq: uniform bound on fourth moment of N} follows by analogy with \eqref{eq: uniform bound on fourth moment of L}.
\qed
\end{proof}

\subsubsection{Martingale relations in the limit}

In order to show that the martingale relations are satisfied under the limiting measure, we employ \cite[Proposition IX.1.12]{JacodLimitTheoremsStochastic2003}, cf.~Theorem \ref{thm:jacod} in the appendix.

\begin{proposition}
For any $\varphi\in H^5(U)$, all processes defined in \eqref{eq: continuous martingale problem} are continuous square-integrable $\Q^*$–martingales.
\end{proposition}

\begin{proof}
We only show the proof for $M^{(n),u}(\varphi)$ for a fixed $\varphi\in H^5(U)$. The proof for other processes is analogous. To this end, we want to apply Theorem \ref{thm:jacod}. Denote by $X$ the canonical process on $D([0,T], \R^3)$ and $X^{(n)} := \left(Z^{(n), B}, Z^{(n), Y}, \left\langle Z^{(n), u}, \varphi\right\rangle \right)$. Then, it holds that $\mathcal L^{\Q^{(n)}}(X^{(n)})\Rightarrow \mathcal L^{\Q^*}(X)$ and condition (ii) of Theorem \ref{thm:jacod} is trivially satisfied. The uniform integrability of the family $\left(M^{(n),u}_t(\varphi)\right)_{n\in\N, t\in[0,T]}$ follows from Proposition \ref{prop: martingale relations convergence and bounds}(i) and the de-la-Vallée-Poussin criterion. Thus, condition (i) is also satisfied. Note that by Proposition \ref{prop: tightness}, the law $\mathcal L^{\Q^*}(Y)$ is supported on $C([0,T], \R^3)$, which allows to deduce the validity of condition (iii) from the definition of $M^u(\varphi)$. Moreover, this also implies the continuity of the limiting process $M^{u}(\varphi)$. Finally, condition (iv) is satisfied by the virtue of Proposition \ref{prop: martingale relations convergence and bounds}(ii). Square-integrability follows by Proposition \ref{prop: martingale relations convergence and bounds} and Fatou's lemma.\qed
\end{proof}

%%%%%%%%%%%%%%%%%%%%%%%%%%%%%%%%%%%%%%%%%%%%%%%%%%%%%%%%%%%%
%%%%%%%%%%%%%%%%%%%%%%%%%%%%%%%%%%%%%%%%%%%%%%%%%%%%%%%%%%%%
%%%%%%%%%%%%%%%%%%%%%%%%%%%%%%%%%%%%%%%%%%%%%%%%%%%%%%%%%%%%

\subsection{Proof of the main theorem}
\label{sec: weak solution}

In this subsection we show that the limiting dynamics can be obtained as a weak solution of the evolution equation \eqref{eq: abstract spde} by a martingale representation argument. To this end, we construct a cylindrical Brownian motion on the probability space satisfying the martingale relations \eqref{eq: continuous martingale problem} using an appropriate representation theorem. We first need to ensure that the variance operator is positive and trace class as required for the quadratic variation of a Hilbert space-valued martingale. This can be directly deduced from the obtained moment estimates of Section \ref{sec: moment estimates}.

\begin{proposition}
\label{prop: properties of the variance and volatility operators}

For all $t\in[0,T]$, the operator $\Sigma(t)$ is non-negative and $\int_0^t \Sigma(s)ds$ is a trace-class operator. In particular, the square root $\sigma(t)$ of $\Sigma(t)$ exists, i.e.~$\sigma(t)\sigma(t)' = \Sigma(t)$, and $\sigma(t)$ is a Hilbert-Schmidt operator for each $t\in[0,T]$ s.t.~$\int_0^T \|\sigma(t)\|^2_2 \d t < \infty$.

\end{proposition}

\begin{proof}

Note that for $(a,\varphi) \in \R\times L^2(U)$, we have
\[
\begin{aligned}
(a, \varphi) \Sigma(t) (a, \varphi)' &= \lim_{\delta\rightarrow 0} \frac1\delta \int_t^{t+\delta} (a, \varphi)\Sigma(s)(a, \varphi)' \d s \\
&= \lim_{\delta\rightarrow 0} \frac1\delta \E^{\Q^*}\left[((aM^B_{t+\delta} + M^u_{t+\delta} (\varphi)) - (aM^B_t + M^u_t(\varphi)))^2\right] \geq 0.
\end{aligned}
\]
Thus, the operator $\Sigma(t)$ is non-negative for all $t\in[0,T]$. Let us consider the orthonormal basis $\{e_k(x) = (2L)^{-1/2} e^{\pi ikx/L}: k\in\Z\}$ of $L^2(U)$. We have for $s>\frac 12$,
\begin{equation}
\label{eq: bound on <g, e_k^2> with sobolev norm}
\begin{split}
&\sum_{k=-\infty}^\infty \langle g, e_k^2 \rangle  = \frac{1}{\sqrt{2L}} \sum_{k=-\infty}^\infty \langle g, e_{2k} \rangle \\ 
&\qquad \leq \frac{1}{\sqrt{2L}} \left(\sum_{k=-\infty}^\infty \langle g, e_{2k} \rangle ^2 \left(1+(2k)^2\right)^{s} \right)^{1/2} \left(\sum_{k=-\infty}^\infty \left(1+(2k)^2\right)^{-s} \right)^{1/2}  \lesssim \|g\|^2_{H^s}.
\end{split}
\end{equation}
Using non-negativity of $R(t)$, Assumptions \ref{assum: p}~\ref{assum: p: sup}, \ref{assum: f}~\ref{assum: f: sup of L^2 norms of f}, \ref{assum: g}\ref{assum: g: fractional sobolev norm}, \eqref{eq: bound on <g, e_k^2> with sobolev norm}, and Young inequality, we obtain
\begin{equation}
\label{eq: bound on sum of R(t)(e_k, e_k)}
\begin{aligned}
&\sum_{k=-\infty}^\infty \langle R(t)e_k, e_k \rangle  \leq \sum_{k=-\infty}^\infty \left\{\sigma_0^2 \langle \partial_x u, e_k \rangle ^2 + 2p \langle f, e_k\rangle \langle \partial_x u, e_k \rangle + \langle g, e_k^2 \rangle  + \langle f, e_k \rangle ^2 \right\} \\
&\qquad\qquad\lesssim \sup_{(b,y)\in\R^2}\left\{ 1 + |p(b,y)| + \sigma_0(b, y)^2 \right\} \left\{ \|f\|_{L^2}^2 + \|\partial_x u\|_{L^2}^2 + \|g\|^2_{H^s} \right\} < \infty.
\end{aligned}
\end{equation}
Therefore, considering an orthonormal basis $\{(1,0)\}\cup\{(0, e_k):k\in\N\}$ of $\R\times L^2(U)$, we have
\[
\|\Sigma(t)\|_1 = \operatorname{Tr}(\Sigma(t)) = \sigma_0^2 + \sum_{k=-\infty}^\infty \langle R(t)e_k, e_k \rangle < +\infty
\]
uniformly in $t\in[0,T]$, i.e.~$\Sigma(t)$ and thus also $\int_0^t \Sigma(s)ds$ are trace class operators. Moreover,
\[
\|\sigma(t)\|_2 = \sum_{k=-\infty}^\infty \langle \sigma(t)'e_k, \sigma(t)'e_k \rangle = \sum_{k=-\infty}^\infty \langle \Sigma(t)e_k, e_k \rangle = \|\Sigma(t)\|_1 < +\infty.
\]
Thus, $\sigma(t)$ is a Hilbert-Schmidt operator, and since the bound is uniform in $t\in[0,T]$ it follows that 
\[
    \int_0^T \|\sigma(t)\|^2_2 \d t < \infty.
\]\qed
\end{proof}

We are now ready to prove the main result of the paper.

\paragraph{Proof of Theorem \ref{thm: main theorem}} We want to apply Theorem \ref{thm: representation theorem} with $U=H=L^2(U)$. Consider $M^u$ as a linear map
\[
M^u: C^\infty_0(U) \rightarrow \mathcal M^2_T(\R).
\]
Using non-negativity of $R(s)$, Hölder inequality, Assumptions \ref{assum: p}~\ref{assum: p: sup}, \ref{assum: f}~\ref{assum: f: sup of L^2 norms of f}, \ref{assum: g}\ref{assum: g: sup}, and Proposition \ref{prop: law of large numbers}, it follows that
\begin{equation}
\label{eq: bound on the R norm in varphi}
\begin{split}
\int_0^T \langle R(s)\varphi, \varphi \rangle \d s \leq \int_0^T\sigma_0^2 \langle \partial_x u, \varphi \rangle ^2 + 2 p \langle f, \varphi\rangle \langle \partial_x u, \varphi \rangle + \langle g, \varphi^2 \rangle + \langle f, \varphi \rangle ^2 \d s \lesssim \|\varphi\|^2_{L^2}.
\end{split}
\end{equation}
Using Doob's inequality, It\^o's isometry, and \eqref{eq: bound on the R norm in varphi}, we have for all $\varphi\in C^\infty_0$,
\begin{equation}
\label{eq: M^2-norm of M^u_t}
\E^{\Q^*} \left[\sup_{t\in[0,T]}|M^u_t(\varphi)|^2\right] \leq 2 \E^{\Q^*} \left[|M^u_T(\varphi)|^2\right] = 2 \int_0^T \langle R(s)\varphi, \varphi \rangle \d s \lesssim  \|\varphi\|^2_{L^2}.
\end{equation}
Since $M^u$ is a linear operator and $\mathcal M^2_T(\R)$, the space of all continuous, $\R$-valued, square-integrable martingales, is complete, by density we can extend $M^u$ to $L^2(U)$ such that \eqref{eq: M^2-norm of M^u_t} holds for all $\varphi\in L^2(U)$, i.e.~$M^u(\varphi) \in \mathcal M _T^2(\R)$ for all $\varphi\in L^2(U)$. For an orthonormal basis $\{e_k: k\in\N\}$ of $L^2(U)$, we define a process $M^u:=\sum _{k\in\Z} M^u(e_k) e_k$ and its approximations $M^{(n),u}:=\sum_{k=-n}^{n} M^u(e_k) e_k$ for all $n\in\N$. Obviously, $M^{(n),u}\in \mathcal M^2_T(L^2(U))$, i.e.~each $M^{(n),u}$ is a continuous, $L^2(U)$-valued, square-integrable martingale. Using Parseval's identity, monotone convergence, Doob's inequality, and \eqref{eq: bound on sum of R(t)(e_k, e_k)}, we obtain
\[
\begin{aligned}
&\left\| M^{u} - M^{(n),u} \right\|^2_{\mathcal M_T^2(L^2(U))} = \E^{\Q^*}\left[\sup_{t\in[0,T]} \left\|\sum_{k\in\Z \setminus \{-n, \dots, n\}} M^u_t(e_k) e_k \right\|^2_{L^2}\right] \\
&\qquad\leq 2\sum_{k\in\Z \setminus \{-n, \dots, n\}} \E^{\Q^*} \left[|M^u_T(e_k)|^2\right]  = 2\sum_{k\in\Z \setminus \{-n, \dots, n\}} \int_0^T \langle R(s)e_k, e_k\rangle \d s \longrightarrow 0.
\end{aligned}
\]
Since $\mathcal M^2_T(L^2(U))$ is a Banach space, we have $M^u\in \mathcal M^2_T(L^2(U))$. By Proposition \ref{prop: properties of the variance and volatility operators}, for all $t\in[0,T]$ the operator $\int_0^t\Sigma(s)ds\in \mathcal L_1(\mathcal H)$ is trace class. Thus we can define an $\mathcal H$-valued martingale $M_T := (M^B_T, M^u_T)\in \mathcal M ^2_T(\mathcal H)$ with quadratic variation given by
\[
\langle M \rangle_t = \int_0^t \Sigma(s) \d s = \int_0^t \sigma(s)\sigma(s)^* \d s.
\]
By Proposition \ref{prop: properties of the variance and volatility operators}, for all $t\in[0,T]$ the operator $\sigma(t)$ is Hilbert-Schmidt. By Theorem \ref{thm: representation theorem}, there exists an extended filtered probability space $\left(\Omega\times\widetilde\Omega, \F\otimes \widetilde{\F}, \Q\otimes \widetilde{\Q}\right)$ which supports an $\mathcal H$-valued cylindrical Brownian motion $W$ such that $\Q\otimes \widetilde \Q$-a.s.
\begin{equation*}
M_t =\int_{0}^{t} \sigma(s) \d W_s,\quad t\in[0,T].
\end{equation*}
Moreover, since the martingales defined in \eqref{eq: continuous martingale problem} are linear in $\varphi\in H^5(U)$, it holds $\Q\otimes \widetilde\Q$-a.s.~for any $\ell \in \R\times H^5(U)$ and $t\in[0,T]$ that
\begin{equation}
\label{eq: weak solution satisfied}
\left\langle \ell, \int_0^t \sigma(s) \d W_s \right\rangle = \langle \ell, M_t \rangle = \langle \ell, Z_t \rangle - \langle \ell, z_0 \rangle - \int _0^t \langle A(s)^* \ell, Z_s \rangle \d s - \int_0^t \langle \ell, F(s) Z_s \rangle \d s.
\end{equation}
Since $W$ is a cylindrical Brownian motion over $\mathcal H$, by pathwise uniqueness of $Z$, cf.~Proposition \ref{prop: well-posedness}, the relation \eqref{eq: weak solution satisfied} holds for all $\ell \in \R\times H^1(U)$, which concludes the proof.
\qed

%%%%%%%%%%%%%%%%%%%%%%%%%%%%%%%%%%%%%%%%%%%%%%%%%%%%%%%%%%%%
%%%%%%%%%%%%%%%%%%%%%%%%%%%%%%%%%%%%%%%%%%%%%%%%%%%%%%%%%%%%
%%%%%%%%%%%%%%%%%%%%%%%%%%%%%%%%%%%%%%%%%%%%%%%%%%%%%%%%%%%%

\begin{appendix}

\section{Growth estimates of the discrete LOB dynamics}
\label{appendix: proof of discrete dynamics}

In the proof of the moment estimates in Section \ref{sec: moment estimates}, we need the following uniform growth estimates on the discrete state dynamics. 

\begin{proposition}
\label{prop: uniform bound of discrete dynamics}
For any $m\in 2\N\cup\{1\}$, it holds that
\begin{equation*}
\begin{aligned}
\sup_{n\in \N}\sup_{k\leq T_n} \E \left[ \left\|u^{(n)}_k\right\|^m_{L^2} \right] & < \infty, \qquad
\sup_{n\in \N}\sup_{k\leq T_n} \E \left[ \left|B^{(n)}_k\right|^m \right] &< \infty.
\end{aligned}
\end{equation*}
\end{proposition}

\begin{proof}

We only show the proof for the volume dynamics. The estimate for the price dynamics is obtained similarly, but the proof for the volume dynamics is more involved. Since only events of type $C$ contribute to the change in the volume profile (other events merely shift it), it is more convenient in this context to work with absolute volume density functions
\[
v^{(n)}_k(x) := u^{(n)}_k \left(x-B^{(n)}_k \right) = v^{(n)}_0(x) + \sum_{i=1}^k \Delta^{(n)} J^{(n)}_i\left(x-B^{(n)}_{i-1}\right),\quad k=0,\dots,T_n,
\]
so that the events of type $A$ and $B$ can be omitted, while  $\left\|u^{(n)}_k\right\|_{L^2} = \left\|v^{(n)}_k\right\|_{L^2}$.

\textbf{Preliminaries.} Note that with
\[
S(n, k, \ell) := \E \left[ \left(\int \left(\sum_{i=1}^k \Delta^{(n)} J^{(n)}_i\left(x-B^{(n)}_{i-1}\right) \right)^2 \d x \right)^\ell \right]
\]
it holds that
\begin{equation}
\label{eq: triangle inequality in the proof of the volume dynamics estimate}
\begin{aligned}
\E \left[ \left\|v^{(n)}_k\right\|^{2\ell}_{L^2} \right] &\lesssim_{\,\ell} \E\left[\left\|v^{(n)}_0\right\|^{2\ell}_{L^2}\right] + S(n, k, \ell).
\end{aligned}
\end{equation}
Denote $K(i, j) := \left(\Delta^{(n)}\right)^2 \int J^{(n),C}_i\left(x-B^{(n)}_{i-1}\right)  J^{(n),C}_j \left(x-B^{(n)}_{j-1}\right) \d x$. Then,
\begin{equation}
\label{eq: expression for the 2l-th moment}
S(n, k, \ell) = \E \left[ \left(\sum_{i,j=1}^k K(i,j) \right)^\ell \right]= \E \left[ \sum_{i_1, \dots i_{2\ell}=1}^k \prod_{j=1}^\ell K(i_{2j-1}, i_{2j}) \right].
\end{equation}
We list some estimates that we will use below. First, note that
\begin{equation}
\label{eq: K(i, j) bound by M}
\begin{aligned}
\E [K(i,j)] &= \left(\Delta^{(n)}\right)^2 \E \left[\int J^{(n),C}_i\left(x-B^{(n)}_{i-1}\right) J^{(n),C}_j\left(x-B^{(n)}_{j-1}\right) \d x \right] \\
& \leq M \Delta^{(n)} \E \left[\int J^{(n),C}_i\left(x-B^{(n)}_{i-1}\right) \d x \right] \leq \Delta^{(n)} M^2.
\end{aligned}
\end{equation}
Also note that for $i < j$, using the tower rule and the fact that $f^{(n)}$ is supported on a compactum, we can write
\begin{equation}
\label{eq: K(i, j) for i neq j}
\begin{aligned}
\E \left[ K(i, j) \right] &= \left(\Delta^{(n)}\right)^2 \, \E \left[ \int J^{(n),C}_i\left(x-B^{(n)}_{i-1}\right) f^{(n)}\left(S^{(n)}_{j-1}; x-B^{(n)}_{j-1}\right) \d x \right] \\
&\leq \left(\Delta^{(n)}\right)^2 \sup_{(b,y)\in\R^2} \left\|f^{(n)}(b,y)\right\|_{L^\infty}  \E \left[\int J^{(n),C}_i\left(x-B^{(n)}_{i-1}\right) \d x\right]  \leq \left(\Delta^{(n)}\right)^2 \sup_{(b,y)\in\R^2} \left\|f^{(n)}(b,y)\right\|^2_{L^\infty}.
\end{aligned}
\end{equation}
Finally, note that
\begin{equation}
\label{eq: int I(i, x) dx}
\E \int \Delta^{(n)} J^{(n),C}_i\left(x-B^{(n)}_{i-1}\right) \d x = \Delta^{(n)} \E \int f^{(n)}\left(S^{(n)}_{i-1}; x-B^{(n)}_{i-1}\right) \d x \leq \Delta^{(n)} \sup_{(b, y)\in\R^2} \left\|f^{(n)}(b,y)\right\|_{L^1}.
\end{equation}

\textbf{Induction proof.} We want to show by induction that
\begin{equation*}
S(n, k, \ell) = O(\ell^{2\ell}).
\end{equation*}
For $\ell=1$, using \eqref{eq: K(i, j) bound by M}, \eqref{eq: K(i, j) for i neq j} and $k\Delta^{(n)}\leq T$, we have
\[
\begin{aligned}
S(n, k, 1) &= \E \left[ \sum_{i_1, i_2=1}^k K(i_1, i_2) \right]= \E \left[ \sum_{\substack{i_1, i_2=1\\i_1 \neq i_2}}^k K(i_1, i_2)  +  \sum_{\substack{i_1, i_2=1\\i_1 = i_2}}^k K(i_1, i_2) \right]\\
&\leq \sum_{\substack{i_1, i_2=1\\i_1 \neq i_2}}^k \left(\Delta^{(n)}\right)^2 \sup_{(b,y)\in\R^2}\left\|f^{(n)}(b,y)\right\|^2_{L^1} + \sum_{\substack{i_1, i_2=1\\i_1 = i_2}}^k \Delta^{(n)} M^2 = O(1).
\end{aligned}
\]
Now we proceed to the induction step. Using \eqref{eq: expression for the 2l-th moment}, we can write
\[
\begin{aligned}
S(n, k, \ell+1) &= \E \left[ \sum_{i_1, \dots i_{2(\ell+1)}=1}^k \prod_{j=1}^{\ell+1} K(i_{2j-1}, i_{2j}) \right]= \E \left[ \sum_{i_1, \dots i_{2\ell}=1}^k \prod_{j=1}^{\ell} K(i_{2j-1}, i_{2j}) \sum_{i_{2\ell+1}, i_{2\ell+2}=1}^k K(i_{2\ell+1}, i_{2\ell+2})\right]\\
&= \E \left[ \sum_{i_1, \dots i_{2\ell}=1}^k \prod_{j=1}^{\ell} K(i_{2j-1}, i_{2j}) \left(\sum_{i\in I_1\cup I_2 \cup I_3 \cup I_4} K(i_{2\ell+1}, i_{2\ell+2})\right) \right],
\end{aligned}
\]
where $I_1\cup I_2 \cup I_3 \cup I_4$ is a partition of $\{1, \dots k\}^2$ which depends on $\mathcal J:=\{i_1, \dots i_{2\ell}\}$, such that
\[
\begin{aligned}
I_1 &= \{ (i_{2\ell+1}, i_{2\ell+2}): i_{2\ell+1}, i_{2\ell+2} \notin \mathcal J \text{ and } i_{2\ell+1} \neq i_{2\ell+2} \}, \\
I_2 &= \{ (i_{2\ell+1}, i_{2\ell+2}): i_{2\ell+1}, i_{2\ell+2} \notin \mathcal J \text{ and } i_{2\ell+1} = i_{2\ell+2} \}, \\
I_3 &= \{ (i_{2\ell+1}, i_{2\ell+2}): i_{2\ell+1} \in \mathcal J, i_{2\ell+2} \notin \mathcal J \text{ or } i_{2\ell+1} \notin J, i_{2\ell+2} \in J \}, \\
I_4 &= \{ (i_{2\ell+1}, i_{2\ell+2}): i_{2\ell+1}, i_{2\ell+2} \in \mathcal J \}.
\end{aligned}
\]
For $I_1$ and $I_2$, using the tower rule, \eqref{eq: K(i, j) bound by M}, and \eqref{eq: K(i, j) for i neq j}, we get that
\begin{align*}
\E \left[ \sum_{i_1, \dots i_{2\ell}=1}^k \prod_{j=1}^{\ell} K(i_{2j-1}, i_{2j}) \left(\sum_{i\in I_1} K(i_{2\ell+1}, i_{2\ell+2})\right) \right] &\leq S(n,k,\ell) \sum_{i\in I_1} \left(\Delta^{(n)}\right)^2 \sup_{(b,y)\in\R^2}\left\|f^{(n)}(b,y)\right\|^2_{L^2},\\
\E \left[ \sum_{i_1, \dots i_{2\ell}=1}^k \prod_{j=1}^{\ell} K(i_{2j-1}, i_{2j}) \left(\sum_{i\in I_2} K(i_{2\ell+1}, i_{2\ell+2})\right) \right] &\leq S(n,k,\ell) \sum_{i\in I_2} \Delta^{(n)} M^2.
\end{align*}
Furthermore, using the tower rule, \eqref{eq: int I(i, x) dx}, and \eqref{eq: K(i, j) bound by M}, we obtain for $I_3$
\[
\E \left[ \sum_{i_1, \dots i_{2\ell}=1}^k \prod_{j=1}^{\ell} K(i_{2j-1}, i_{2j}) \left(\sum_{i\in I_3} K(i_{2\ell+1}, i_{2\ell+2})\right) \right] \leq S(n,k,\ell) \sum_{i\in I_3} \Delta^{(n)} M \sup_{(b,y)\in\R^2}\left\|f^{(n)}(b,y)\right\|_{L^1}.
\]
Moreover, using \eqref{eq: K(i, j) bound by M} we obtain for $I_4$
\[
\begin{aligned}
\E \left[ \sum_{i_1, \dots i_{2\ell}=1}^k \prod_{j=1}^{\ell} K(i_{2j-1}, i_{2j}) \left(\sum_{i\in I_4} K(i_{2\ell+1}, i_{2\ell+2})\right) \right] &\leq S(n,k,\ell) \sum_{i\in I_4} \Delta^{(n)} M^2.
\end{aligned}
\]
Thus, it follows that
\begin{equation}
\label{eq: S(n,k,l) inductive inequality}
\begin{aligned}
&S(n, k, \ell+1) \leq S(n, k, \ell)\cdot\\
& \left\{ \sum_{i\in I_1} \left(\Delta^{(n)}\right)^2 \sup_{(b,y)\in\R^2}\left\|f^{(n)}(b,y)\right\|^2_{L^2} + \sum_{i\in I_3} \Delta^{(n)} M \sup_{(b,y)\in\R^2}\left\|f^{(n)}(b,y)\right\|_{L^1} + \sum_{i\in I_2\cup I_4} \Delta^{(n)} M^2\right\}.
\end{aligned}
\end{equation}
Using that $|I_1|= O(k^2)$, $|I_2|= O(k)$, $|I_3| = O(k\ell)$, $|I_4|= O(\ell^2)$, the constraint $k\Delta^{(n)} \leq T$, and \cref{assum: f}\ref{assum: f: sup of L^infty norm of f^n}, we obtain
\begin{equation}
\label{eq: estimate for the partition I}
\begin{aligned}
\sum_{i\in I_1} \left(\Delta^{(n)}\right)^2 \sup_{(b,y)\in\R^2}\left\|f^{(n)}(b,y)\right\|^2_{L^2} + \sum_{i\in I_2\cup I_4} \Delta^{(n)} M^2 + \sum_{i\in I_3} \Delta^{(n)} M \sup_{(b,y)\in\R^2}\left\|f^{(n)}(b,y)\right\|_{L^1}  = O(\ell^2).
\end{aligned}
\end{equation}
By the induction hypothesis, \eqref{eq: S(n,k,l) inductive inequality} and \eqref{eq: estimate for the partition I}, it follows that
\[
S(n,k,\ell+1) = O(\ell^{2\ell+2}).
\]

\textbf{Conclusion.} Since we consider $\ell=m$ finite, due to \eqref{eq: triangle inequality in the proof of the volume dynamics estimate} it follows that
\[
\E \left[ \left\|v^{(n)}_k\right\|^{m}_{L^2} \right] \lesssim \E\left[ \left\|v^{(n)}_0\right\|^{m}_{L^2} \right] + S(n, k, m/2) = O(1).
\]
Since the asymptotic estimate does not depend on $k$ and $n$, the bound holds uniformly in $k$ and $n$. \qed
\end{proof}

%%%%%%%%%%%%%%%%%%%%%%%%%%%%%%%%%%%%%%%%%%%%%%%%%%%%%%%%%%%%
%%%%%%%%%%%%%%%%%%%%%%%%%%%%%%%%%%%%%%%%%%%%%%%%%%%%%%%%%%%%
%%%%%%%%%%%%%%%%%%%%%%%%%%%%%%%%%%%%%%%%%%%%%%%%%%%%%%%%%%%%

\section{Auxiliary results}

\begin{theorem}[\cite{JacodLimitTheoremsStochastic2003}, Proposition IX.1.12]\label{thm:jacod}
For each $n\in\N$, consider a stochastic basis $\left(\Omega^n, \mathscr{F}^n, \mathbb {F}^n, P^n\right)$, on which there are defined an $\R^m$-valued, càdlàg process $X^{(n)}$ and a martingale $M^{(n)}$. Furthermore, let $M$ be a càdlàg adapted process on the space $D([0,T], \R^m)$, endowed with the usual augmentation of the filtration generated by the canonical process $X$. Let $D$ be a dense subset of $[0,T]$. Assume that:
\begin{enumerate}[label=(\roman*),topsep=0pt, itemsep=0pt]
\item the family $\left(M^{(n)}_t\right)_{n\in\N, t\in[0,T]}$ is uniformly integrable;
\item $X^{(n)}\Rightarrow X$, where $P=\mathcal{L}(X)$;
\item for all $t\in D$, the map $X \mapsto M_t(X)$ is $P$-a.s. continuous on $D([0,T], \R^m)$;
\item $M^{(n)}_t - M_t \big(X^{(n)}\big) \rightarrow 0$ in $P$-probability for all $t\in D$.
\end{enumerate}
Then the process $M\circ X$ is a martingale with respect to the canonical filtration generated by $X$.
\end{theorem}

\begin{theorem}[\cite{DaPratoStochasticEquationsInfinite2014}, Theorem 8.2]
\label{thm: representation theorem}

Let $U$ and $H$ be separable Hilbert spaces. Assume that $M$ is an $H$-valued, continuous, square-integrable martingale on a filtered probability space $(\Omega,\mathcal{F},\mathbb{F},\Q)$, starting from $M(0)=0$, with
\[
\langle M \rangle_t = \int_0^t \sigma(s) \sigma(s)' \d s, \quad t\in[0,T],\quad\text{and}\quad \E\left[\sup_{t\in[0,T]}\| M_t\|^2\right]<\infty,
\]
for a predictable $L_2(U, H)$-valued process $\sigma$. Then, there exists a filtered probability space $\left(\widetilde\Omega, \widetilde{\F}, \widetilde{\mathbb{F}}, \widetilde{\Q}\right)$ and a cylindrical Brownian motion $W$ with values in $U$, defined on $\left(\Omega\times\widetilde\Omega, \F\times \widetilde{\F}, \mathbb{F}\times\widetilde{\mathbb{F}},\Q\times \widetilde{\Q}\right)$, adapted to $\mathbb{F}\times \widetilde{\mathbb{F}}$, such that
\[
M(t, \omega, \widetilde{\omega})=\int_{0}^{t} \sigma(s, \omega, \widetilde{\omega}) \d W(s, \omega, \widetilde{\omega}), \quad t \in[0, T],\quad (\omega, \widetilde{\omega}) \in \Omega \times \widetilde{\Omega},
\]
where
\[
M(t, \omega, \widetilde{\omega})=M(t, \omega), \quad \sigma(t, \omega, \widetilde{\omega})=\sigma(t, \omega), \quad t \in[0, T],\quad (\omega, \widetilde{\omega}) \in \Omega \times \widetilde{\Omega}.
\]

\end{theorem}

%%%%%%%%%%%%%%%%%%%%%%%%%%%%%%%%%%%%%%%%%%%%%%%%
%%%%%%%%%%%%%%%%%%%%%%%%%%%%%%%%%%%%%%%%%%%%%%%%

\section{Empirical results}\label{app:empirics}

In this section, we report the ridge regression coefficients estimated for the linear model \eqref{eq: regression model} together with 95\% bootstrap confidence intervals for $N=100$ resamples. 

\begin{table}[htbp]
\centering
\caption{Ridge regression estimates and 95\% bootstrap CIs for MSFT}
\label{tab:MSFT}

\begin{tabular}{
l
S[table-format=-1.2]
@{\,[}
S[table-format=-1.2]
@{,\,}
S[table-format=-1.2]
@{]\hspace{2em}}
S[table-format=1.3]
@{\,[}
S[table-format=1.3]
@{,\,}
S[table-format=1.3]
@{]\hspace{2em}}
S[table-format=1.3]
@{\,[}
S[table-format=-1.3]
@{,\,}
S[table-format=1.3]
@{]}
}
\toprule
& \multicolumn{3}{c}{$\hat p^i_c$} & \multicolumn{3}{c}{$\hat p^i_b$} & \multicolumn{3}{c}{$\hat p^i_y$} \\
\midrule
$p^A$ & -6.59 & -7.74 & -5.38 & 0.016 & 0.013 & 0.019 & 0.025 & 0.013 & 0.038 \\
$p^B$ & -7.58 & -8.77 & -6.64 & 0.019 & 0.017 & 0.022 & 0.007 & -0.004 & 0.023 \\
\bottomrule
\end{tabular}

\vskip12pt

\begin{tabular}{
l
S[table-format=-1.3]
@{\,[}
S[table-format=-1.3]
@{,\,}
S[table-format=-1.3]
@{]\hspace{2em}}
S[table-format=1.3]
@{\,[}
S[table-format=-1.3]
@{,\,}
S[table-format=1.3]
@{]\hspace{2em}}
S[table-format=-1.3]
@{\,[}
S[table-format=-1.3]
@{,\,}
S[table-format=-1.3]
@{]}
}
\toprule
& \multicolumn{3}{c}{$\hat f_c^k$}
& \multicolumn{3}{c}{$\hat f_b^k$}
& \multicolumn{3}{c}{$\hat f_y^k$} \\
\midrule
$f^0$ & -3.501 & -6.007 & -1.650 & 0.009 &  0.005 & 0.015 & -0.346 & -0.399 & -0.297 \\
$f^1$ & -0.209 & -2.914 &  2.147 & 0.001 & -0.005 & 0.007 &  0.004 & -0.026 &  0.033 \\
$f^2$ & -0.336 & -1.666 &  1.057 & 0.001 & -0.003 & 0.004 & -0.010 & -0.028 &  0.005 \\
$f^3$ & -0.116 & -1.756 &  1.984 & 0.000 & -0.005 & 0.004 & -0.007 & -0.025 &  0.011 \\
$f^4$ & -0.408 & -1.961 &  1.425 & 0.001 & -0.003 & 0.005 & -0.022 & -0.045 &  0.011 \\
$f^5$ & -0.202 & -1.532 &  1.232 & 0.001 & -0.003 & 0.004 & -0.010 & -0.026 &  0.012 \\
$f^6$ & -0.269 & -2.116 &  1.539 & 0.001 & -0.004 & 0.005 &  0.007 & -0.010 &  0.026 \\
$f^7$ & -0.261 & -2.487 &  1.311 & 0.001 & -0.003 & 0.006 &  0.011 & -0.004 &  0.034 \\
$f^8$ & -0.112 & -1.882 &  1.637 & 0.000 & -0.004 & 0.005 &  0.013 & -0.005 &  0.033 \\
$f^9$ & -0.043 & -2.753 &  2.268 & 0.000 & -0.006 & 0.007 &  0.001 & -0.026 &  0.027 \\
\bottomrule
\end{tabular}

\vskip12pt

\begin{tabular}{
l
S[table-format=-2.2]
@{\,[}
S[table-format=-2.2]
@{,\,}
S[table-format=-2.2]
@{]\hspace{1em}}
S[table-format=-1.3]
@{\,[}
S[table-format=-1.3]
@{,\,}
S[table-format=1.3]
@{]\hspace{1em}}
S[table-format=-1.3]
@{\,[}
S[table-format=-1.3]
@{,\,}
S[table-format=1.3]
@{]}
}
\toprule
& \multicolumn{3}{c}{$\hat g_c^k$}
& \multicolumn{3}{c}{$\hat g_b^k$}
& \multicolumn{3}{c}{$\hat g_y^k$} \\
\midrule
$g^0$ & -38.72 & -52.90 & -24.56 & 0.096 &  0.061 & 0.130 &  0.154 &  0.092 & 0.264 \\
$g^1$ & -32.34 & -74.75 &  21.22 & 0.081 & -0.048 & 0.184 &  0.037 & -0.001 & 0.130 \\
$g^2$ & -14.62 & -18.06 & -11.84 & 0.036 &  0.030 & 0.045 & -0.002 & -0.004 & 0.003 \\
$g^3$ & -21.30 & -25.43 & -17.06 & 0.053 &  0.042 & 0.063 &  0.000 & -0.004 & 0.005 \\
$g^4$ &  -9.90 & -20.37 &  24.96 & 0.025 & -0.059 & 0.051 &  0.009 & -0.006 & 0.096 \\
$g^5$ & -11.88 & -14.86 &  -9.34 & 0.030 &  0.024 & 0.037 & -0.001 & -0.004 & 0.009 \\
$g^6$ & -18.64 & -22.39 & -14.06 & 0.046 &  0.035 & 0.056 &  0.000 & -0.004 & 0.012 \\
$g^7$ & -23.65 & -32.68 & -13.33 & 0.059 &  0.034 & 0.081 &  0.006 & -0.005 & 0.030 \\
$g^8$ & -14.64 & -20.59 & -11.08 & 0.037 &  0.028 & 0.051 & -0.001 & -0.003 & 0.010 \\
$g^9$ & -53.79 & -84.08 & -12.67 & 0.133 &  0.034 & 0.206 &  0.024 & -0.011 & 0.088 \\
\bottomrule
\end{tabular}

\end{table}

\begin{table}

\centering
\caption{Ridge regression estimates and 95\% bootstrap CIs for VZ}
\label{tab:VZ}

\begin{tabular}{
l
S[table-format=-1.2]
@{\,[}
S[table-format=-1.2]
@{,\,}
S[table-format=-1.2]
@{]\hspace{2em}}
S[table-format=1.3]
@{\,[}
S[table-format=1.3]
@{,\,}
S[table-format=1.3]
@{]\hspace{2em}}
S[table-format=1.3]
@{\,[}
S[table-format=-1.3]
@{,\,}
S[table-format=1.3]
@{]}
}
\toprule
& \multicolumn{3}{c}{$\hat p_c^i$} & \multicolumn{3}{c}{$\hat p_b^i$} & \multicolumn{3}{c}{$\hat p_y^i$} \\
\midrule
$p^A$ & -3.26 & -4.02 & -2.45 & 0.084 & 0.063 & 0.103 & 0.017 & 0.012 & 0.024 \\
$p^B$ & -5.83 & -6.85 & -4.76 & 0.150 & 0.123 & 0.176 & 0.001 & -0.004 & 0.007 \\
\bottomrule
\end{tabular}

\vskip12pt

\begin{tabular}{
l
S[table-format=-2.2]
@{\,[}
S[table-format=-2.2]
@{,\,}
S[table-format=-2.2]
@{]\hspace{2em}}
S[table-format=-1.3]
@{\,[}
S[table-format=-1.3]
@{,\,}
S[table-format=-1.3]
@{]\hspace{2em}}
S[table-format=-1.3]
@{\,[}
S[table-format=-1.3]
@{,\,}
S[table-format=-1.3]
@{]}
}
\toprule
& \multicolumn{3}{c}{$\hat f_c^k$} & \multicolumn{3}{c}{$\hat f_b^k$} & \multicolumn{3}{c}{$\hat f_y^k$} \\
\midrule
$f^0$ & -23.78 & -27.97 & -19.36 &  0.617 &  0.504 &  0.725 & -0.312 & -0.349 & -0.273 \\
$f^1$ &   0.29 &  -3.65 &   4.41 & -0.007 & -0.113 &  0.094 & -0.005 & -0.037 &  0.027 \\
$f^2$ &  -1.62 &  -7.11 &   3.49 &  0.042 & -0.089 &  0.183 & -0.021 & -0.057 &  0.019 \\
$f^3$ &  -8.63 & -19.14 &   3.35 &  0.223 & -0.083 &  0.494 & -0.094 & -0.190 & -0.008 \\
$f^4$ &   3.97 &  -6.66 &  16.52 & -0.103 & -0.425 &  0.170 &  0.055 & -0.023 &  0.142 \\
$f^5$ &   4.10 &  -8.02 &  13.72 & -0.106 & -0.352 &  0.205 &  0.038 & -0.020 &  0.099 \\
$f^6$ &   5.02 &  -3.25 &  13.62 & -0.130 & -0.351 &  0.082 &  0.072 &  0.029 &  0.122 \\
$f^7$ &   4.76 &   0.84 &   8.52 & -0.123 & -0.219 & -0.023 &  0.054 &  0.026 &  0.077 \\
$f^8$ &   5.10 &   2.81 &   8.10 & -0.132 & -0.209 & -0.073 &  0.052 &  0.025 &  0.073 \\
$f^9$ &   4.40 &   2.13 &   6.64 & -0.114 & -0.171 & -0.055 &  0.044 &  0.026 &  0.059 \\
\bottomrule
\end{tabular}

\vskip12pt

\begin{tabular}{
l
S[table-format=-4.2]
@{\,[}
S[table-format=-4.2]
@{,\,}
S[table-format=-3.2]
@{]\hspace{2em}}
S[table-format=2.2]
@{\,[}
S[table-format=2.2]
@{,\,}
S[table-format=2.2]
@{]\hspace{2em}}
S[table-format=1.3]
@{\,[}
S[table-format=-1.3]
@{,\,}
S[table-format=1.3]
@{]}
}
\toprule
& \multicolumn{3}{c}{$\hat g_c^k$} & \multicolumn{3}{c}{$\hat g_b^k$} & \multicolumn{3}{c}{$\hat g_y^k$} \\
\midrule
$g^0$ &  -97.05 & -120.66 &  -76.77 &  2.48 &  1.96 &  3.09 &  0.163 &  0.116 &  0.220 \\
$g^1$ & -112.48 & -134.33 &  -90.51 &  2.89 &  2.32 &  3.45 &  0.067 &  0.031 &  0.099 \\
$g^2$ & -304.18 & -393.12 & -237.47 &  7.82 &  6.10 & 10.10 &  0.047 & -0.021 &  0.104 \\
$g^3$ & -883.37 & -1100.07 & -675.36 & 22.68 & 17.36 & 28.23 &  0.201 &  0.082 &  0.367 \\
$g^4$ & -931.98 & -1077.84 & -771.29 & 23.94 & 19.83 & 27.69 &  0.126 &  0.001 &  0.244 \\
$g^5$ & -883.11 & -1025.21 & -699.85 & 22.67 & 17.96 & 26.33 &  0.102 & -0.041 &  0.246 \\
$g^6$ & -512.23 &  -632.04 & -356.97 & 13.15 &  9.17 & 16.23 &  0.026 & -0.055 &  0.116 \\
$g^7$ & -260.27 &  -361.27 & -162.35 &  6.68 &  4.17 &  9.28 & -0.022 & -0.050 &  0.011 \\
$g^8$ &  -95.18 &  -115.83 &  -75.98 &  2.44 &  1.95 &  2.97 &  0.007 & -0.006 &  0.020 \\
$g^9$ &  -58.03 &   -68.52 &  -45.39 &  1.49 &  1.17 &  1.76 &  0.001 & -0.007 &  0.009 \\
\bottomrule
\end{tabular}

\end{table}

\end{appendix}

\begin{acknowledgements}
Horst and Kreher gratefully acknowledge financial support from DFG CRC/TRR 388 “Rough Analysis, Stochastic Dynamics and Related Fields” - Project ID 516748464 (Project B02).
\end{acknowledgements}

\bibliographystyle{spmpsci}      % mathematics and physical sciences
\bibliography{bibliography.bib}   % name your BibTeX data base

@article{CHEN201889,
author = {Chen, Yuanyuan and Gao, Xuefeng Gao and Li, Duan},
title = {Optimal order execution using hidden orders},
journal = {Journal of Economic Dynamics and Control},
volume = {94},
pages = {89-116},
year = {2018},
}

@article{CH,
  author  = {Cebiroglu, Gökhan and Horst, Ulrich},
  title   = {Optimal Order Display in Limit Order Markets with Liquidity Competition},
  journal = {Journal of Economic Dynamics and Control},
  year    = {2015},
  volume  = {58},
  pages   = {81--100}
}

@article{EsserMoench2007,
  author  = {Esser, André and M{\"o}nch, Bernd},
  title   = {The Navigation of an Iceberg: The Optimal Use of Hidden Orders},
  journal = {Finance Research Letters},
  year    = {2007},
  volume  = {4},
  pages   = {68--81}
}

@article{HollifieldMillerSandasSlive2006,
  author  = {Hollifield, Burton and Miller, Robert A. and Sand{\aa}s, Patrik and Slive, Joshua},
  title   = {Estimating the Gains from Trade in Limit-Order Markets},
  journal = {The Journal of Finance},
  year    = {2006},
  volume  = {61},
  pages   = {2753--2804}
}

@article{CaoHanschWang2009,
  author  = {Cao, Charles and Hansch, Oliver and Wang, Xiaoyan},
  title   = {The Information Content of an Open Limit-Order Book},
  journal = {Journal of Futures Markets},
  year    = {2009},
  volume  = {29},
  pages   = {16--41}
}

@article{HautschHuang2012,
  author  = {Hautsch, Nikolaus and Huang, Ruihong},
  title   = {The Market Impact of a Limit Order},
  journal = {Journal of Economic Dynamics and Control},
  year    = {2012},
  volume  = {36},
  pages   = {501--522}
}

@article{Griffiths2000,
  author  = {Griffiths, Mark D. and Smith, Brian F. and Turnbull, Stuart M. and White, R. W.},
  title   = {The Costs and Determinants of Order Aggressiveness},
  journal = {Journal of Financial Economics},
  year    = {2000},
  volume  = {56},
  pages   = {65--88}
}

@article{Biais1995,
  author  = {Biais, Bruno and Hillion, Pierre and Spatt, Chester},
  title   = {An Empirical Analysis of the Limit Order Book and the Order Flow in the Paris Bourse},
  journal = {The Journal of Finance},
  year    = {1995},
  volume  = {50},
  pages   = {1655--1689}
}

@article{Ranaldo2004,
  author  = {Ranaldo, Angelo},
  title   = {Order Aggressiveness in Limit Order Book Markets},
  journal = {Journal of Financial Markets},
  year    = {2004},
  volume  = {7},
  pages   = {53--74}
}

@Article{Farmer,
  author    = {Farmer, J.~Doyne and Gillemot Laszlo and Fabrizio Lillo and Mike Szabolcs and Anindya Sen},
  title     = {What Really Causes Large Price Changes?},
  journal   = {Quantitative Finance},
  year      = {2004},
  volume    = {4(4)},
  pages     = {383-397},
}

@misc{ZhengZ,
	author = {Zheng, Z.},
	title = {Stochastic {Stefan} Problems: Existence, Uniqueness and Modeling of Market Limit Orders},
	howpublished = {PhD thesis, Graduate College of the University of Illinois at Urbana-Champaign},
	year = {2012},
}

@article{Cont2,
    Author = {Rama {Cont} and Sasha {Stoikov} and Rishi {Talreja}},
    Title = {{A Stochastic Model for Order Book Dynamics}},
    FJournal = {{Operations Research}},
    Journal = {{Oper. Res.}},
    Volume = {58},
    Number = {3},
    Pages = {549--563},
    Year = {2010}
}

@article{BayerFunctionalLimitTheorem2017,
  title = {A Functional Limit Theorem for Limit Order Books with State Dependent Price Dynamics},
  author = {Bayer, Christian and Horst, Ulrich and Qiu, Jinniao},
  year = {2017},
  journal = {The Annals of Applied Probability},
  shortjournal = {Ann. Appl. Probab.},
  volume = {27},
  number = {5},
  pages={2753--2806},
}

@article{ContPriceDynamicsMarkovian2013,
  title = {Price {{Dynamics}} in a {{Markovian Limit Order Market}}},
  author = {Cont, Rama and de Larrard, Adrien},
  options = {useprefix=true},
  year = {2013},
  journal = {SIAM Journal on Financial Mathematics},
  shortjournal = {SIAM J. Finan. Math.},
  volume = {4},
  number = {1},
  pages = {1--25},
}

@book{DaPratoStochasticEquationsInfinite2014,
  title = {Stochastic {{Equations}} in {{Infinite Dimensions}}},
  author = {Da Prato, Giuseppe and Zabczyk, Jerzy},
  year = {2014},
  publisher = {{Cambridge University Press}},
  location = {{Cambridge}},
}

@article{GaoHydrodynamicLimitOrderbook2018,
  title = {Hydrodynamic {{Limit}} of {{Order-book Dynamics}}},
  author = {Gao, Xuefeng and Deng, S.J.},
  year = {2018},
  journal = {Probability in the Engineering and Informational Sciences},
  shortjournal = {Prob. Eng. Inf. Sci.},
  volume = {32},
  number = {1},
  pages = {96--125},
}

@article{HorstDiffusionApproximationLimit2019,
  title = {A Diffusion Approximation for Limit Order Book Models},
  author = {Horst, Ulrich and Kreher, Dörte},
  year = {2019},
  journal = {Stochastic Processes and their Applications},
  shortjournal = {Stochastic Processes and their Applications},
  volume = {129},
  number = {11},
  pages = {4431--4479},
}

@article{HorstLawLargeNumbers2017,
  title = {A {{Law}} of {{Large Numbers}} for {{Limit Order Books}}},
  author = {Horst, Ulrich and Paulsen, Michael},
  year = {2017},
  journal = {Mathematics of Operations Research},
  shortjournal = {Mathematics of OR},
  volume = {42},
  number = {4},
  pages = {1280--1312},
}

@article{ContPriceImpactOrder2014,
  title = {The {{Price Impact}} of {{Order Book Events}}},
  author = {Cont, R. and Kukanov, A. and Stoikov, S.},
  year = {2014},
  journal = {Journal of Financial Econometrics},
  volume = {12},
  number = {1},
  pages = {47--88},
}

@article{Keller-ResselStefantypeStochasticMoving2016,
  title = {A {{Stefan-type}} Stochastic Moving Boundary Problem},
  author = {{Keller-Ressel}, Martin and M{\"u}ller, Marvin S.},
  year = {2016},
  month = dec,
  journal = {Stochastics and Partial Differential Equations: Analysis and Computations},
  volume = {4},
  number = {4},
  pages = {746--790},
}

@article{HorstScalingLimitLimit2019,
  title = {A {{Scaling Limit}} for {{Limit Order Books Driven}} by {{Hawkes Processes}}},
  author = {Horst, Ulrich and Xu, Wei},
  year = {2019},
  journal = {SIAM Journal on Financial Mathematics},
  shortjournal = {SIAM J. Finan. Math.},
  volume = {10},
  number = {2},
  pages = {350--393},
}

@article{HorstSecondOrderApproximations2018,
  title = {Second Order Approximations for Limit Order Books},
  author = {Horst, Ulrich and Kreher, D\"orte},
  year = {2018},
  journal = {Finance and Stochastics},
  shortjournal = {Finance Stoch.},
  volume = {22},
  number = {4},
  pages = {827--877},
}

@article{HorstWeakLawLarge2017,
  title = {A {{Weak Law}} of {{Large Numbers}} for a {{Limit Order Book Model}} with {{Fully State Dependent Order Dynamics}}},
  author = {Horst, Ulrich and Kreher, D\"orte},
  year = {2017},
  journal = {SIAM Journal on Financial Mathematics},
  shortjournal = {SIAM J. Finan. Math.},
  volume = {8},
  number = {1},
  pages = {314--343},
}

@book{JacodLimitTheoremsStochastic2003,
  title = {Limit {{Theorems}} for {{Stochastic Processes}}},
  author = {Jacod, Jean and Shiryaev, Albert N.},
  options = {useprefix=true},
  year = {2003},
  publisher = {{Springer Berlin Heidelberg}},
  location = {{Berlin, Heidelberg}},
  editorb = {Chenciner, A. and de la Harpe, P. and Hörmander, L. and Lebeau, G. and Sinai, Ya G. and Vershik, A. and Chern, S. S. and Hirzebruch, F. and Knus, M.-A. and Ratner, M. and Sloane, N. J. A. and Waldschmidt, M. and Eckmann, B. and Hitchin, N. and Kupiainen, A. and Serre, D. and Totaro, B. and Berger, M. and Coates, J. and Varadhan, S. R. S.},
  editorbtype = {redactor},
}

@article{JedidiStabilityPriceScaling2013,
  title={On the Stability and Price Scaling Limit of a Hawkes Process-Based Order Book Model},
  author={Aymen Jedidi and Fr{\'e}d{\'e}ric Abergel},
  journal={Capital Markets: Market Microstructure eJournal},
  year={2013}
}

@InProceedings{WalshIntroductionStochasticPartial1986,
author="Walsh, John B.",
editor="Hennequin, P. L.",
title="An introduction to stochastic partial differential equations",
booktitle="{\'E}cole d'{\'E}t{\'e} de Probabilit{\'e}s de Saint Flour XIV - 1984",
year="1986",
publisher="Springer Berlin Heidelberg",
address="Berlin, Heidelberg",
pages="265--439",
}

@article{HuangErgodicityDiffusivityMarkovian2015a,
author = {Huang, Weibing and Rosenbaum, Mathieu},
title = {Ergodicity and Diffusivity of Markovian Order Book Models: A General Framework},
journal = {SIAM Journal on Financial Mathematics},
volume = {8},
number = {1},
pages = {874-900},
year = {2017},
}

@article{DonoghueTheoremMaurin1964,
  title = {On a Theorem of {{K}}. {{Maurin}}},
  author = {Donoghue, W.},
  year = {1964},
  journal = {Studia Mathematica},
  volume = {24},
  number = {1},
  pages = {1--5},
}

@article{VeraarNONAUTONOMOUSSTOCHASTICCAUCHY,
  title={Non-autonomous Stochastic Cauchy Problems in Banach Spaces},
  author={Veraar, MC and Zimmerschied, Jan},
  journal={Studia Mathematica},
  volume = {185},
  number = {1},
  pages = {1--34},
  year={2008},
  publisher={Polish Academy of Sciences, Institute of Mathematics}
}

@article{KreherJumpDiffusionApproximation2022,
author = {Kreher, D\"{o}rte and Milbradt, Cassandra},
title = {Jump Diffusion Approximation for the Price Dynamics of a Fully State Dependent Limit Order Book Model},
journal = {SIAM Journal on Financial Mathematics},
volume = {14},
number = {1},
pages = {1--51},
year = {2023},
}

@article{hambly2020limit,
  title={Limit Order Books, Diffusion Approximations and Reflected SPDEs: from Microscopic to Macroscopic Models},
  author={Hambly, Ben and Kalsi, Jasdeep and Newbury, James},
  journal={Applied Mathematical Finance},
  volume={27},
  number={1-2},
  pages={132--170},
  year={2020},
  publisher={Taylor \& Francis}
}

@article{marinelli2016,
title = {On the maximal inequalities of Burkholder, Davis and Gundy},
journal = {Expositiones Mathematicae},
volume = {34},
number = {1},
pages = {1-26},
year = {2016},
author = {Carlo Marinelli and Michael Röckner},
}

\end{document}